\documentstyle[preprint,aps,eqsecnum,epsfig]{revtex}

\newcommand{\bra}[1]{\mbox{$\langle {#1} |$}}
\newcommand{\ket}[1]{\mbox{$| {#1} \rangle$}}
\newcommand{\bracket}[2]{\mbox{$\langle {#1} \mid {#2} \rangle$}} 
\newcommand{\ketbra}[2]{\mbox{$| {#1} \rangle\langle {#2} |$}} 
\newcommand{\melt}[3]{\mbox{$\langle {#1} | {#2} | {#3} \rangle$}}
\newcommand{\Bra}[1]{\mbox{$\langle {#1} \|$}}
\newcommand{\Ket}[1]{\mbox{$\| {#1} \rangle$}}
\newcommand{\Bracket}[2]{\mbox{$\langle {#1} \parallel {#2} \rangle$}} 
\newcommand{\Ketbra}[2]{\mbox{$\| {#1} \rangle\langle {#2} \|$}} 
\newcommand{\Melt}[3]{\mbox{$\langle {#1} \| {#2} \| {#3} \rangle$}}

\newcommand{\ketl}[1]{\mbox{$| {#1} \rangle$}}
\newcommand{\bracketl}[2]{\mbox{$\langle {#1} \mid {#2} \rangle$}} 
\newcommand{\ketbral}[2]{\mbox{$| {#1} \rangle\langle {#2} |$}} 
\newcommand{\meltl}[3]{\mbox{$\langle {#1} | {#2} | {#3} \rangle$}}

\newcommand{\Ketl}[1]{\mbox{$\| {#1} \rangle$}}
\newcommand{\Bracketl}[2]{\mbox{$\langle {#1} \parallel {#2} \rangle$}} 
\newcommand{\Ketbral}[2]{\mbox{$\| {#1} \rangle\langle {#2} \|$}} 
\newcommand{\Meltl}[3]{\mbox{$\langle {#1} \| {#2} \| {#3} \rangle$}}

\newcommand\Span{\mathop{\rm span}}

\newcommand{\nosum}{\mbox{$\mathop{\not\kern-0.4em\sum}$}}
\newcommand{\rhalpha}{\mbox{$\rho_{\alpha}$}}
\newcommand{\rhomega}{\mbox{$\rho_{\omega}$}}
\newcommand{\ralpha}{\mbox{$r_{\alpha}$}}
\newcommand{\romega}{\mbox{$r_{\omega}$}}
\newcommand{\system}{\mbox{$S$}}
\newcommand{\Hilbert}{\mbox{$H$}}
\newcommand{\HilbertS}{\mbox{$H_S$}}

\newcommand{\hilbert}{\mbox{${\cal H}$}}
\newcommand{\hilbertbar}{\underline{\hilbert}}
\newcommand{\hilbertsubd}{\mbox{${\cal H}_{d}$}}
\newcommand{\hilbertif}{\mbox{${\cal H}_{\alpha\omega}$}}
\newcommand{\hilbertsubS}{\mbox{${\cal H}_{{\scriptstyle {\cal S}}}$}}
\newcommand{\hilbertsubSsup}[1]{
     \mbox{${\cal H}_{{\scriptstyle {\cal S}}}^{ {#1} }$}}

\newcommand{\histories}{\mbox{${\cal U}$}}

\newcommand{\temporalsub}[1]{\mbox{${\cal T}_{#1}$}}
\newcommand{\linearops}{\mbox{${\cal L}$}}

\newcommand{\projlattice}{\mbox{${\cal P}$}}
\newcommand{\consistentset}{\mbox{${\cal S}$}}

\newcommand{\consistentsetsubd}{\mbox{${\cal S}_{d}$}}
\newcommand{\consistentsetsub}[1]{\mbox{${\cal S}_{#1}$}}

\newcommand{\ball}{\mbox{${\cal B}$}}
\newcommand{\op}{\mbox{${\cal O}$}}

\newcommand{\rep}{\mbox{${\cal R}$}}
\newcommand{\repbar}{\mbox{$\overline{{\cal R}}$}}
\newcommand{\projrep}{\mbox{${\cal R}_{{\cal P}}$}}
\newcommand{\projrepbar}{\mbox{$\overline{{\cal R}}_{{\cal P}}$}}
\newcommand{\classoprep}{\mbox{${\cal R}_{{\cal C}}$}}
\newcommand{\classoprepbar}{\mbox{$\overline{{\cal R}}_{{\cal C}}$}}

\newcommand{\decoheringsubd}{\mbox{${\cal D}_{d}$}}
\newcommand{\decoheringsub}[1]{\mbox{${\cal D}_{ {#1} }$}}
\newcommand{\consistentops}{\mbox{${\cal O}_{d}$}}

\newcommand{\consistentopsets}{\mbox{${\cal Q}_{d}$}}

\newcommand{\consistentsubd}{\mbox{${\cal C}_{d}$}}
\newcommand{\consistentsub}[1]{\mbox{${\cal C}_{ {#1} }$}}
\newcommand{\nullspace}{\mbox{${\cal N}$}}
\newcommand{\nullspacesubd}{\mbox{${\cal N}_{d}$}}
\newcommand{\nullspaceperp}{\mbox{${\cal N}^{\perp}$}}
\newcommand{\nullspaceperpsubd}{\mbox{${\cal N}^{\perp}_{d}$}}

\newcommand{\nullspaceperpif}{\mbox{${\cal N}^{\perp}_{\alpha\omega}$}}
\newcommand{\nullspaceif}{\mbox{${\cal N}_{\alpha\omega}$}}

\newcommand{\factorspacesubd}{\mbox{$\hilbert/\nullspacesubd$}}

\newcommand{\OneH}{\mbox{$1_{{\scriptscriptstyle H  }}$}}
\newcommand{\Oneh}{\mbox{$1_{{\scriptscriptstyle {\cal H}  }}$}}
\newcommand{\Onewiggle}{\mbox{$\tilde{1}$}}
\newcommand{\OnehS}{
     \mbox{$1_{{\scriptscriptstyle {\cal H}_{\cal S}}}$}}

\newcommand{\abssubh}[1]{\mbox{$|{#1}|_{{\scriptscriptstyle {\cal H}}}$}}
\newcommand{\abssubhbar}[1]{
      \mbox{$|{#1}|_{{\scriptscriptstyle \underline{ {\cal H} }}}$}}
\newcommand{\abssubhbarsup}[2]{
   \mbox{$|{#1}|_{{\scriptscriptstyle\underline{ {\cal H} }}}^{ {#2} }$}}

\newcommand{\betasub}[1]{\mbox{$ \beta_{#1} $}}

\newcommand{\tsub}[1]{\mbox{$ t_{#1} $}}

\newcommand{\h}{\mbox{$h$}}

\newcommand{\Projsub}[1]{\mbox{$P_{\mbox{\scriptsize{${#1}$}}}$}}
\newcommand{\Projsupb}[2]{\mbox{$P^{{#1}}_{\mbox{\scriptsize{${#2}$}}}$}}

\newcommand{\tr}[1]{\mbox{${\rm tr}[ {#1} ]$}}
\newcommand{\Tr}{\mbox{${\rm tr\, }$}}

\newcommand{\trif}[2]{\mbox{${\rm tr}[\rhomega {#1}^{\dag} \rhalpha {#2} ]$}} 

\newcommand{\trH}{\mbox{$\mathop{\rm tr}
     \limits_{\scriptscriptstyle H}$}}
\newcommand{\trHH}{\mbox{$\mathop{\rm tr}
     \limits_{\scriptscriptstyle H\otimes H}$}}  
  
\newcommand{\trh}{\mbox{$\mathop{\rm tr}
     \limits_{\scriptscriptstyle {\cal H} }$}}

\newcommand{\decoh}[2]{\mbox{$d({#1},{#2})$}}
\newcommand{\dcanonical}{\mbox{$d_{\alpha\omega}$}}
\newcommand{\Dcanonical}{\mbox{$D_{\alpha\omega}$}}
\newcommand{\dcanonicalsup}[1]{\mbox{$d_{\alpha\omega}^{ {#1} }$}}
\newcommand{\Dcanonicalsup}[1]{\mbox{$D_{\alpha\omega}^{ {#1} }$}}

\newcommand{\decohif}[2]{\mbox{$d_{\alpha\omega}({#1},{#2})$}}
\newcommand{\dhS}{
     \mbox{$d|_{{\scriptscriptstyle {\cal H}_{\cal S}}}$}}

\def\gtwid{\mathrel{\raise.3ex\hbox{$>$\kern-.75em\lower1ex\hbox{$\sim$}}}}  
\def\ltwid{\mathrel{\raise.3ex\hbox{$<$\kern-.75em\lower1ex\hbox{$\sim$}}}}

\def\Complex{{\mathchoice
{\setbox0=\hbox{$\displaystyle\rm C$}\hbox{\hbox to0pt
{\kern0.4\wd0\vrule height0.9\ht0\hss}\box0}}
{\setbox0=\hbox{$\textstyle\rm C$}\hbox{\hbox to0pt
{\kern0.4\wd0\vrule height0.9\ht0\hss}\box0}}
{\setbox0=\hbox{$\scriptstyle\rm C$}\hbox{\hbox to0pt
{\kern0.4\wd0\vrule height0.9\ht0\hss}\box0}}
{\setbox0=\hbox{$\scriptscriptstyle\rm C$}\hbox{\hbox to0pt
{\kern0.4\wd0\vrule height0.9\ht0\hss}\box0}}}}
\def\Co{{\mathchoice
{\setbox0=\hbox{$\displaystyle\rm C$}\hbox{\hbox to0pt
{\kern0.4\wd0\vrule height0.9\ht0\hss}\box0}}
{\setbox0=\hbox{$\textstyle\rm C$}\hbox{\hbox to0pt
{\kern0.4\wd0\vrule height0.9\ht0\hss}\box0}}
{\setbox0=\hbox{$\scriptstyle\rm C$}\hbox{\hbox to0pt
{\kern0.4\wd0\vrule height0.9\ht0\hss}\box0}}
{\setbox0=\hbox{$\scriptscriptstyle\rm C$}\hbox{\hbox to0pt
{\kern0.4\wd0\vrule height0.9\ht0\hss}\box0}}}}
\def\Real{{\mathchoice
{\setbox0=\hbox{$\displaystyle\rm R$}\hbox{\hbox to0pt
{\kern0.4\wd0\vrule height0.9\ht0\hss}\box0}}
{\setbox0=\hbox{$\textstyle\rm R$}\hbox{\hbox to0pt
{\kern0.4\wd0\vrule height0.9\ht0\hss}\box0}}
{\setbox0=\hbox{$\scriptstyle\rm R$}\hbox{\hbox to0pt
{\kern0.4\wd0\vrule height0.9\ht0\hss}\box0}}
{\setbox0=\hbox{$\scriptscriptstyle\rm R$}\hbox{\hbox to0pt
{\kern0.4\wd0\vrule height0.9\ht0\hss}\box0}}}}

\renewcommand{\linearops}[1]{\mbox{${\cal L}({#1})$}}
\renewcommand{\projlattice}[1]{\mbox{${\cal P}({#1})$}}

\def\be{\begin{equation}}
\def\ee{\end{equation}}
\def\bea{\begin{eqnarray}}
\def\eea{\end{eqnarray}}

\begin{document}

\draft
\tighten

\preprint{\vbox{\hfill ALBERTA-THY-04-97 \\
          \vbox{\hfill UCSB-TH-95-10} \\
          \vbox{\hfill quant-ph/9704031} \\
          \vbox{\hfill May 1996} \\
          \vbox{\vskip0.50in}
         }}

\title{The Geometry of Consistency: \\
       Decohering Histories in Generalized Quantum Theory}

\author{David A. Craig\footnote{CITA National Research Fellow.  
E-mail: dcraig@hawking.phys.ualberta.ca}}
%dac@cosmic.physics.ucsb.edu}}
%E--mail: {\tt dac@cosmic.physics.ucsb.edu}}
\vskip .13 in
%\address{CIAR Cosmology Program, Institute for Theoretical Physics \\
\address{Theoretical Physics Institute,
         Department of Physics, University of Alberta \\
         412 Avadh Bhatia Laboratory, Edmonton, Alberta T6G 2J1\\ 
         and \\
         Department of Physics, University of California,
         Santa Barbara, CA 93106-9530
         \\ \vskip0.5in}

\date{\today}

\maketitle

\begin{abstract}
The geometry of decoherence in generalized ``consistent histories" 
quantum theory is explored, revealing properties of the theory
that are independent of any particular application of it.  
%(Conventional quantum mechanics is a special case of this study.)  
Attention is focussed on the case of quantum mechanics in a 
finite dimensional Hilbert space \Hilbert.  
It is shown how the domain of definition of a general decoherence 
functional naturally extends to the entire space of linear operators
$\hilbert = \linearops{\Hilbert}$ on \Hilbert, on which the decoherence
functional is an Hermitian form.  This identification makes manifest a
number of structural properties of decoherence functionals.  For example,
a bound on the maximum number of histories in a consistent set is
determined.  In the case of  the ``canonical" decoherence functional 
with boundary conditions at the initial and final 
times, it is shown that the maximum number of histories with non-zero 
probability in any decohering set is $r_{\alpha}r_{\omega}$, 
where $r_{\alpha}$ and $r_{\omega}$ are the ranks of the initial 
and final density operators.  
This is one reason that some coarse graining is generally necessary for
decoherence.  When the decoherence functional is positive -- as in, for 
instance, conventional quantum mechanics on a Hilbert space \HilbertS, 
where the histories of a closed physical system \system\ are represented by
``class operators" in $\hilbert = \linearops{\HilbertS}$ -- it %actually
defines a semi-inner product in \hilbert.  This shows that consistent
sets of histories are just orthogonal sets in this inner product.
It further implies the existence in general of Cauchy-Schwarz and
triangle inequalities for positive decoherence functionals.
The geometrical significance of the important ILS theorem classifying
all possible decoherence functionals is illuminated, and a version
of the ILS theorem for decoherence functionals on class operators
is given.  Additionally, the class of history operators consistent 
according to a given decoherence functional is found (most explicitly
for positive decoherence functionals), and the problem of determining all 
the decoherence functionals according to which a given set of histories is
consistent is addressed.  In particular, it is shown how to construct 
explicitly these decoherence functionals,  thus showing how the
decoherence functional of a closed quantum system is constrained 
by observations.
Finally, the conditions 
under which a general decoherence functional on class operators 
is canonical are determined.
%The issue of equivalences between boundary
%conditions at one and two times is discussed.  (This equivalence is
%complete in classical physics.)  
%Further, the relation between physical 
%histories in \HilbertS\ and elements of \hilbert\ is explored.   
%The paper concludes with a brief discussion of the question of experimental 
%determination of the decoherence functional of our universe.
More generally, the ``geometric" point of view here developed 
supplies a powerful unified language with which to solve 
problems in generalized quantum theory.

\end{abstract}

%\pacs{}

\newpage

\tableofcontents

%\newpage
\vskip 1.5in

\section{Introduction}
\label{sec:introgoc}

%tradition of algebraic analysis (deconstruction) 
%of the structure of quantum theory.

This paper studies 
the geometry of decoherence in finite dimensional
Hilbert spaces, with the aim of illuminating the basic formal
structures of the quantum mechanics of closed systems.   
The properties studied are those that arise only from the
axioms of the formalism, and hence are universal in the
theory, independent of any particular application of it.
The framework is that of the generalized quantum  
theory of closed systems, as first defined by J.B. Hartle in 
\cite{Jerusalem,Lesh}, and much expanded upon by Isham, Linden, 
and others \cite[see \cite{IshamIntro}\ 
for an introduction to their work]{Isham,IL1,IL2,ILS,S,Wright}.
The central structural element of this 
formalism is the ``decoherence functional'',
an Hermitian functional on pairs of ``histories'' 
of a physical system \system.  
The decoherence functional both provides a measure of interference 
between histories, thereby determining which sets of histories 
are ``consistent'' in the sense that probabilities may consistently 
be assigned to the various histories in the set, and fixes those 
probabilities within consistent (or ``decohering'') sets.  
For closed quantum systems (such as the universe), 
in which there are no external 
classical observers around to measure anything, the decoherence 
functional thus replaces the usual rule of ``Copenhagen'' quantum 
mechanics that probabilities may only be assigned to histories which are 
measured\cite{Jerusalem,GMH}.\footnote{Though I will be employing the
formalism of Gell-Mann and Hartle throughout, it should be noted
that similar notions were introduced earlier, but independently, by
Griffiths and by Omn\`es \cite[see \cite{GriffithsQM} and \cite{OmnesQM}
for recent reviews of their work.]{Griffiths,Omnes}.}   The decoherence
functional is a natural generalization to closed systems of the notion 
of ``quantum state", as the term is used in quantum logic and in
algebraic quantization \cite{IL1},  
a connection that will be described in section \ref{sec:generalities}.

In this investigation, consideration is restricted to 
systems whose observables live on a finite dimensional
Hilbert space \HilbertS.
The question of decoherence is approached from the point of view
that the alternative histories of the system \system\ can be taken as 
living in $\hilbert = \linearops{\HilbertS},$ the linear operators on 
\HilbertS,
and that decoherence functionals define Hermitian forms on 
\hilbert\ \cite{Wright}.
Indeed, a wide class of decoherence functionals may be thought of
%(including in particular the canonical \cite{GMH,Jerusalem,Lesh} time 
%neutral decoherence functional) 
as positive (but not necessarily positive definite) inner products on 
this vector space; these include all of those used in practical
applications of generalized quantum theory.  
Decohering, or consistent,  histories are then
orthogonal vectors under this inner product, and the ``length'' (squared)
of a history consistent in this inner product is its probability.  The study
of the ``geometry of decoherence'' is thus the study of the Hermitian
forms and semi-inner
products on \hilbert. 
The canonical decoherence functionals over \HilbertS, which correspond to
situations in which boundary conditions are imposed on \system\ at
initial and 
possibly also at 
final times, are examples of such positive 
decoherence functionals of particular importance in theories with an
externally supplied notion of time; they are the decoherence 
functionals of ordinary quantum mechanics.  

(A parallel treatment of these ideas is also given for the case where
quantum histories are represented by projection operators in a bigger
Hilbert space $\otimes^k H_S$, as in the quantum-logical formulation
of Isham {\it et al.}.)

The focus of this work is thus
the {\it structure} of generalized quantum theory. Particular physical
applications of it are not addressed.  
Rather, I aim to reveal general and universal properties
of the formalism that apply to any particular application of it.
Apart from the general structural results, 
the physical problem lurking in the background  
is the issue of 
what additional physical constraints may reasonably be imposed
on decoherence functionals, and in particular, of
how one infers %properties of 
the decoherence functional of a closed quantum system from experiments;
the analogous problem in ordinary quantum mechanics is how to reconstruct 
from observations the density matrix of a quantum system.
% from observations of it.)
This is a subject which will be approached more directly in future work,
using the tools developed here.  (Indeed, I address the {\it mathematical}
side of this problem in section \ref{sec:functional?}, leaving the
difficult physical questions for another venue.)

%Vague; make this paragraph *useful* to the reader.
Taking a close 
look at what the basic assumptions of the
theory do and do not imply 
supplies
information concerning the question of which aspects of 
generalized quantum theory %in Hilbert space
follow from its basic framework, and which are properties of a 
specific choice of decoherence functional and method for representing 
quantum histories.  This knowledge in turn illuminates the potential 
physical significance of the choices that are made, and supplies
a clear structure within which to study the implications of any
further physical conditions %that might be 
imposed on the theory.
In addition, a number of useful calculational tools are developed.
The ``geometric'' point of view %interpretation 
developed here also has the value of building intuition for the 
mathematics of decoherence.

A number of useful properties of decoherence functionals 
(particularly the positive ones, which include the decoherence
functional of ordinary quantum mechanics) 
are derived here in a very general setting.  These include a bound on
the maximum number of histories in any consistent set, and a
Cauchy-Schwarz inequality for positive decoherence functionals.
Some of these results were known previously for the special case of the
decoherence functional that arises in ordinary quantum mechanics, but
the present work illuminates the generality of their geometric origins.  
I also show how to characterize algebraically the canonical
decoherence functionals that are used in conventional 
applications of generalized quantum theory.

This work is thus complementary in spirit to the sophisticated
quantum-logical investigations of 
Isham, Linden, {\it et al.\ }\cite{Isham,IL1,IL2,ILS,S,Wright},
in some respects building a bridge between their approach to
generalized quantum theory and the more conventional one.
Among their most significant results is the important ILS theorem
\cite{ILS,Wright,Rudolph}\ classifying decoherence functionals, in direct
analogy with the Gleason theorem classifying algebraic quantum states in
ordinary quantum mechanics.  (We will see the ILS theorem emerge quite
naturally out of the present formulation of generalized quantum theory, 
illuminating its %essentially
geometrical significance.  In addition, a version of the ILS theorem for
decoherence functionals defined on ``class operators'' appears.)
While the program of Isham and
Linden is unquestionably appropriate for the mathematically rigorous
formulation of generalized quantum theory, and in particular is 
important to the program of generalizing quantum mechanics sufficiently
to encompass the requirements of quantum gravity and other theories
without an {\it a priori} definition of time, the present work makes more
direct contact with the way in which decoherence functionals are actually
used in computations, and the elucidation of the geometry of decoherence
in \hilbert\ has its own intrinsic interest and usefulness.  

After these introductory words, the succeeding sections proceed in the
following fashion: section \ref{sec:GQTgoc} defines the notion of a 
generalized quantum theory, realizes this notion in a finite 
dimensional Hilbert space \HilbertS\ by exhibiting the canonical 
decoherence functional \decohif{\cdot}{\cdot} over \HilbertS, 
and discusses quantum histories and their operator representations.
Section \ref{sec:hermitianform}\ explains how to extend the domain of
definition of a general decoherence functional on a Hilbert space
\Hilbert\ to the entire space of
linear operators \linearops{\Hilbert}\ on the Hilbert space \Hilbert,
showing that it is an Hermitian form on $\hilbert= \linearops{\Hilbert}$
 \cite{Wright}.
(In the quantum-logical approach of Isham and Linden, \Hilbert\ is
not the same as \HilbertS, as will be explained in the sequel.)
This observation is then exploited to bound 
the maximum number of histories in a consistent set, and show
how positive decoherence functionals are actually semi-inner products 
on \hilbert.  Section \ref{sec:notation}\ develops 
some useful tools for calculating with a decoherence functional on
\hilbert.  Section \ref{sec:questions}\ lists some basic structural issues 
in the program of generalized quantum theory, the solutions to which are
described in the balance of the paper. 
Section \ref{sec:history?}\ shows how to find %explicitly
all the histories (operators) consistent according to a given
decoherence functional, and section \ref{sec:functional?}\ addresses the
converse problem.  Finally, section \ref{sec:dcanonical?}\ solves the
problem of determining when an arbitrary decoherence functional on class
operator histories is canonical.  Section \ref{sec:answers}\ is reserved
for summary and some final discussion.

Appendix %\ref{sec:dictionary} 
B summarizes the important notation 
for ease of reference.

This is a long document, 
so I offer occasional suggestions about sections
that may be skipped on a first reading.  Those interested
in an overview of the ``geometric'' perspective on generalized
quantum theory employed here should read through sections
\ref{sec:generalities}, \ref{sec:hilbertGQT}, \ref{sec:canonicald},
\ref{sec:extension}, \ref{sec:notation}, and \ref{sec:example} first.

\section{Generalized Quantum Theory}
\label{sec:GQTgoc}

This portion of the paper is devoted to a description of the 
generalized quantum theory and its Hilbert space implementation
that will be the subject of the remainder of the paper.  
Section \ref{sec:generalities}\ defines in the abstract the idea of a
generalized quantum theory.  Section \ref{sec:ILSW}\ describes
the realization of this idea in quantum temporal logic, and 
discusses the ILS theorem which classifies the possible
decoherence functionals.  Sections \ref{sec:hilbertGQT}\ and
\ref{sec:historyreconstruction}\ begin a close look at the 
representation of quantum histories by operators in Hilbert
space, making it possible to define decoherence functionals
as operators on those histories.  Finally, section
\ref{sec:canonicald}\ defines the canonical decoherence functional,
the decoherence functional that arises out of conventional
quantum theory.  The rest of the paper goes on to discuss 
in more detail the idea of the decoherence functional as an 
operator on histories.

\subsection{Generalities}
\label{sec:generalities}

A ``generalized quantum mechanics'' is defined by Hartle 
\cite{Jerusalem,Lesh} to consist in:     

I. {\bf Fine-Grained Histories}: Sets of exhaustive collections of 
alternative physical histories of \system.  By definition, a 
fine-grained history is the most refined description of a history possible.  
The empty history $\emptyset$ (or 0) is always a member of a
fine grained set.  Familiar examples of a ``history'' include 
a particle's path in non-relativistic 
quantum mechanics, and the spacetime configuration of a quantum field in 
quantum field theory.\footnote{Note, however, that no notion of time is 
required {\it a priori} to define a ``history'' of \system, and may
indeed emerge only obliquely (or not at all!) from other physical 
considerations.  For instance, the collection of all four-geometries on
some set of manifolds might be taken as the set of fine-grained
histories in a quantum theory of gravitation.}  It is possible, and 
indeed common, for there to be many allowed complete sets of fine-grained 
histories.  The democracy of Dirac's transformation theory in ordinary 
quantum mechanics is a common illustration of this.

II. {\bf Allowed Coarse-Grainings}:  The allowed partitions of a 
%complete 
set of fine-grained histories into mutually exclusive and jointly 
exhaustive classes.  Each class of an allowed partition is defined 
to be a coarse-grained history.  The trivial coarse grainings where the
partition is merely the identity (no coarse graining at all), and 
where all of the histories in an exhaustive fine grained set are collected 
together into the completely coarse-grained history $u$, are always taken 
to be allowed coarse grainings.  ($u$ is assumed to be the common
complete coarse graining of all exhaustive fine grained sets.  Thus,
specifying $u$ is equivalent to defining what is meant by a
``complete" or ``exhaustive" set of histories.  Depending on the
context, $u$ is often instead denoted 1.)  

A coarse-graining of a more finely-grained (exclusive, exhaustive) set 
of histories $\consistentset = \{h\}$ is often denoted 
$\overline{\consistentset} = \{\overline{h}\}$.
(To be explicit, $\overline{h}$ denotes the class containing $h.)$
I will denote the collection of {\it all} histories, fine or coarse
grained, by \histories .

III. {\bf Decoherence Functional}:  
%A complex valued functional 
%\decoh{h}{h'} on pairs of histories in exhaustive coarse-grained sets
%$\{h\}$.  It possesses the following properties:
A functional $d:\ \histories \times \histories \rightarrow \Co $
%\decoh{h}{h'} 
%on pairs of histories in exhaustive coarse-grained sets
%$\consistentset = \{h\} \subset  \histories.$  It 
possessing the following properties:

  \begin{enumerate}

        \item {\it Hermiticity} 
          \begin{equation}   \label{eq:hermiticity}
                     \decoh{h}{h'} = \decoh{h'}{h}^{*}
          \end{equation}

        \item {\it Positivity}
          \begin{equation}   \label{eq:positivity}
                     \decoh{h}{h} \geq 0
          \end{equation}

        \item {\it Additivity (``Principle of Superposition")}
          \begin{equation}   \label{eq:additivity}
             \decoh{\overline{h}}{\overline{h}'} =
             \sum_{h \in \overline{h},\ h' \in \overline{h}'} \decoh{h}{h'}
          \end{equation}

        \item {\it Normalization}
          \begin{equation}   \label{eq:normalization}
               \sum_{h,h' \in {\scriptstyle {\cal S} } } \decoh{h}{h'} = 1
          \end{equation}

  \end{enumerate}
for all exclusive, exhaustive sets $\consistentset \subset \histories.$

IV. {\bf Decoherence Condition}:  This condition determines the kind and
degree of interference between histories that will be tolerated in sets of 
alternative coarse grained histories in which probabilities may be 
assigned, {\it i.e.} determines which sets of histories are ``consistent'', 
or ``decohere''.\footnote{There is some tension in the literature
concerning the proper usage of these terms.  It is of particular note
that the term ``decoherence" is becoming entrenched in much of the literature 
as referring to the {\it dynamical process} of the decay of the
off-diagonal elements of the density matrix in appropriate
semi-classical bases \cite{Giulini}.  While this notion is not the same 
as the consistency of a set of histories it is not unrelated either.  (See
\cite{GMHCQM} and \cite[note added in proof]{tsa} for discussion.)
Though I have some sympathy with the utility of making this distinction, 
by using these terms interchangeably, I do nothing here to quiet the 
discussion.}
A good decoherence condition must guarantee that probabilities may be
assigned consistently in the decohering sets which it defines.

What is the precise meaning of this?  
The probabilities for various histories in 
%an exclusive and 
an exhaustive set of mutually decohering histories are defined to be the 
diagonal elements of the decoherence functional,
\begin{equation}   \label{eq:p(h)}
                         p(h) = \decoh{h}{h}.
\end{equation}
In order to permit interpretation as a probability, these numbers must
satisfy the standard Kolmogorov rules:  they must be real numbers between 
0 and 1, must satisfy $p(\emptyset) = 0$ and $p(u) = 1$, and be additive 
under coarse-graining,
\begin{equation}  \label{eq:additiveprobs}
       p(\overline{h}) = \sum_{h \in \overline{h} } p(h)
\end{equation}
for any coarse graining 
$\overline{\consistentset } = \{ \overline{h} \}$ of the exclusive and
exhaustive set $\consistentset = \{ h \}$.
In order for the ``probabilities'' of (\ref{eq:p(h)}) to be
physically interpretable as such, the decoherence condition should be 
%such that there are at least some complete coarse-grained sets, 
%d(1,1) would seem to work, always, ....
%all of whose histories decohere according to 
%the given decoherence functional, and such that in all such sets these
sufficient for these rules of probability theory to be satisfied.  
Histories in decohering sets are
said to be ``consistent with'' the decoherence functional; 
it will become clear below that there is indeed no difficulty 
in thinking of {\it individual} histories as being ``consistent'', 
as well as sets of histories.  The collection of histories which 
decohere according to the decoherence functional $d$ ({\it i.e.} 
are in {\it some} consistent set) will be
denoted $\decoheringsub{d}$.  The class of exclusive, 
exhaustive {\it sets} of histories $\consistentset_{d}$
consistent according to $d$ will be denoted $\consistentsub{d}$.
(The structure of these spaces is a recurring theme of this work.)
%(partially explored in sections
%\ref{sec:history?}\ and \ref{sec:functional?}, and in appendix
%\ref{sec:raycomplete}.)

An appendix summarizing most of the notation is 
provided at the end.   

If {\it arbitrary} partitions are allowed coarse-grainings,
so-called ``weak decoherence''
\begin{equation}   
%\begin{array}{cc} 
%       Re\, \decoh{h}{h'} = 0,   &   h \neq h'
%\end{array}
       {\rm Re}\, \decoh{h}{h'} = 0,\ \ \       h \neq h'
\label{eq:weakdecoherence}
\end{equation}
is necessary and sufficient for the correct probability sum rules to be
satfisfied.  However, the stronger condition of ``medium decoherence''
\begin{equation}   
%\begin{array}{cc} 
%      \decoh{h}{h'} = 0,   &   h \ne h'
%\end{array}
      \decoh{h}{h'} = 0,\ \ \       h \ne h'
\label{eq:mediumdecoherence}
\end{equation}
is much more mathematically convenient and I will take it to be the
decoherence condition IV unless otherwise specified; 
(\ref{eq:mediumdecoherence}) both implies (\ref{eq:weakdecoherence}), 
and arises naturally in many physical problems.  (Decoherence conditions 
both stronger and weaker than these may also find application,
% in physical problems, 
but will not be considered here.  They are discussed for instance 
in \cite{GMH,Griffiths,Omnes,GMHCQM}.)  Decoherence might not
be required to hold exactly, but only to the accuracy to which the
probabilities are used.  The physical meaning of an approximate probability in 
generalized quantum theory is discussed in \cite{Jerusalem}, and
various mathematical aspects treated in \cite{DH,DKsummary,DK,McElwaine}.

\subsubsection*{The Decoherence Functional as Generalized Quantum State}
\addcontentsline{toc}{subsubsection}{\numberline{}%
The Decoherence Functional as Generalized Quantum State}
%cf. The LaTeX companion section 2.4.2

As first observed by Isham and Linden in their quantum-logical
formulation of generalized quantum theory, the decoherence functional is
a natural generalization of the notion of (algebraic) quantum state to closed
systems \cite{IL1}.  This may seem an odd perspective upon first encounter,
but, after all, a quantum state has physical significance only in virtue
of the questions asked of it.  That is the reason for the emphasis on
$d$ as a functional of quantum histories.

(In fact, those familiar with algebraic quantization will recognize the
great similarity between the defining conditions of a decoherence
functional and those of an algebraic quantum state.  Given a $C^{*}$
algebra of observables $\cal{A}$, a quantum state is defined to be a
linear map $\omega: \cal{A} \rightarrow \Co$ which satisfies
$\omega(A^{*}A) \geq 0\ \ \forall\  A \in \cal{A},$ and $\omega(1) = 1;$
see, for example \cite{Haag,Wald}.  Alternately, in quantum logic,
a state on, for instance, the space (lattice) \projlattice{\Hilbert} 
of projections over a Hilbert space
\Hilbert\ -- the projections correspond physically to single time
propositions about the system -- is defined as a positive, normalized,
real valued map which is additive on disjoint projections:
$\sigma(P \oplus Q) = \sigma(P) + \sigma(Q)$ if $P$ and $Q$ are
disjoint \cite{BC,Hughes,vF}.)

The new element the decoherence functional brings to the 
quantum-theoretic picture is an internally consistent identification 
of those histories which may sensibly be assigned probabilities.  
This measure is concrete and quantitative, circumventing %obviating
the need to forge an unambiguously applicable notion of ``measurement" 
(the elusiveness of which is positively legendary \cite{WZ}.)  What it
does {\it not} do is supply the ontological significance of these
%ontology: the philosophy of what it means "to be".  
%Fay prefers "scientific".  Coward.  
probabilities.  This (generalized quantum theory) remains, after all, 
quantum mechanics \cite{JBH}.  (See \cite{dE1,dE2} for entr\'{e}e into 
this knotty realm.)

The decoherence functional of a physical system therefore not only 
determines the sets of histories about which predictions can be made,
but also what is the probability of each such physically realizable
quantum history.  It is therefore very appropriate 
to view the decoherence functional as the ``quantum state'' 
of the system in this generalized sense.

\subsection[Quantum Temporal Logic and the ILS Theorem]{Quantum Temporal 
Logic and the ILS Theorem\protect\footnote{This section may be skipped 
on a first reading.}}
\label{sec:ILSW}

As already mentioned, the definition of a generalized quantum theory 
given above has been formulated as a rigorous quantum logical structure 
by Isham and collaborators \cite[an excellent lightning introduction is 
\cite{IshamIntro}]{Isham,IL1,IL2,ILS,S,Wright}.   The role of the space 
\histories\ is played by a so-called orthoalgebra $\cal{UP}$ of history
propositions.  This is merely a space with: a partial ordering
(expressing notions of relative fine or coarse graining); a 
``biggest" and a ``smallest" element (1 and 0) such that
$0 \leq P \leq 1 \ \forall\ P \in \cal{UP};$ a negation 
(or ``complementation") operation $\neg$ (expressing the 
history ``not $h$") such that $\neg (\neg h) = h;$
a ``disjointness" relation $\perp $ between histories (corresponding 
to the idea of mutually exclusive histories); and a (commutative,
associative) ``join" operation $\oplus$ on {\it disjoint} histories 
(expressing the logical notion ``$h$ or $h'$") 
for which $h \perp h'$ implies $h \oplus h' \in \cal{UP}.$  Finally,
there are a couple of sensible identities interrelating these relations,  
namely, that $\neg h$ is the unique element for which both $h \perp \neg h$
and $h \oplus \neg h = 1,$ and that $h \leq h''$ iff there is some
$h'$ such that $h'' = h \oplus h'.$  %(This is then sufficient to imply 
%the important result that $h \perp h'$ iff $h \leq \neg h'.$)
While it is often the case that $\cal{UP}$ has {\it more} structure 
than that of an orthoalgebra ({\it e.g.\ }that of a lattice \cite{BC,Hughes}, 
in which the ``join" operation is extended to {\it all} pairs of elements,
and in addition a ``meet" operation is introduced which encodes the
logical ``and" for all pairs of propositions),
the orthoalgebraic structure seems to be the minimum one can
usefully get away with.  (\cite{Hughes}\ is a straightforward introduction.)
It should come as no surprise that on \projlattice{\Hilbert},
the lattice of projections on a Hilbert space \Hilbert,
$\neg P = 1-P$, that $P \perp Q$ iff $PQ = 0$, and that $\oplus$
is just the usual operator sum when $P$ and $Q$ are disjoint.

A decoherence functional is
then defined just as above, namely, as an Hermitian, positive, normalized
map $d: \cal{UP} \times \cal{UP} \rightarrow \Co$ which is bi-additive on
disjoint history propositions.  Isham, Linden, and Shreckenberg
\cite{ILS} have given a complete classification of the decoherence
functionals when $\cal{UP}$ is the lattice %space (lattice \cite{BC}) 
\projlattice{\Hilbert} of projections on a Hilbert space \Hilbert\ with
$\rm{dim\, }\Hilbert > 2,$ and Wright \cite{Wright} has
extended the theorem to an (almost) general von Neumann algebra (the 
``almost" is directly related to the dimension requirement.)  The 
ILS theorem is the natural generalization to generalized quantum theory
of Gleason's famous theorem \cite[see \cite{Hughes,Peres} for discussion 
and \cite{Hughes} for a glossed proof]{Gleason,CKM} 
classifying the states on Hilbert space.  (The Gleason theorem puts 
algebraic states in one to one correspondence with density matrices 
on \Hilbert, so long as ${\rm dim}\, \Hilbert > 2.$)  Indeed, 
according to the ILS theorem, (bi-continuous)
decoherence functionals over \projlattice{\Hilbert} are, when 
$\rm{dim}\, \Hilbert > 2,$ in correspondence with operators $X$ 
on $\Hilbert \otimes \Hilbert$ via
\begin{equation} \label{eq:dILS}
    d(h,h') = {\rm tr}[h \otimes h'X].
\end{equation}
In order to fulfill the Hermiticity requirement, 
the operators $X$ satisfy $ X^{\dagger} = MXM$, where $M$ is the
operator on $\Hilbert \otimes \Hilbert$ which switches vectors,
$M(u \otimes v) = v \otimes u$, so that for operators $O$ and $O'$ on
\Hilbert, $M(O \otimes O')M = O' \otimes O$.  Further, $X$ is ``positive",
${\rm tr}[P\otimes PX]  \geq 0$ on all projections $P$, and normalized,
${\rm tr}[X] = 1$.   
(Actually, Hermiticity implies that ${\rm tr}[P\otimes PX_{I}]
= 0$, where $X = X_{R} + i X_{I}$ is the decomposition  of $X$ into
Hermitian parts.  Thus, the last two requirements need only be
imposed on $X_{R}.$)  Note that positivity on all projections is {\it not} 
sufficient to guarantee that 
${\rm tr}[L \otimes LX] \geq 0\ \ \forall\ L \in \linearops{\Hilbert}.$  
As all operators may be written as a linear
combination of some collection of projections (via, for example, its 
decomposition into self-adjoint pieces), 
necessary and sufficient for positivity is what Isham \cite{Isham}  
has called a ``positive kernel" condition,
\begin{equation} \label{eq:positivekernel}
    \sum_{i,j} c^*_{i}\, c_{j}\ d(\Projsub{i},\Projsub{j}) \geq 0
\end{equation}
for all sets of projections \Projsub{i} and complex numbers $c_{i}$.  
%It is worth
%noting here that the canonical decoherence functionals used in most
%traditional applications of quantum theory (and defined in
%(\ref{eq:dcanonical}) below) satisy this positive kernel condition 
%by construction.
%Is this true even on \otimes^2k\HilbertS?  I wuz thinking of \classoprep.
%Nope!

We shall see $X$ resurface later, in section \ref{sec:ILSrevisited}.

Considerable further discussion regarding the notion of a generalized
quantum theory may be found in 
\cite{Jerusalem,Lesh,Isham,IL1,IL2,IshamIntro,DK}.

%\vfill

\subsection[Generalized Quantum Theory in Hilbert Space: 
Histories and Their Representation]{Generalized Quantum Theory 
in Hilbert Space:\protect\\ Histories and Their Representation}
\label{sec:hilbertGQT}

There are two methods of representing quantum histories of 
a physical system $\system$
as operators in a Hilbert space: as ``class operators'' on the Hilbert
space $\HilbertS$, or as projection operators on a larger space
$\otimes^k \HilbertS$.  The resulting collections of operators 
are called \classoprep\ and \projrep.  After describing these two
strategies for representing histories, I go on to discuss two
important categories of permissible coarse grainings, homogeneous
and inhomogeneous, and the associated spaces of history operators
\classoprepbar\ and \projrepbar.  Before moving on to discuss
the decoherence functional as an Hermitian form on these operator
spaces in section \ref{sec:hermitianform}, I briefly address 
in section \ref{sec:historyreconstruction}\ the problem of 
reconstructing histories from their operator representatives,
and exhibit the 
decoherence functional that is used in all conventional applications 
of generalized quantum theory in section \ref{sec:canonicald}.  

\subsubsection*{Quantum Histories}
\addcontentsline{toc}{subsubsection}{\numberline{}%
Quantum Histories}
%cf. The LaTeX companion section 2.4.2

A canonical example of a generalized quantum mechanics is a direct
generalization to closed systems of ordinary quantum mechanics in
a Hilbert space \HilbertS.  Physical alternatives are described by projection
operators \Projsub{a} onto ranges $a$ of eigenvalues of observables $A$.
Properly, a fine-grained history $h$ is then a time-ordered sequence 
of  Schr\"{o}dinger projection operators,
\begin{equation} \label{eq:historysequence}
h = (\Projsupb{1}{\betasub{1}},\cdots,\Projsupb{k-1}{\betasub{k-1}},%
\Projsupb{k}{\betasub{k}}),
\end{equation}
at, for simplicity, some finite, fixed set 
of times $\{ \tsub{1} < \cdots < \tsub{k-1} < \tsub{k} \}$ called the 
``temporal support" \temporalsub{k}\ of the fine grained histories.
Here, the superscript labels which observable is taken at time \tsub{i}, 
the subscript labelling the chosen range of eigenvalues of that observable.  
$h$ thus corresponds to the history in which \system\ is in the
range of eigenvalues \betasub{1} of observable $1$ at time \tsub{1}, 
the range \betasub{k-1} of observable $k-1$ at time \tsub{k-1}, and so on. 
Now, we obviously do not want the definition of the theory to depend on $k$
in any essential way.  As simply setting \Projsupb{i}{\betasub{i}} always
to 1 effectively reduces $k$ to $k-1,$ $k$ can just be imagined to be
bigger than any conceivably required value. (\cite{Isham,IL1,IL2}\ begin
to treat the case where ${\cal T}$ is a segment of $\Real.$  I shan't deal
with that complication here; 
%complications: \inf dim Hilbert spaces in \projrep; \inf op products
%in \classoprep
so long as $k$ is finite,  the
expressions below will be mathematically well defined.  From a practical
point of view, unless it is actually desired to consider alternatives at
a continuous sequence of times, merely imagining $k$ to be as large as
necessary is equivalent anyhow.)

In accord with the customary equity of transformation theory
in conventional Hilbert space quantum mechanics, {\it all} 
projections on \HilbertS\ will be allowed choices at each 
time \tsub{i}.  

The notion of disjointness should correspond to the physical idea that
if $h$ is realized, then it would be contradictory to say that $h'$ is
realized also.  Therefore two histories (sequences of projection
operators) will be said to be {\it disjoint} if, at one of the times 
\tsub{i}, $\Projsupb{i}{\betasub{i}}\Projsupb{i}{\beta_i'} = 0.$  
Similarly, a collection of mutually exclusive ({\it i.e.} pairwise 
disjoint) histories will be said to be {\it exhaustive} if, at each \tsub{i},
the set of projections $\{ \Projsupb{i}{\betasub{i}} \}$ taken from each
history constitute a complete set, 
$\sum_{\betasub{i}}\Projsupb{i}{\betasub{i}} = 1$ (thereby exhausting all
possibilities: it can be said with certainty that {\it something}
``happened" at \tsub{i}.)

\subsubsection*{Representation of Quantum Histories by Projection Operators}
\addcontentsline{toc}{subsubsection}{\numberline{}%
Representation of Quantum Histories by Projection Operators}
%cf. The LaTeX companion section 2.4.2

For the purposes of quantum logic, the best way \cite{Isham}\
to represent the history proposition
$(\Projsupb{1}{\betasub{1}},\cdots,\Projsupb{k-1}{\betasub{k-1}},%
\Projsupb{k}{\betasub{k}})$ 
is as the tensor product
\begin{equation}  \label{eq:historyproduct}
\Projsupb{1}{\betasub{1}} \otimes \cdots \otimes \Projsupb{k-1}{\betasub{k-1}}
\otimes \Projsupb{k}{\betasub{k}}.
\end{equation}
This is desirable because these operators on $\otimes^{k}\HilbertS$ are
in one-to-one correspondence with the $k\rm{-fold}$ sequences of 
projection operators on \HilbertS.  (The rigorous definition of 
time-ordered products over a possibly continuous sequence of times 
is treated in \cite{Isham,IL1,IL2}.  This complication will 
%prove inessential to the purposes of 
not be considered here.)       
Two such history representatives are obviously disjoint iff their 
operator product on $\otimes^{k}\HilbertS$ is 0.  A set of them is
exhaustive if 
\begin{equation} \label{eq:exhaustiveproduct}
    \sum_{\beta_{1}\cdots\beta_{k}}\
    \Projsupb{1}{\betasub{1}} \otimes \cdots \otimes 
    \Projsupb{k-1}{\betasub{k-1}} \otimes \Projsupb{k}{\betasub{k}}
    = \otimes^{k}1.
\end{equation}
(The fully coarse grained history $u$ is thereby represented by 
$\otimes^{k}1,$ where 1 is the unit operator on \HilbertS.)
An exhaustive, exclusive set of histories is thus a partition of unity
in $\otimes^{k}\HilbertS.$
%; the converse is obviously false.

\subsubsection*{Representation of Quantum Histories by Class Operators}
\addcontentsline{toc}{subsubsection}{\numberline{}%
Representation of Quantum Histories by Class Operators}
%cf. The LaTeX companion section 2.4.2

On the other hand, in everyday 
quantum mechanics a natural representative of 
the history $h$ is the so-called ``class operator"
on \HilbertS\ given by
\begin{equation}   \label{eq:classop}
          C_{\beta} = \Projsupb{1}{\betasub{1}}(\tsub{1}) \cdots
                      \Projsupb{k-1}{\betasub{k-1}}(\tsub{k-1})
                      \Projsupb{k}{\betasub{k}}(\tsub{k}).
\end{equation}
(Note the choice of time ordering is the reverse of that generally
preferred by Gell-Mann and Hartle, but it is much more convenient for
present purposes.)  Unfortunately these are {\it not} in one-to-one 
correspondence with sequences of projections, as simple dimensional 
counting is sufficient to reveal: with $\rm{dim}\, \HilbertS = N,$ 
there are $N^{k}$ linearly independent operators on \HilbertS\
of the form (\ref{eq:historyproduct}), but only $N^{2}$
linearly independent 
%dimension counting doesn't exclude possibility that worrying about
%overall coeffs => 1-1, but it doesn't, and don't bother bringing up.
%Reference ray-completeness appendix for examples if people bother me
%with it.
class operators (\ref{eq:classop}).\footnote{In (\ref{eq:classop}), 
the dynamics is encoded in the time evolution 
of the Heisenberg projections.  In the program of Isham {\it et al.}, 
where everything is written in terms of Schr\"{o}dinger projections on 
$\otimes^{k}\HilbertS$ as in (\ref{eq:historyproduct}), the dynamics is 
unfolded from the histories and instead bound up into the single 
operator $X$ of (\ref{eq:dILS}), which describes the boundary conditions 
as well. (See \cite{Isham} or \cite{ILS} for how to do so in the case of
a canonical decoherence functional with 
$\rhomega =1;$ {\it cf.\ }(\ref{eq:dcanonical}) below.  
It is not difficult to generalize the 
algorithm given there to more general cases, for instance, 
to the case $\rhomega \neq 1.$) 
%Is this process invertible?  I don't remember.  Check.
%Nope.  The partial trace that removes the bc and folds them into
%$X$ is not invertible.
In this work, however, where I deal with histories in \HilbertS\
in the abstract, the projections will be
arbitrary, and the question of dynamics need never arise. 
\label{foot:X}} 
Correspondingly, there is no simple reflection in \linearops{\HilbertS}
of the disjointness criterion in \histories, to which we must resort.
%on which we must [therefore] rely.
%\projrep isn't immune either; P^{\dagger}P=P, sure, but is it a sum of
%a disjoint sequence of P's?
To emphasize the point, given {\it only} the two class operators
$C_{\beta}$ and $C_{\beta'}$,  {\it there is
no way in general to tell} whether they correspond to physically
disjoint histories.                       %\footnote{
(Related issues are taken up in
section \ref{sec:historyreconstruction}.)   %}  
%($C_{\beta}^{\dagger}C_{\beta'}=0$ or 
%$C_{\beta}C_{\beta'}^{\dagger}=0$ are sufficient, but not
%necessary, conditions for $h\perp h'.$)  
However, it {\it is} obviously
true that a mutually disjoint set \consistentset\ is exhaustive iff
$\sum_{\beta}C_{\beta} = 1.$  (Thus, $u$ is represented by the unit
operator on \HilbertS.)
%Incidentally, note that it is also true
%that $\sum\ C_{\beta}^{\dagger}C_{\beta}=1.$)
It is not strictly correct to say that $\{ C_{\beta}\}$ then constitutes
a partition of unity in \linearops{\HilbertS} because two logically
disjoint histories are not necessarily linearly independent as operators.
However, as I will show in section \ref{sec:maxnumhistories}, 
if the set $\{ C_{\beta} \}$
is {\it consistent} according to some decoherence functional $d$ 
({\it i.e.\ }$\{ C_{\beta} \} \in \consistentsub{d}$),
then all the $C_{\beta}$ with non-zero probability must be linearly 
independent in \linearops{\HilbertS}.  Suppose therefore that all
linearly dependent subsets of $\{ C_{\beta} \}$ are coarse grained down
to one history.  In that case, then, a consistent $\{ C_{\beta} \}$ is a
genuine partition of the unit operator on \HilbertS.
%0 histories?
%Sadly, linear independence of a collection of $ C_{\beta} $ does 
%{\it not} imply they are disjoint; the construction of counterexamples
%is very easy.

The origin of 
the representation of \histories\ by class operators can be most 
easily understood by considering 
$C_{\beta}^{\dagger}$ applied to some pure state $\ket{\psi}$,
$C_{\beta}^{\dagger}\ket{\psi}$.  Writing it out, it is easy to see
that up to normalization, this object represents a state $\ket{\psi}$
that evolves unitarily to time $\tsub{1}$, at which alternative
\Projsupb{1}{\betasub{1}} is realized, and so on up to time $\tsub{k}$.
Thus this object expresses the two forms of dynamical evolution that 
quantum states in ordinary ``Copenhagen" quantum mechanics can undergo, 
wave function collapse upon ``measurement", and unitary evolution in
between \cite{JBH}, and is therefore a natural choice of representation,
with apparent physical significance,
in ordinary quantum mechanics.\footnote{Notice that no notion of measurement 
is required, however, it may be recovered in appropriate circumstances
\cite{Jerusalem,Lesh,HalliwellRecords}.  Nevertheless, other potentially
serious ambiguities arise.  
For discussion, see \cite{DK,Kent1,Kent2,GH,GMHEH,GMHSD}.}
This fact alone makes it worthwhile to study the representation of
histories by class operators independently of their representation as
projection operators.
For more detail, see for instance \cite{Lesh}.
%We shall see at the end of this subsection how this choice
%reproduces the probability formula we expect in familiar situations.

%So the loss of information represents the quantum "branching" ...
%How does IL's QTL based on P's do this? 

In any event, while the formulation of quantum temporal logic on
$\otimes^{k}\HilbertS$ has certain mathematical advantages, it has the
feature that (unless $X$ has a very special form; 
see footnote \ref{foot:X})
there is no immediate analogue other than consistency itself of the 
``wave function collapse"-like behaviour so characteristic of 
ordinary quantum mechanics.  
%you get a reduction when a msmt is made; one branch is thereafter
%selected
This is of no particular concern to Isham and Linden, 
whose aims include {\it generalizing} quantum theory to timeless
theories, but it is not always desirable to toss out {\it a priori} 
such familiar and important 
physics. 
Considerable attention will therefore be given 
to decoherence functionals on the representation of
histories by class operators.  We thus can study the effect of retaining
this ``quantum branching" behaviour, while generalizing other aspects
of quantum theory.  

%As I am not a mathematician 
To control the proliferation of notation I 
will show no shame in often denoting by $h$ 
an operator representative (\ref{eq:historyproduct}) or (\ref{eq:classop}) 
of the history (\ref{eq:historysequence}).  (In section
\ref{sec:innerproduct} this abuse will be extended even further to
include linear combinations of such representatives.)
%$(\Projsupb{1}{\betasub{1}},\cdots,\Projsupb{k-1}{\betasub{k-1}},%
%\Projsupb{k}{\betasub{k}})$.  
As the saying goes, which is meant should be evident from the context.

\subsubsection*{The History Representation Spaces \projrep\ and \classoprep}
\addcontentsline{toc}{subsubsection}{\numberline{}%
The History Representation Spaces \projrep\ and \classoprep}
%cf. The LaTeX companion section 2.4.2

When $k>1$ (which I shall always assume), there are thus two distinct
possible choices for the collection of operator representatives of the
fine grained histories
$(\Projsupb{1}{\betasub{1}},\cdots,\Projsupb{k-1}{\betasub{k-1}},%
\Projsupb{k}{\betasub{k}})$
with temporal support \temporalsub{k},  the projections
(\ref{eq:historyproduct}) or the class operators (\ref{eq:classop}).
I shall call the former choice the (faithful) {\it projection representation}
\projrep(\histories) of the fine grained histories in \histories,
the latter the (unfaithful) {\it class operator representation}, or
sometimes, the {\it conventional} representation, 
\classoprep(\histories).
%\footnote{As \histories\ is not a group, the use of the term 
%``representation" should
%not be overinterpreted. 
%be interpreted as applying to the reflection of the orthoalgebraic
%structure of \histories\ in \projrep\ or \classoprep.
%Nevertheless, the orthoalgebraic %lattice?
%structure of \histories\ is reflected to a certain extent in operator 
%relations in \projrep\ or \classoprep.  
%For example, in either case
(In either case, $\neg h$ is represented by $1-h,$ $h \oplus h'$ by $h+h',$ 
and so on.  Warning: The definitions of \projrep\ and \classoprep\ will
be extended shortly, after consideration has been given to the allowed
coarse grainings.)
%See aforementioned saying if the notation is confusing!)
%However, the operator relations corresponding to the application of the
%full {\it lattice} operations on histories which are not disjoint are not
%usually so clear.  See \cite{Isham} for some related discussion.}
%cf BEH p547, eg.
This distinction is useful because while it may be most proper for
the purposes of rigorous analysis or quantum logic to employ
\projrep, it is in fact the case that it is as the class operators
(\ref{eq:classop}) that histories which arise in actual physical 
problems enter.  
(This is more or less true even for the decoherence
functionals which have been suggested for spacetimes which contain
closed timelike curves \cite{HCTC,ArleyCTC}.  The ``more or less" refers
to the way in which a nonunitary time evolution enters into the problem.
This does not however change the fact that histories end up being
represented by a collection of operators in \linearops{\HilbertS}\ with
all essential properties the same as \classoprep; {\it cf.\ }the beginning 
of section \ref{sec:innerproduct}.)
%so far studied.  
%Namely, contain a basis for \linearops{\Hilbert}; ray-completeness of
%\classoprepbar\ and ray-incompleteness of \classoprep.
%I refer to this in a more specific way below.  Perhaps leave there.
Put another way, the decoherence functionals that 
have been discussed in the literature so far 
are sensitive only to the information about histories in \histories\
contained in \classoprep.
%Except disjointness ... but \decohif can't see that and doesn't seem to
%care much ... except comment above re: linear independence of a
%consistent set?
While the roots of this in ordinary quantum mechanics (as described above) 
are clear, I think it is safe to say that the meaning %implications
of this fact %,if any,
is not fully understood.

%Note hilbert space canonical D corresponds to path integral/propagator
%connection.  Must paths move forward in time for the equivalence?
%Think so, yes.

(For this reason, much of the work that follows is done with
\classoprep\ foremost in mind.  This applies particularly to 
sections \ref{sec:history?}\ and \ref{sec:dcanonical?}.  Nonetheless,
the general structural results of the present investigation are applicable to 
either method of representing the histories $h$, though a few are
more immediately {\it useful} for decoherence functionals on class
operators.  
%One important 
%difference will arise when it comes time to utilize the positivity
%conditions on the decoherence functional.  As will become clear in
%section \ref{sec:innerproduct},
%just how much one is allowed to infer from the positivity requirement
%depends on whether the domain of definition of $d$ is 
%\projrep\ or \classoprep, and %in the latter case
%even on what are the allowed coarse grainings.
%%imagined to be
%%only some lattice of projections \projlattice{\HilbertS} of some Hilbert
%%space \HilbertS, or the collection of history representatives 
%%$\{ C_{\beta}, \rm{possible\ additional\ operators\ representing\
%%certain\ coarse\ grained\ histories} \}.$  
%In any event, many of the strongest conclusions of this paper apply to
%any ``postive kernel" decoherence functional.  This is nice, not least
%because I will show shortly that the choice of the class operator
%representation, as in, for example, ordinary quantum mechanics,
%inevitably implies the decoherence functional is positive {\it if} 
%inhomogeneous coarse grainings (see the next paragraph) are permitted. 
%Moreover, the canonical decoherence functional which arises in ordinary
%Hamiltonian quantum mechanics is positive.
I will try and make a note of it whenever a result has a usefulness
more specific to one representation than the other.)

\subsubsection*{Homogeneous and Inhomogeneous Coarse Grainings: 
\projrepbar\ and \classoprepbar}
\addcontentsline{toc}{subsubsection}{\numberline{}%
Homogeneous and Inhomogeneous Coarse Grainings: 
\projrepbar\ and \classoprepbar}
%cf. The LaTeX companion section 2.4.2

Next, the permissible coarse grainings must be specified.  Recall that
a coarse graining is a partition of an exhaustive and exclusive
collection of fine grained histories in \histories\ into exclusive classes.
What is the meaning of this in \projrep\ and \classoprep?
Partitions of histories disjoint in \histories\ clearly correspond to 
operator sums in either representation.  Such operator sums divide
themselves into two distinct classes.
%The allowed coarse-grainings will be taken as either of two types.
In either representation, sums of projections {\it at one time} still 
have the form (\ref{eq:historyproduct}) or (\ref{eq:classop}), except
the projections are no longer required to be one dimensional.
Following Isham, any such histories are called ``homogeneous".
More generally, sums of homogeneous histories are {\it not} in general
again of the form (\ref{eq:historyproduct}) or (\ref{eq:classop}), but
are sums of such terms.
Such histories are called ``inhomogeneous".  (Neither homogeneous 
nor inhomogeneous class operators are projection operators in general,
but sums of {\it disjoint} projections (\ref{eq:historyproduct}) 
in \projrep\ are
always again projection operators.)  There are therefore two obvious
choices for the classes of permissible coarse grainings:  
``homogeneous" coarse grainings, where the allowed partitions are 
those for which the operator representatives $\overline{h}$ of coarse
grained histories are all required to be again homogeneous; or more 
generally, {\it arbitrary} partitions of exclusive, exhaustive sets 
might be allowed, corresponding to admitting arbitrary sums of operators
in exclusive, exhaustive sets $\consistentset \subset \projrep$ or
$\consistentset \subset \classoprep$ (``inhomogeneous" coarse
grainings, naturally.)   

From the point of view of generalized quantum
theory, there seems no obvious physical reason to exclude inhomogeneous 
coarse grainings.  Indeed, through the course of this investigation
it will become apparent that from the point of view of the quantum
mechanics of history, it appears unnatural to restrict to only
the homogeneous coarse grainings.  
%i)Ray-completeness => positivity => cauchy-schwarz => no interference
%from zero probability histories.
%ii)1-h is not in general a physical history unless inhomogeneous
%coarse-grainings are admitted.
%iii)
However, as emphasized by Hartle in \cite[section IV.2]{Lesh}, 
restricting to homogeneous coarse grainings ensures that we can maintain 
in ordinary quantum mechanics 
a description of the system \system\ 
by a single state vector evolving unitarily in time between ``wave function
reductions".  Admitting inhomogeneous coarse grainings is therefore to be
regarded as an extension of ordinary Hamiltonian quantum mechanics.
%Work out this picture explicitly.  Perpetually mixed states?  
%Note: inhomo => pure -> mixed generically.  Argument against, or for? :-)
%It is always possible to define a \rho_eff = h^\dagger\rho h/tr (ditto),
%but do you run into difficulties recovering the conditional probabilities
%required for its interpretation as a state?
%He also emphasizes the relationship between incogh's and branch dependence.
%Note that in either case we are admitting on the basis of democratic 
%principles {\it arbitrary} partitions the sets of fine grained histories.
Moreover, as will become evident, admitting inhomogeneous coarse grainings 
has stronger mathematical consequences.  I will therefore not gloss over 
the distinction between homogeneous and inhomogeneous coarse grainings.

In order to avoid wanton proliferation of 
notation,\footnote{Appendix B  %\ref{sec:dictionary} 
summarizes the important notation for ease of reference.} 
the definition of the history representation spaces \projrep\ 
and \classoprep\ is extended to include not only the fine grained
histories (\ref{eq:historyproduct}) and (\ref{eq:classop}), but also
all the histories that may be obtained from them by homogeneous 
coarse grainings.
%histories that can be obtained by operator sums of
%disjoint histories in them, 
\projrepbar\ and \classoprepbar\ will be used to denote the even
larger operator spaces obtained by including 
%in \projrep\ and \classoprep\ 
the inhomogeneously coarse grained histories as well.
These spaces are then Hilbert space realizations of the
abstract space of all histories \histories.  

It should be reasonably obvious that
\begin{equation} \label{eq:subsetsequenceproj}
\projrep(\otimes^k\HilbertS)\subset\projrepbar(\otimes^k\HilbertS)
\subset\projlattice{\otimes^k\HilbertS}
\subset\linearops{\otimes^k\HilbertS}
\end{equation}
and that
\begin{equation} \label{eq:subsetsequenceclassop}
\projlattice{\HilbertS}\subset\classoprep(\HilbertS)
\subset\classoprepbar(\HilbertS)\subset\linearops{\HilbertS},
\end{equation}
where $\projlattice{\Hilbert}$ is the collection (lattice) of
projections on \Hilbert.  
All of these inclusions are proper so long as $k>1$. 

Note that 
$\projrepbar(\otimes^{k}\HilbertS) \neq \projlattice{\otimes^{k}\HilbertS}$ 
because \projrepbar\ includes only sums of {\it disjoint} projection
operators, and it is {\it not} true that all projection operators on
$\otimes^{k}\HilbertS$ can be written as sums of disjoint 
homogeneous projections
\cite{IL1}; those which cannot are termed ``exotic".  
Their physical interpretation remains at best unclear.
(Isham, Linden, and Schreckenberg \cite{ILS} understandably
circumvent this irritation by fiat, choosing to work 
with $\projlattice{\otimes^{k}\HilbertS}$ exclusively.)

The meaning of the inclusions (\ref{eq:subsetsequenceproj}) and
(\ref{eq:subsetsequenceclassop}) is that \classoprep\ is ``larger''
than \projrep, in the sense that \classoprep\ ``fills up'' more
of the space of operators ${\cal L}$ in which it resides than does
\projrep.  The most striking manifestation of this observation is
the ``ray-completeness'' property of $\classoprepbar(\HilbertS)$ in
$\linearops{\HilbertS}$ that I discuss just below.  This property
actually implies that a decoherence functional on \classoprepbar\
is a positive functional, as will be seen in section
\ref{sec:innerproduct}.  Taking the representation space to be
\classoprepbar\ therefore has stronger mathematical consequences
than taking the representation space to be anything else.

To avoid needless repetition, \rep\ will frequently stand in for
either of \projrep\ or \classoprep, and \repbar\ similarly for
\projrepbar\ or \classoprepbar.  (I will not always bother to
say ``\rep\ {\it or} \repbar'' when it is reasonably clear
that either will do.)  $\Hilbert$ will similarly stand for either
$\HilbertS$ or $\otimes^k\HilbertS$, as appropriate.
The reason it is useful to consolidate
the discussion of decoherence functionals on these representation 
spaces is that most of the structural results for decoherence
functionals discussed in sections \ref{sec:hermitianform}\ and
beyond depend in no way on the details of the chosen representation
space \rep.  It is enough that all of these contain a basis for
the operator space in which they are contained, as I now discuss.

\subsubsection*{How Much of $\linearops{\Hilbert}$ 
Does $\rep(\Hilbert)$ Fill?  The Ray-Completeness of \classoprepbar}
\addcontentsline{toc}{subsubsection}{\numberline{}%
How Much of $\linearops{\Hilbert}$ 
Does $\rep(\Hilbert)$ Fill?  The Ray-Completeness of \classoprepbar}
%cf. The LaTeX companion section 2.4.2

There are two points that call for emphasis.  First, by definition
it is only sums of {\it disjoint} histories which are defined in 
\rep\ or \repbar\ by coarse graining.  
The second, related, point is that the \rep's {\it are not linear spaces};  
linear combinations of histories in \rep\ are not usually again histories.  
Nevertheless, in spite of not being linear spaces,
$\projrep(\otimes^{k}\HilbertS)$ contains a basis for
$\linearops{\otimes^{k}\HilbertS},$ 
and $\classoprep(\HilbertS)$ likewise contains a basis of 
\linearops{\HilbertS}.  In other words, {\it the linear, or vector space, 
completion of any of the spaces
$\rep(\Hilbert)$ is the full space of operators \linearops{\Hilbert}.}

This is true  
whatever the admitted coarse grainings; under our democratic assumptions a 
basis is already contained in the full collection of fine grained histories 
(\ref{eq:historyproduct}) or (\ref{eq:classop}).  Moreover, so long as 
$k>3$ (which I shall henceforth always assume), and inhomogeneous coarse 
grainings are permitted, \classoprepbar\ {\it contains a vector 
(operator) along every ray in} \linearops{\HilbertS}.  This is
the ``ray-completeness" property of \classoprepbar(\HilbertS) 
in \linearops{\HilbertS}.  (This statement is not meant to be obvious.  
A proof appears in appendix \ref{sec:raycomplete}.)
In this sense, \classoprepbar(\Hilbert) contains, up to normalization, 
 {\it every} basis for \linearops{\HilbertS}, 
The ray-completeness property is {\it not} true of \projrepbar, 
or even \classoprep, {\it i.e.} if only homogeneous coarse grainings 
are allowed.  

The fact that the linear completion of each representation space
\rep\ is the full space of operators ${\cal L}$
will become the basis for extending the domain of 
definition of $d$ to all of ${\cal L}$ in the next section.
The ray-completeness of \classoprepbar\ is strong enough to imply
that a decoherence functional defined defined on \classoprepbar\ 
is always a positive functional when extended to ${\cal L}.$
(That is not true for any of the other \rep's.)

The rest of this section, and all of the next 
(section \ref{sec:historyreconstruction}), 
should probably be skipped on a first reading.

To close this subsection, let me emphasize the fact that the \rep's 
are not linear spaces by relaying some observations
about the structure of \projrep\ and \classoprep\ that will be 
of some importance as we go along.
First, for finite temporal 
supports (finite $k$), being all projection operators,  
$\projrepbar(\otimes^{k}\HilbertS) \subset 
       \ball_{1}(\linearops{\otimes^{k}\HilbertS}).$
That is, \projrepbar\ is contained in the unit ball in the space of linear
operators on $\otimes^{k}\HilbertS$ (in the usual norm, or uniform operator,
topology.  The norm here is the standard operator norm on a Hilbert
space \Hilbert, 
$\| {\cal O} \| = \sup_{ \| x \|_H = 1 } \| {\cal O}x \|_H.$)
Similarly, as each homogeneous class operator (\ref{eq:classop})
is merely a product of projection operators, the norm of each 
such homogeneous class operator is %dominated 
bounded by 1, $\| C_{\beta} \| \leq 1,$ so that
$\classoprep \subset \ball_{1}(\linearops{\HilbertS}).$
%proper subset not least because \classoprep is not ray complete.
However, \classoprepbar\ can be considerably larger, as inhomogeneous
coarse grainings permit histories which are sums of many class operators
which point in almost the same ``direction".  An upper bound in finite
dimensions comes from the observation that a coarse grained history is a
sum of fine grained ones, and with ${\rm dim\, }\HilbertS = N,$ there
are at most $N^{k}$ such fine grained histories.  Therefore, the
triangle inequality tells us that 
%physically disjoint histories in any
%exclusive, exhaustive set \consistentset\ (that is, there is a choice 
%of at most $N$ orthogonal projections at each time $t_{i}$), via the
%triangle inequality the most coarse grained history possible has norm
%less than or equal to the most number of terms in the sum defining it,
%$N^{k}.$  Thus 
$\classoprepbar(\HilbertS)\subset\ball_{N^{k}}(\linearops{\HilbertS}).$
%\footnote{}???
(One might wonder if the fact that each $ h \in \classoprepbar $ appears
in some ``partition" of unity, $ 1 = \sum h,$ can strengthen this bound
much.  At least in some specific cases, it can.  Suppose 
$1 = h + \overline{h},$ supposing further that $h$ $(\overline{h})$
is a sum of $n$ $(\overline{n})$ fine grained histories, where of course
$n + \overline{n} = N,$ so that 
$\| h \| \leq n$ $(\| \overline{h} \| \leq \overline{n}).$  Then
$\| h \| = \| 1- \overline{h} \| \leq 1 + \|\overline{h}\| 
\leq 1 + \overline{n}.$  This doesn't seem to be of much help in
general.   An exception is cases where $\overline{h}$ is homogeneous,
so that $\| \overline{h} \| \leq 1$ (despite $\overline{n} > 1)$ and
therefore $\| h \| \leq 2.$)

%\newpage

\subsection[Reconstruction of History Sequences from History 
Operators]{Reconstruction of History Sequences 
from History Operators\protect\footnote{This section
should be skipped on a first reading.  It is placed
here for logical coherence.}}
\label{sec:historyreconstruction}

%Role of times tk and the U's?

It was remarked in the paragraph following (\ref{eq:classop}) that given
only two homogeneous class operators, there is no way in general to
determine whether or not they represent disjoint physical histories
(\ref{eq:historysequence}).  Continuing this discussion, 
it is interesting to enquire, given an operator in \projrep\ or
\classoprep, to what history (\ref{eq:historysequence}) does it
correspond?  A closely related problem, which will be of some 
importance later on, is, how does one determine whether an operator
$h\in\linearops{\HilbertS}$ ($h\in\linearops{\otimes^{k}\HilbertS}$ )
is actually a member of \classoprep\ (\projrep)?

Unfortunately, at present there do not appear to be easy answers to
these questions.

\subsubsection*{Physical History Operators and \projrep}
\addcontentsline{toc}{subsubsection}{\numberline{}%
Physical History Operators and \projrep}
%cf. The LaTeX companion section 2.4.2

First consider \projrep.  It is of course true that every element $h$ 
of \projrep\ is a projection operator, $h^{\dagger}h=h.$  However, as noted
above, not every projection operator on $\otimes^{k}\HilbertS$ can be
written as a sum of disjoint projections of the form
(\ref{eq:historyproduct}).  Those projections which cannot have been
termed ``exotic" histories by Isham; their physical interpretation, if
any, remains obscure.  (This is one reason to remember, at least, that
$\projrepbar(\otimes^{k}\HilbertS)$ is actually only a proper subset of
$\projlattice{\otimes^{k}\HilbertS}.$)  
An extension of the methods of section \ref{sec:dcanonical?} should
make it possible to determine algebraically whether or not a history is
a simple product of projections as in (\ref{eq:historyproduct}), that is,
is a member of \projrep.
%Really?!  Carry this prgram through.  Yucks.
Moreover, as in section \ref{sec:dcanonical?}, it should be possible to
construct the individual projections explicitly.  However, there is 
no obvious technique for determining whether or not more general
projection operators can be written as sums of {\it disjoint} projections, 
that is, are actually members of \projrepbar.

\subsubsection*{Physical History Operators and \classoprep}
\addcontentsline{toc}{subsubsection}{\numberline{}%
Physical History Operators and \classoprep}
%cf. The LaTeX companion section 2.4.2

The situation is considerably worse for \classoprep.  As noted 
(and seen explicitly in appendix A), %\ref{sec:raycomplete}), 
there are in 
general {\it many} histories (\ref{eq:historysequence}) which might give
rise to a given class operator.  Further, at present there are no
techniques for constructing this family of histories given only the
class operator.  
%though the first and last $P(t)$'s can be inferred from
%${\rm Ran}\, C_{\beta}$ and ${\rm Ran}\, C_{\beta}^{\dagger}.$
Even worse,
unless a class operator is actually just a projection, there does not
appear to be any easy (practical) way to tell in general whether or not
an arbitrary operator in \linearops{\HilbertS}\ is actually a class
operator at all, {\it i.e.\ }is a member of \classoprep\ or \classoprepbar.  
This will become an issue in section \ref{sec:history?}\, after I have
shown how to extend the domain of definition of a general decoherence
functional from \rep(\Hilbert) to $\hilbert = \linearops{\Hilbert}.$
There I show how to solve the problem of determining what are all the
``histories" (operators in \hilbert) which are consistent according to a
given decoherence functional.  In \classoprep\ or \classoprepbar\ it
remains to determine which of these operators actually correspond to
physical histories.

\subsubsection*{Bounds on the Norm of Physical History Operators}
\addcontentsline{toc}{subsubsection}{\numberline{}%
Bounds on the Norm of Physical History Operators}
%cf. The LaTeX companion section 2.4.2

The best that can be made of the situation is the following.

By assumption (\ref{eq:positivity}), $d(h,h) \geq 0$ if $h \in \rep.$
%(\rep\ is obviously {\it not} obligated to be \classopprep\ in this
%discussion, and that of the following paragpraph.)
Therefore, if some operator $n$ is found for which $d(n,n) < 0,$ it
obviously does not correspond to any physical history.\footnote{In 
passing I note that it is true that decoherence functionals are 
always positive on the {\it span} of each of their (physical) consistent 
sets of histories in \hilbert; see section \ref{sec:innerproduct}.}  
Similarly, suppose $ h \in \decoheringsubd,$ {\it i.e.\ }$h$ 
appears in at least one set of histories consistent according 
to $d.$  (In section \ref{sec:history?} I show that
$\decoheringsubd = \{ h \in \rep \|\ d(h,1-h) = 0 \}.$)  Then
(\ref{eq:positivity}) and (\ref{eq:normalization}) are sufficient to
imply that $d(h,h) \leq 1.$  (That $0 \leq d(h,h) \leq 1$ for any
consistent $ h \in \rep$ is, of course, much of the motivation for the
precise form of the definition of the decoherence functional.)  
Thus, any operator $n$ for which $d(n,n) > 1$ is not an element of 
\decoheringsubd; though it is possible for {\it in}consistent
$h \in \rep$ to have $d(h,h) > 1,$ precisely for this reason such 
histories are not usually of interest.   

On the other hand, $0\leq d(h,h) \leq 1$ is not (of course) sufficient
to imply that $h \in \rep.$  Indeed, $0 \leq d(h,h) \leq 1$ for {\it any}
``consistent" operator $h \in \hilbert = \linearops{\Hilbert}$ if $d$
is positive on \hilbert.   This includes in particular the important
case of the canonical decoherence functional.  

(Given the definition of
\decoheringsubd\ noted in the previous paragraph, it will come as no
surprise that the usage of the term ``consistent" here merely means that
$h \in \consistentops,$ where \consistentops\ is defined in the same way
as \decoheringsubd, namely,
$\consistentops = \{ h' \in \hilbert \|\ d(h',1-h') = 0 \}.$
In other words, \consistentops\ is the generalization of
\decoheringsubd\ to {\it any} operator in \hilbert, not just
those that correspond to physical histories.
If $d$ is positive and $h \in \consistentops,$ $d(h,h) \leq 1$ for the
same reason that is true if $h \in \decoheringsubd.$  Similarly, it is
useful to define \consistentopsets, an operator analogue of 
\consistentsubd, as the collection of exhaustive ($\sum h = \OneH$),
mutually consistent sets of operators.  Clearly,
%$\decoheringsubd \subset \consistentops$ and 
%$\consistentsubd \subset \consistentopsets.$)
$\decoheringsubd = \consistentops\cap\rep$ and 
$\consistentsubd = \consistentopsets\cap \{ \consistentsetsubd \}.$)

Thus, if only homogeneous class operators are permissible 
($\rep = \classoprep$), the ``probability" of an operator
$p(h) = d(h,h)$ is insufficient to tell us whether or not
$h \in \classoprep.$

In \classoprepbar, the situation is slightly better.  To see why,
remember the ray-completeness of \classoprepbar.   
In fact, as shown in appendix A, %\ref{sec:raycomplete}, 
$\ball_{\frac{1}{8}}(\linearops{\HilbertS})\subset\classoprepbar(\HilbertS),$
so that if $\|h\|\leq\frac{1}{8}$ we know that it is {\it some} class
operator, even if we can't reconstruct all of its associated histories
(\ref{eq:historysequence}).\footnote{Now, one might wonder whether or
not it is possible to bound $\| h \|$ by $p(h)$ in some fashion.
Unfortunately, this does not appear to be the case.  Instead, one can
demonstrate inequalities like
$\| h \|_{d} \leq C_{d}\, \| h \|$ and
$\| h \|_{d} \leq C_{d2}\, \| h \|_{2},$ where 
$\| h \|_{d} = \sqrt{|d(h,h)|},$ $\| h \|_{2}^{2} = \trH h^{\dagger}h,$
and the $C_{d}$'s are constants depending only on $d.$  Working with a
general decoherence functional in the form (\ref{eq:diracd}), these are
simple consequences of standard trace-norm inequalities that can 
be found in, for instance, \cite[chapter VI]{RS}.}  
%As they don't appear to offer us much of use, I leave them as
%exercises for the bored.}
As noted in the first appendix, it is possible to enlarge this ball as $k$ 
or $N = {\rm dim}\, \HilbertS$ increases.  Recall on the other hand that 
$\classoprepbar(\HilbertS)\subseteq\ball_{N^{k}}(\linearops{\HilbertS})$
(see the end of the previous section), so that the ball containing
$\classoprepbar(\HilbertS)$ increases with $k$ and $N$ as well.
Thus, the operator norm of $h$ is not in general of much use in fixing
whether or not $h \in \classoprepbar$.

\subsubsection*{Why it Matters}
\addcontentsline{toc}{subsubsection}{\numberline{}%
Why it Matters}
%cf. The LaTeX companion section 2.4.2

The difficulty here is analogous %not much different than 
to the inability in conventional applications of quantum theory 
to decide to what physical quantity an arbitrary self-adjoint 
operator corresponds.  However, the inability to reconstruct histories 
from class operators is a somewhat more severe complication.
For, suppose all of the ``histories" (operators) consistent according 
to a decoherence functional over 
$\classoprep(\HilbertS)$  -- that is, \consistentops\ -- have been found.  
(In section \ref{sec:history?}\ I show how to do this.)
What good is this information if we cannot determine to what {\it physical}
histories these operators correspond?  
Formally, of course, to find \decoheringsubd\ we can just
intersect \consistentops\ with $\classoprep(\HilbertS)$, that is, 
$\decoheringsub{d} = \consistentops\cap\rep,$
but that is not very comforting to a physicist.
%if we cannot determine exactly what \rep\ is.
At present, we can only make much {\it practical} use of the converse 
problem of determining whether a given set physical histories is consistent.

%And in \projrep, just what is the physical significance of the subspace
%we are projecting into?

Thus, there are at least two important open problems.  First, 
is there a practical test for determining whether or not 
$h \in \rep$?  This is non-trivial
so long as $\rep(\Hilbert)\neq\projlattice{\Hilbert}.$  Second,
given a class operator (homogeneous or inhomogeneous), how does one
construct the family of histories which it represents?  
%These remain projects for the future.  
It remains a project for the future to see if this situation can be
improved; one intriguing possibility is that the operators in
\consistentops\ not in \classoprep\ may have some physical
interpretation after all.\footnote{I would like to thank J.B. Hartle 
for emphasizing this possibility to me.}
I will not permit such
obstacles to deter me from analyzing other algebraic properties of
decoherence functionals and their consistent sets.  Nor do they otherwise
mitigate the usefulness of the methods and results introduced here.

\subsection{The Canonical Decoherence Functional}
\label{sec:canonicald}

%Why not make this section IIB?  Because I haven't defined
%class operators yet.

Finally, to end this section, it is time to exhibit the ``canonical" 
decoherence functional of ordinary quantum mechanics, in which boundary 
conditions are imposed at some initial and final times.  Define
$d_{\alpha\omega}:\classoprepbar\times\classoprepbar \rightarrow \Complex$ 
by 
\begin{equation}\label{eq:dcanonical}
    \decohif{h}{h'} = \trif{h}{h'},
\end{equation}
where \rhalpha\ and \rhomega\ are positive Hermitian operators on
\HilbertS\ whose product is required to be trace-class, and normalized,
${\rm tr}[\rhomega\rhalpha] = 1.$  (Out of laziness it is hard
not to refer to  the initial and final boundary conditions
\rhalpha\ and \rhomega\ as ``density operators",
even though they may not {\it individually} be 
trace class, {\it e.g.\ }in the familiar case where 
$\rhomega=1$ on an infinite dimensional
space.  Also, it is of course possible to consider restricting the
domain of $d_{\alpha\omega}$ to $\classoprep\times\classoprep$, however,
there appears to be no particular profit in this constraint unless
$\rhomega=1$, in which case the restriction to \classoprep\ 
guarantees that the familiar notion of ``quantum state at a moment
of time'' may be defined \cite[section IV.2]{Lesh}.)

The origin of the definition (\ref{eq:dcanonical}) 
lies in the old \cite{Groenewold,Wigner}\
formula for the probability of measuring the eigenvalue $a$ of
observable $A$, $p(a) = {\rm tr}[\rho P_{a}],$ where $\rho$ is the
initial density matrix.  Using the fact that projections
satisfy $P_{a}P_{a'}= \delta_{aa'},$ writing it out
it is not difficult to see
that ${\rm tr}[C_{\beta}^{\dagger}\rho C_{\beta}]$ is the joint
probability for the alternatives in the history represented 
by $C_{\beta}$ ({\it cf.\ }(\ref{eq:classop})) to be realized
upon a sequence of measurements.   Some thirty years ago,
\cite{ablts} extended this formula to the case where there 
is also a nontrivial final boundary condition \rhomega.  
Finally, Gell-Mann and Hartle \cite{GMH}\ generalized the
probability formula to measure the interference between histories, 
as well as determine their probabilities.  

%(As an aside, let me very briefly describe how one constructs the ILS
%operator $X_{\alpha\omega}$ on $\otimes^{2k}\HilbertS$ corresponding
%to the decoherence functional (\ref{eq:dcanonical}).   Using the fact
%that $\trH\, A\trH\, B = \trHH\, A\otimes B,$  inserting 1 in between
%the projection operators appearing in the class operators and making
%judicious use of ``shift" and ``reversal" operators on the prodcut
%hilbert space allows one to write (\ref{eq:dcanonical}) as a trace
%on $\otimes^{2k}\HilbertS.$  The histories $h$ and $h'$ now appear
%in the form
%\begin{equation}
%\Projsupb{1}{\betasub{1}}\otimes\cdots\otimes\Projsupb{k-1}{\betasub{k-1}}
%\otimes\Projsupb{k}{\betasub{k}}\otimes\Projsupb{1}{\betaprimesub{1}}
%\otimes \cdots\otimes\Projsupb{k-1}{\betaprimesub{k-1}}\otimes
%\Projsupb{k}{\betaprimesub{k}},
%\end{equation}
%the boundary conditions and the unitary time evolutions having 
%been ``factored out".  The messy factor that remains is the ILS
%operator $X_{\alpha\omega}.$)
%Is this process invertible?  Check, durnit.
%Nope.  The partial trace that folds the bc into $X$ is not invertible.

For culture, I note that it is possible by the standard methods
to rewrite (\ref{eq:dcanonical})
(or, indeed, any decoherence functional on \classoprepbar ) 
as a functional integral, but that is not a 
topic that will be explored here.

In spite of the rather general tone of this work, the canonical
decoherence functional is obviously a most important case to consider,
not least because (with $\rhomega = 1$) it is the decoherence functional 
of ordinary quantum mechanics.
It has several nice properties, among the most useful of which is that
it is positive on all of $\hilbert = \linearops{\hilbert}$ (see section
\ref{sec:example}), which in turn implies other useful properties such
as the Cauchy-Schwarz inequality (\ref{eq:cauchyschwarz}).

\section{The Decoherence Functional as Hermitian Form}
\label{sec:hermitianform}                                  

This portion of the paper
is devoted to a discussion of the decoherence functional
as an operator on histories.  Section \ref{sec:extension}\ shows how 
a decoherence functional naturally defines an Hermitian form on
the full space of operators in which the histories live.  Sections
\ref{sec:maxnumhistories}\ and \ref{sec:innerproduct}\ discuss
some immediate consequences of this observation, a bound on the
maximum number of histories in a consistent set, and a general
Cauchy-Schwarz inequality for positive decoherence functionals.
Section \ref{sec:notation}\ makes use of the point of view that
$d$ is an operator on histories by introducing some familiar and
useful notation in the context of the quantum mechanics of history, 
and develops some powerful %pertinent calculational 
tools for calculating with a general decoherence functional.  
Section \ref{sec:ILSrevisited}\ shows how the
ILS theorem classifying decoherence functionals pops out almost
magically from this formalism, and section \ref{sec:example}\
exhibits the canonical decoherence functional as an %illustrative 
example illustrative of the preceeding sections.

\subsection{Extending the Domain of $d$ from $\rep(\Hilbert)$
to \hilbert = \linearops{\Hilbert}}
\label{sec:extension}

%\projrep or \classoprep - don't have to arise from a theory w/a time,
% or usual causal relations.  Generalize the discussion by not asking
%about the {\it origin} of the operator representations \projrep\
%or \classoprep\ of \histories.  Let us merely consider the consequences
%of making the alternate choices 
%$d:\projrep\times\projrep \rightarrow\ \Complex$ 
%(the general quantum logical case considered by Isham {\it et al.})
%or the case which arises in conventional quantum mechanics,
%$d:\classoprep\times\classoprep \rightarrow\ \Complex.$ 

In this section I show how -- relying on a result of Wright 
\cite{Wright} -- to extend straightforwardly the domain of
definition of  a decoherence functional $d$ to all of the operators
on the pertinent Hilbert space (\HilbertS\ or $\otimes^{k}\HilbertS,$
in the preceeding section.)  $d$ is then an inner-product like form,
and the positive $d$'s (like the ones arising in ordinary quantum
mechanics) actually {\it are} semi-inner products on ${\cal L}.$  
As evidence of the utility of this geometric perspective, some consequences 
of this identification are discussed in the subsequent subsections.  
These include a Cauchy-Schwarz inequality for 
positive decoherence functionals, a bound on
the maximum number of histories in any consistent set (therefore
implying in general a fact which has long been observed in practice,
that some coarse graining is inevitably required for decoherence),
and more generally, a very useful geometric point of view on the 
nature of consistent
sets of histories and the decoherence functionals which define them.

The properties of decoherence functionals studied in the
sequel do not generally depend on the specific structure
of the space of operators chosen to represent quantum
histories, but rather depend only on the algebraic
properties of the decoherence functional embodied
in (\ref{eq:hermiticity}-\ref{eq:normalization}).
The only property of the history representation spaces 
that is truly essential is that their linear completion
is the entire space of operators in which they are 
contained.\footnote{An exception to the rule that the specific 
choice of \rep\ does not matter very much concerns the positivity
of $d$.  As emphasized in section \ref{sec:innerproduct},
the ray-completeness property of \classoprepbar\ ({\it cf.\ }the
end of section \ref{sec:hilbertGQT}) is strong enough to imply 
that a decoherence
functional on \classoprepbar\ is actually positive on all
of \linearops{\HilbertS}.  The equivalent statement is {\it not}
true for any of the other \rep's, which as noted in
(\ref{eq:subsetsequenceproj}-\ref{eq:subsetsequenceclassop}) 
do not ``fill up'' as much of \linearops{\Hilbert} as does
$\classoprepbar(\Hilbert)$.}

(Actually, as we shall see in a moment, rigor currently
requires that $\projlattice{\Hilbert}\subseteq\rep(\Hilbert)$ 
for the results of this paper to be rigorously  
guaranteed to apply to any $d$ on \rep.)

Let me therefore condense and unify the notation and treatment a bit.
To be specific, I am going to consider decoherence functionals
$d:\projrep\times\projrep \rightarrow\ \Complex$ or
$d:\classoprep\times\classoprep \rightarrow\ \Complex$ 
(or $\projrepbar\times\projrepbar$ or
$\classoprepbar\times\classoprepbar$).
Simplify, as before, by letting \rep\ stand for
\projrep\ {\it or} \classoprep, and \repbar\ for
\projrepbar\ or \classoprepbar.  
Similarly, \Hilbert\ will stand for either
of \HilbertS\ or $\otimes^k\HilbertS$, 
as appropriate.\footnote{As mentioned 
in section \ref{sec:generalities}, 
no {\it a priori} notion of time is required for the 
definition of a generalized quantum theory.  In the
general case, therefore, there will be no immediate analog
of the representation spaces \projrep\ or \classoprep\
that are present in theories with a time.  However, though 
there of course are no general rules for what collections 
of operators can represent the ``histories'' of an
arbitrary generalized quantum theory, from the
point of view of quantum logic it would not be unusual 
to assume that there is a representation in terms
of something like the lattice of projections on
a Hilbert space.  In any case, as just argued, very
little of what follows depends on the details of
the choice of \Hilbert\ -- for most purposes, all that
really matters is that the linear completion of
\rep(\Hilbert) is \linearops{\Hilbert} -- 
so that most of the 
discussion should be applicable 
to any Hilbert space quantum theory.}
The discussion will therefore be about a general 
decoherence functional $d:\rep\times\rep\rightarrow\ \Complex$
(or $d:\repbar\times\repbar\rightarrow\ \Complex$) whose
histories are operators on some Hilbert space \Hilbert.

With this simplification in hand, let me proceed.

The goal is to extend a decoherence functional $d$ defined on \rep\ 
(or \repbar)\footnote{An addendum I shall from here 
on often not bother to repeat.} 
to a sesquilinear form defined on all of 
$\hilbert = \linearops{\Hilbert}.$  
Because the linear completion of any
of the spaces $\rep(\Hilbert)$ is \linearops{\Hilbert}, 
this is, on the face of it, straightforward.
%\footnote{The attentive reader may
%have noticed that $d$ was defined in section \ref{sec:generalities}\
%to be a functional on any pair of histories in \rep.  As was pointed
%out to me by Jim Hartle, this is a slight
%generalization from the original definition \cite{Jerusalem,Lesh}, which
%defined $d$ only on pairs of histories in exclusive, exhaustive sets.
%The generalization is of no particular consequence, however, because
%there are in general many exclusive, exhaustive sets which contain a
%basis for \hilbert, and the discussion therefore proceeds
%substantially unaltered.}  %Well, that ain't true for \projrep, 
%%though of course \projrep does contain a basis.
We are given an Hermitian, positive,
normalized function $d: \rep\times\rep \rightarrow \Complex$ which is
bi-additive on disjoint histories.  
%i.e. a positive, Hermitian quantum bi-measure.
Leaving aside positivity and
normalization for the moment, begin by defining a bilinear functional
$b:\hilbert\times\hilbert \rightarrow \Complex$ by
\begin{equation} \label{eq:sesquilinearity1}
b(h,\alpha h' + \beta h'') = 
     \alpha\, \decoh{h^{\dagger}}{h'} + \beta\, \decoh{h^{\dagger}}{h''}
\end{equation}
%$\forall\ \alpha, \beta \in \Complex$ 
for all complex numbers $\alpha$ and $\beta$ and for all histories
$h, h', h'' \in \rep$  (whether or not they are disjoint; the sum on the
left hand side is well defined in \hilbert\ irrespective, and this 
definition thereby extends additivity of $d$ to {\it all} 
pairs of 
%histories\footnote{The daggers are clearly 
%only relevant to the case $\rep = \classoprep.$  
%Well, they are necessary in the {\it next} equation.  So don't bother
%with this; they'll figure it out.
histories),
%\footnote{It is perhaps worthwhile mentioning %also 
%%nah
%that if $\rep = \projrep,$ the sum on the left hand side is {\it not} 
%the lattice ``join" operation, but the operator sum in \hilbert.  
%If $h'$ and $h''$ are not disjoint these will not in general agree.}), 
and similarly
%The definition is self-consistent because when \alpha=\beta=1
%and h' and h'' are disjoint, the join {\it is} just the operator sum.
%Possibly this could be interpreted as small cheat - depending
%on how you interpret the wording of my definition, d is only 
%supposed to be defined on pairs of disjoint histories.
%why torture yourself with small stupidities?  I changed the wording.  
\begin{equation} \label{eq:sesquilinearity2}
b(\alpha h' + \beta h'',h) = 
   \alpha\, \decoh{{h'}^{\dagger}}{h} + \beta\, \decoh{{h''}^{\dagger}}{h}.
\end{equation}
By obvious abuses of notation, finally define  %d AND h's, don't ya know
$d:\hilbert\times\hilbert \rightarrow \Complex$ by
\begin{equation} \label{eq:sesquilinear}
\decoh{h}{h'} = b(h^{\dagger},h').
\end{equation}
The $d$ defined by (\ref{eq:sesquilinear}) is then an Hermitian, 
normalized, sesquilinear form on \hilbert\
that equals $d$ when restricted to $\rep\times\rep.$

Recall that all the \rep's contain at least one basis of \hilbert, so the
extension is defined on all of \rep.  
Moreover, as finite linear combinations of 
vectors in \rep\ generate \hilbert\ in finite dimensions 
-- the collection of finite linear combinations is actually dense 
in \hilbert\ in infinite dimensions --
this extension is essentially unique.
%by continuity \leftrightarrow boundedness
%Comment out that remark?  Yep ....
In order to guarantee that the extension is consistently defined,
%-- in other words, to guarantee that it never occurs that
%$d(\cdot,h'+h'') \neq d(\cdot,h') + d(\cdot,h')$ when
%$h'$ and $h''$ are not disjoint -- 
we must appeal to a result of Wright \cite{Wright}, 
which amounts even in infinite dimensions to a rigorous 
demonstration that a bounded decoherence functional 
on \projlattice{\Hilbert}\ can be linearly extended
in this way.\footnote{It is
important not to underestimate the importance of this highly non-trivial
result to the present work; it is the key to the general applicability
of the results obtained here to decoherence functionals other than
the (manifestly sesquilinear) canonical one.  
Many of these ideas were developed before the 
paper by Wright appeared, but it is only his results that guarantee 
that the extension of $d$ is well defined.}
%In particular, Wright showed that a general decoherence
%functional may be written as a difference of two semi-inner products 
%on \hilbert.  (From the present point of view, that simply amounts
%to the observation that any Hermitian form may be written as a
%sum of positive and negative semi-definite pieces.)
%%Note the extensions guaranteed by Wright must agree with rhs of
%%(\ref{eq:sesquilinearity1-2}); just expand an arbitrary
%%$h\in\hilbert$ until it is a linear combination of pieces in \rep.

%Concession speech:

We must be somewhat careful here.  Wright guarantees for us that the
extension described above is consistent for decoherence functionals
defined on \projlattice{\Hilbert}.  However, \projrep\ and \projrepbar\ 
are proper subsets of \projlattice{\Hilbert}; it is not obvious 
(at least to me) whether or not the same guarantee holds as well 
for decoherence functionals defined {\it a priori} only on 
\projrep\ or \projrepbar.  I shall assume that it does.
%If that assumption is ever found to be false, then we will have
%found an unusual class of decoherence functionals with properties
%that are likely to be unphysical.
%Otherwise, we must identify the class of decoherence functionals
%on which the assumption fails, and exclude them from the
%present discussion.
(This is one reason ILS are willing to tolerate the presence in
\projlattice{\Hilbert}\ of the ``exotic'' projections mentioned
below (\ref{eq:subsetsequenceclassop}).  Failure of (sesqui-)linear 
extendibility on \projrep\ or \projrepbar\ would provide even
stronger motivation to find a physical interpretation
for the ``exotic" projections in \projlattice{\Hilbert}.)
%Isham notes in \cite{ishamintro}\ that he and Linden think that
%any projection can be reached from a homo by the full lattice
%operations, but that the physical significance of this is unclear.

On the other hand,
$\classoprepbar(\Hilbert)\supset\classoprep(\Hilbert)\supset
\projlattice{\Hilbert}.$  
In this sense, the decoherence functionals on \classoprepbar(\Hilbert)\
are a {\it subset} of the decoherence functionals on \classoprep(\Hilbert),
which are in turn a subset of the decoherence functionals on
\projlattice{\Hilbert}.  Wright showed that every decoherence
functional on \projlattice{\Hilbert} may consistently be 
(sesqui-)linearly extended; therefore every decoherence functional
on \classoprep\ and \classoprepbar\ may be as well.
%Wright's first proposition is about quantum bi-measures, which
%is really what I'm talking about here (as decoherence functionals
%are special quantum bi-measures.)   Every qbm on R_c is a qbm
%on R_p, so because the qbms on R_p are linearly extendible,
%so must the qbms on R_c.  
%%No need at all to worry about restricting the domain first.
%It is possible to define a sesquilinear
%form on \linearops{\Hilbert}\ as above by restriction of a $d$
%defined on \classoprep(\Hilbert) to \projlattice{\Hilbert}.
%This restriction procedure not necessary, is it.
%Is it always true that this form agrees with $d$ when evaluated
%on a history which is in \classoprep(\Hilbert) but not in
%\projlattice{\Hilbert}?
%I shall assume that it is; it would be interesting 
%to discover that this assumption fails.
%As the extension guaranteed by Wright is unique, failure would mean 
%that on such $d,$ (\ref{eq:sequilinearity1}-\ref{eq:sesquilinear})
%yield {\it inconsistent} results.

The extension of the domain of $d$ to all of \hilbert\ has the distinct
virtue of making available the full arsenal of algebraic tools for
studying Hermitian operators.  In spite of the issues raised in
section \ref{sec:historyreconstruction}, I will frequently lapse into
the habit of referring to any element of \hilbert\ as a ``history" or
``history operator".

\subsection{The Maximum Number of Histories in a Consistent Set}
\label{sec:maxnumhistories}

It is well known that decoherence in general requires some coarse
graining.  In this section I explain one of the mathematical origins of
this phenomenon, that there is an upper bound on the number of histories
with non-zero probability in any consistent set.  This bound turns
out to be ${\rm dim}\, \hilbert/\nullspacesubd$, where $\nullspacesubd$
is the nullspace of $d$ in \hilbert.  As we shall see, this is a simple 
consequence of the interesting fact (demonstrated below) 
that histories with non-zero
probability must be linearly independent of one another.  For the case of 
a canonical decoherence functional (\ref{eq:dcanonical}) this bound turns
out to be \ralpha\romega, where \ralpha\ and \romega\ are the ranks of
the initial and final boundary conditions.  (In the special case of the
canonical decoherence functional with $\rhomega=1$,
the existence of such a bound was noticed previously by Diosi \cite{Diosi}
and by Dowker and Kent \cite{DK}.)

Let us see how this bound comes about.  

Any decoherence functional will possess, in general, a nonempty nullspace 
\nullspacesubd\ defined as
$\nullspacesubd(\hilbert)=\{ o\in\hilbert \|\ d(\cdot,o) = 0 \}.$
There is then a natural consistency-preserving equivalence 
relation\footnote{While this equivalence relation on histories under $d$
is certainly of {\it mathematical} relevance, there is no apparent 
reason that histories equivalent under this relation
must be in any sense {\it physically} equivalent.}
in \hilbert\ associated with \nullspacesubd, 
$h \sim h+o,$ $o\in\nullspacesubd,$ 
which defines the factor space
$ \hilbertsubd = \hilbert/\nullspacesubd\ $
(which is in turn naturally isomorphic to $\nullspaceperpsubd.$)
%$[h]\rightarrow\Projsub{\perp}h.$
This is a useful space because $d$ is non-degenerate on \hilbertsubd\
(that is, $d(h,h')=0\ \forall\ h\in\hilbertsubd \Rightarrow h'=0$).
%by def no 0 evals on \nullspaceperp.

Now, linearly dependent histories cannot be consistent with one 
another unless they have zero probability, for supposing 
$h'=\alpha\, h,$ $ d(h,h')=\alpha\, d(h,h)$.\footnote{While this statement 
is always {\it true}, it only has interesting {\it consequences} 
for \classoprep, as all disjoint histories in \projrep\ are linearly 
independent.} 

Moreover, any history in a consistent set which is linearly dependent on
the other histories in the set must have zero probability, as must all
the histories which it is dependent upon.  In other words, {\it all the
histories in a linearly dependent subset of a consistent set have zero
probability.}  To see this, consider a consistent set
$\{h_{1},\ldots,h_{n}\}\in\consistentsubd.$
Suppose $h_{1} = \sum_{i=2}^{m} c_{i}\,h_{i}$  (where $m<n$).  Then, for 
%$1 < j \leq m,$ $d(h_{j},h_{1})=\not\kern-0.4em\sum_{j} c_{j}\, p_{j} = 0$
$1 < j \leq m,$ $0 = d(h_{j},h_{1})=\nosum_{j} c_{j}\, p_{j}$
by the assumption of consistency, so that all the $h_{j}$ with
$c_{j}\neq 0$ have zero probability $p_{j},$ and 
$p_{1} = \sum_{i=2}^{m}|c_{i}|^{2}\, p_{i} = 0 $ as well.
%Don't worry; this doesn't conflict with the fact that the histories
%in a set sum to 1, because 1 isn't in the set!

%Thus linear dependence => zero probabilities
%Therefore, =/ 0 probs => linearly independent
%*and* the zero probs must be linearly independent of the non-zeros
%(though *not* necessarily of one another), putting a bound
%on the number of linearly independent zero prob h's as well.

Now consider an exclusive, exhaustive set of histories.  Drop all
components lying in \nullspacesubd, thereby %effectively 
reducing the set from \hilbert\ to 
$\hilbertsubd = \hilbert/\nullspacesubd.$
In order to be consistent, the subset of the remaining non-zero histories
which all have non-zero probability must be linearly independent, and
therefore have no more than 
${\rm dim\, }\hilbertsubd = {\rm dim\, }\nullspaceperpsubd$ 
members.\footnote{It is perhaps worth mentioning that in deriving 
this bound we have made no use of the fact that the histories in a 
consistent set must sum to the unit in \Hilbert.  This will gives rise 
to a consistency condition -- see (\ref{eq:consistentcondition}) -- 
which may further reduce the number of allowed non-zero probability
consistent histories for a given decoherence functional.}

Thus, the maximum number of histories with non-zero probability in 
%an exclusive, exhaustive 
a consistent set is ${\rm dim\, }\hilbert/\nullspacesubd$.
%\footnote{An
%interesting example of the kind of physical restrictions to which
%this bound can give rise is given in the last chapter.}

For positive definite decoherence functionals this is of course obvious,
as such decoherence functionals are genuine inner products on
\hilbert\ %\hilbertsubd\ 
({\it cf.\ }section \ref{sec:innerproduct}).  In fact, in
this case it is clear that the maximum number of histories in a mutually
consistent set is ${\rm dim}\, \hilbert.$ 
%irrespective of their probability.  
%Doofus; without the "definite", there are zero probability histories 
%that can be orthogonal just by being in the nullspace, and there can 
%be as many of them as you like (so long as they are disjoint.)
(A consistent set is just a set of operators 
orthogonal under the inner product defined by $d$.)
The broader generality of the result is nonetheless 
interesting.  
%Moreover, I show in section \ref{sec:history?} that
%all decoherence functionals possess a %linearly independent 
%set of non-zero probability ``histories" (operators) which saturates 
%this bound.

For the canonical decoherence functional (\ref{eq:dcanonical})
this upper bound on the most non-zero probability histories in a
decohering set is just $\ralpha\romega,$ where \ralpha\ and \romega\ are
the ranks of the initial and final boundary operators.  

This is easy to see.  
Pick two bases for \Hilbert\ consisting in the eigenvectors of
\rhalpha\ and \rhomega,
$\{ \ket{\alpha},\ketl{\overline{\alpha}} \}$ and
$\{ \ket{\omega},\ketl{\overline{\omega}} \}.$
Here, all the \ralpha\ (\romega ) \ket{\alpha} (\ket{\omega}) have
non-zero eigenvalues, and all the  
\ketl{\overline{\alpha}} (\ketl{\overline{\omega}}) have zero eigenvalues.
Then 
$\hilbert = \linearops{\Hilbert} = \nullspaceperpif\oplus\nullspaceif$
is spanned by
$\{ \ketbra{\alpha}{\omega} \} \cup 
\{ \ketbral{\overline{\alpha}}{\omega}, 
\ketbral{\overline{\alpha}}{\overline{\omega}}, 
\ketbral{\alpha}{\overline{\omega}} \},$
so that the dimension of 
$\nullspaceperpif\simeq\hilbert/\nullspaceif$
is clearly $\ralpha\romega.$

%In the case of a canonical decoherence functional with $\rhomega=1,$
%DK show how this fact implies an upper bound on the ``length" of a
%consistent homogeneous class operator, that is, the number of $P$'s  
%in a (\ref{eq:historyclassop}) which are not equal to the unit \OneH.
%I have not yet attempted to extend the proof more generally yet.

%For decoherence functionals on class operators.
The physical import of the existence of such an upper bound is
that, in general, decoherence requires some coarse graining, a fact
which has long been observed in explicit calculations.  

This is because
the number of histories in a completely fine-grained exclusive,
exhaustive set  %, ${\rm dim\, }\hilbert,$  ??? only in \projrep!
is in general greater than the maximum number of such histories 
with non-zero probability, ${\rm dim\, }\hilbert/\nullspace.$  
If ${\rm dim\, }\HilbertS= N,$ in \projrepbar\ the former number is 
${\rm dim\, }\hilbert = N^k \geq {\rm dim\, }\hilbert/\nullspace.$  
In \classoprepbar\ the situation is considerably more accute.  
There are at most $N^{2}$ linearly independent histories in 
$\hilbert\simeq \HilbertS\otimes H_{S}^{*},$ while there are $N^{k}$
histories in a completely fine grained exclusive, exhaustive set.
%(Two linearly dependent histories cannot, via sesquilinearity, be
%consistent with one another except in the trivial case of histories which
%lie entirely in \nullspace, that is, histories with zero probability.)
In order for a set of class operators to decohere, most of its
histories must have zero probability, or there must be some coarse
graining.
(In general both will be true.)
%Superselection rules are examples where the first "mechanism"
%enforces decoherence.

%Will this be enough to pacify Fay?

%true for pi d's as well, but we know that there is some
%sort of direct connection between path integral and hilbert space
%transition amplitudes.   Anyhow, worth a comment?

\subsection[Positivity: a Cauchy-Schwarz Inequality, 
and the Decoherence Functional as Inner Product]{Positivity: 
a Cauchy-Schwarz Inequality, \protect\\
and the Decoherence Functional as Inner Product}
\label{sec:innerproduct}

In this section, I show that positive decoherence functionals
obey a Cauchy-Schwarz inequality in \hilbert.

By definition, $d$ is positive on \rep.  This is not, however,
sufficient in general to imply positivity of $d$ on all of \hilbert\
unless $\rep = \classoprepbar.$ 
(Indeed, Isham and Linden \cite{IL1} supply a counterexample in the 
\projlattice{\Hilbert}\ case, and I provide in appendix 
\ref{sec:raycomplete} 
a demonstration of the existence of a 
counterexample in \classoprep.)  A necessary and sufficient condition for
positivity on \hilbert\ is, for instance, the positive kernel condition
(\ref{eq:positivekernel}).
%(This is so because every operator may be written, in general in many ways, 
%as a sum of projectors, and every such sum of course defines an operator.)
%Cauchy-Schwarz also N+S ....
Another is that $d$ possess no negative eigenvalues on \hilbert\
({\it cf.\ }(\ref{eq:diracd})).  

More generally, any decoherence functional on \classoprepbar\ is
positive because of its ray-completeness property:
$\classoprepbar(\Hilbert)$ contains a vector in every ray of
\linearops{\Hilbert}.  Being positive on \classoprepbar\ therefore
implies positivity on ${\cal L}.$  This is {\it not} the case for
\classoprep.   
%(See appendix \ref{sec:raycomplete} for a counterexample.)
Decoherence functionals on \classoprep\ or \projrep\ or even
\projlattice{\Hilbert}\ may have negative eigenvalues (but are
still of course positive on 
pairs of histories in \rep.)

In any case, the important observation is that {\it positive decoherence 
functionals are bona fide semi-inner products on \hilbert} \cite{Wright}.
(``Semi", because a general $d$ will as a rule
possess a nontrivial nullspace \nullspace.)  The canonical decoherence
functional (\ref{eq:dcanonical}), the decoherence functional of ordinary 
quantum mechanics, is obviously the most important example:
%of a positive decoherence functional: 
 {\it the decoherence functional of ordinary quantum mechanics is 
an inner product on class operators.}

A related consequence of the positivity of $d$ %on \hilbert\
is that all positive decoherence functionals satisfy a
Cauchy-Schwarz inequality.  (This was first noted in the case of 
the canonical decoherence functional with $\rhomega = 1$ in \cite{DH}.)
That is,
\begin{equation} \label{eq:cauchyschwarz}
 | \decoh{h}{h'} |^{2} \leq \decoh{h}{h}\, \decoh{h'}{h'}
\end{equation}
for positive decoherence functionals.
%in subspaces of \hilbert\ on which $d$ is positive.
%In particular, decoherence functionals obey a Cauchy-Schwarz 
%inequality on each ${\rm span\, }\consistentsetsubd$.
This may be proved in the standard way: a positive decoherence
functional satisfies $ \decoh{h+\lambda h'}{h+\lambda h'} \geq 0$
for any complex number $\lambda.$  Supposing $\decoh{h'}{h'} \neq 0,$ 
choosing $\lambda = -\decoh{h'}{h}/\decoh{h'}{h'}$ and using 
sesquilinearity and Hermiticity gives (\ref{eq:cauchyschwarz}).  

Because of (\ref{eq:normalization}) ($d(1,1) = 1$), that
(\ref{eq:cauchyschwarz}) hold for every $h,h' \in \hilbert$ 
is not only necessary, but also sufficient 
to imply the positivity of $d$ on \hilbert.     
%(That $d$ is positive on just one history is enough if
%(\ref{eq:cauchyschwarz}) holds on all of \hilbert.)
Further, as the proof of the triangle inequality depends only on
the Cauchy-Schwarz inequality (and not on positive {\it definiteness}),
positive $d$'s obey
\begin{equation} \label{eq:triangle}
\sqrt{d(h+h',h+h')} \leq \sqrt{d(h,h)} + \sqrt{d(h',h')}.
\end{equation}

Note that the fact that $d$ has in general a non-empty nullspace  
\nullspace\ does not impede the proof of either of these inequalities.  
In fact, from (\ref{eq:cauchyschwarz}) it is now clear that 
for positive $d$, $\decoh{h}{h} = 0$ implies %``null triviality",
%this is *not* null triviality, that's d(h,0) = 0.  This is better.
$\decoh{h}{h'} = 0.$ %(a fact implied by sesquilinearity anyhow).
This is good for consistency, because it means that zero probability 
members of exhaustive sets of histories cannot spoil the consistency 
of the set.
%interfere with any other history.
%, a fact also noted by Isham \cite{Isham}.
%fay:  why do we care if such a 0 prob history isn't consistent?
(This could, and perhaps should, be taken as a physical %strong 
argument for including positivity   %the positive kernel condition 
as an additional requirement on physical decoherence 
functionals!)  %Zero probability histories can be (trivially) included 
%in consistent sets for free.   %exclusive, exhaustive, dummy.
%In particular, for $o \in \nullspace$
%and $h \in \nullspace^{\perp},$ $h$ and $h+o$ are naturally equivalent.
%Suppose $p_d(o) = 0$ and $o \in \decoheringsubd.$  $o$ cannot spoil the
%consistency of an ee set in which it is included, say to achieve the
%exhaustive (sum = 1) part.  Note $p_d(o) = 0$ and $o \in \decoheringsubd$
% implies $d(1,o) = 0.$

(In fact, the property that zero probability histories cannot interfere
with other histories is actually equivalent to positivity.  For, suppose
$o$ is a zero probability history, $d(o,o)=0.$  Consider the condition
$d(h,o)=0\ \forall\ h\in\hilbert.$  Without loss of generality we can take
$h, o \in \nullspaceperp.$   As noted previously, decoherence
functionals are non-degenerate when restricted to \nullspaceperp, 
so the condition implies $o=0$ and so $o\in\nullspace$ (trivially) after
all.  But non-positive decoherence functionals do have zero
probability histories which do not live in \nullspace, which therefore
must interfere with at least some other $h\in\nullspaceperp.$)

Finally, it is worth noting that the implications of positivity 
are of more general significance for the following reason:
decoherence functionals are {\it always} positive on the 
span of each of their physical consistent sets 
(${\rm span\, }\consistentset$, where $\consistentset\in\consistentsubd$),
so that, for instance, a decoherence functional which possesses a 
maximally fine grained consistent set ({\it i.e.\ }a consistent set
with ${\rm dim\, }\hilbert$ linearly independent members; 
see the previous section) is positive on all of \hilbert.
In any event, this means that (\ref{eq:cauchyschwarz}) and related 
results hold on each and every ${\rm span\, }\consistentset$, 
regardless of whether $d$ is positive on all of \hilbert.

(It is very easy to see why decoherence functionals are positive on the
span of each of their physical consistent sets, whether or not
$d$ is a positive operator: supposing that we have coarse-grained all the
zero probability histories into one, each consistent set
$\consistentsetsubd = \{h_1, \ldots, h_m \}$ is linearly independent,
and so its members serve as a basis for 
${\rm span\, }\consistentsetsubd.$  But \consistentsetsubd\ is
consistent, so $d(h_i,h_j) = p_i\, \delta_{ij}$.  Because the
$p_i$ are non-negative for physical consistent sets, $d$ is
positive -- though not necessarily positive {\it definite} -- 
on ${\rm span\, }\consistentsetsubd.$ )
%\decoheringsubd is
%the %disjoint union of the \consistentsetsubd, so $d$ is positive
%on ${\rm span\, }\decoheringsubd.$  (Compare section 
%\ref{sec:innerproduct}.)
%Tried to claim that $d$ was therefore positive on \decoheringsubd,
%but that would not appear to be true.  
%Suppose d = a\ketbra{a}{a} - c\ketbra{c}{c}, where \bracket{a}{c}=0.  
%Then d(b,b) is positive so long as 
%a/c > |\bracket{c}{b}|^2/|\bracket{a}{b}|^2.  Then d is postive on
%span a and span b, but not on span{a,b}.  Oh, well.
%positive whenever 

%What is this condition on the eigenvalues?  Hierarchy of pk conditions
%corresponding to the various R's? ....

%(It is perhaps worth emphasizing %again 
%the generality of the \classoprepbar\ case.  The decoherence functionals 
%in practical use are positive, and may trivially be extended to 
%decoherence functionals on the full space of inhomogeneously coarse 
%grained histories \classoprepbar.)

%\projrepbar:  not every projection operator (hence not every ray) 
%can be gotten via sums of disjoint histories (Isham).

%Hierarchy of pk conditions corresponding to various \rep's?
%eg perhaps \sum d(P_i,P_j) >0 if positive just P's ...
%so c_i=c_j, etc

\subsubsection*{A Note on Approximate and Exact Consistency of Histories}
\label{sec:approx}
\addcontentsline{toc}{subsubsection}{\numberline{}%
A Note on Approximate and Exact Consistency of Histories}
%cf. The LaTeX companion section 2.4.2

In passing, I note a conjecture of Dowker and Kent \cite{DK}, that 
``close to'' any approximately consistent set there is an exactly
decohering one.
(The physical meaning of an approximately consistent set of histories
in generalized quantum theory is discussed in \cite{Jerusalem}, and
various mathematical aspects treated in \cite{DH,DKsummary,DK,McElwaine}.)
For positive definite decoherence functionals at least, 
this conjecture can
be verified in the sense that ``close to'' an approximately orthogonal 
set of vectors there is always a precisely orthogonal set.  In particular,
the Gram-Schmidt procedure may be applied to the approximately
orthogonal set in such a way that it disturbs the set ``as little as
possible,'' in the sense that the sum of the lengths of the Gram-Schmidt
``correction'' vectors is minimized \cite{dac97}.  

It is worth observing \cite{McElwaine} that, 
depending on the degree of inconsistency
permitted, there may be {\it more histories in an approximately
consistent set than are allowed in an exactly consistent one}
(section \ref{sec:maxnumhistories}). The conjecture
will therefore always fail for approximately consistent sets which
have too many members.  %, unless there is some coarse graining first.

The complicating issue here is, as ever, the question of whether
the resulting operators do, in fact, correspond to physical histories,
that is, are always in \rep.  
In \projrep\ this is unlikely to be
the case in general, and another strategy for resolving the
conjecture must be applied.  On the other hand, the ray-completeness
of \classoprepbar\ (section \ref{sec:hilbertGQT}) makes it
considerably more likely that the Gram-Schmidt tactic will yield 
another set of physical histories close to the original, approximately
decohering, one.

Until this issue can be resolved, 
the physical validity of the conjecture remains in question.

\subsection{Notation}
\label{sec:notation}

In section \ref{sec:example}\ I exhibit the preceeding 
observations for \dcanonical.  
In this section I introduce some calculational tools, applicable 
to \dcanonical\ as well as in the general case, 
that will be of some use in what follows.          
The ``geometric" perspective on generalized quantum theory
developed here will prove to have 
a number of practical and intuitive advantages for performing
calculations in generalized quantum theory, and is therefore 
worth pursuing in some detail.

%Translator's dictionary?  "There are a number of different perspectives
%on the same structures which will prove useful at various points."

Having extended a general decoherence functional to a sesquilinear
Hermitian operator on the linearly extended space of histories
$\hilbert = \linearops{\Hilbert}$ 
(for some Hilbert space \Hilbert; {\it cf.\ }the remarks 
at the beginning of section \ref{sec:extension}), it is sometimes helpful 
%to us feeble minded physicists
to make use of the mathematical familiarity of this situation by employing 
%a notation
notations adapted to it.

In finite dimensions, $\hilbert \simeq \Hilbert\otimes\Hilbert^{*},$
%isometrically isomorphic
%\inf dimensions:  $\Hilbert\otimes\Hilbert^{*} \simeq
%\system_{2}(\Hilbert),$ the Hilbert-Schmidt operators.  (Wald p191)
which is just the statement that the linear operators on \Hilbert\ can
be built from operators of the form  ( $M = (im)$)
\begin{equation}\label{eq:RsubM}
R_{M} \equiv R_{im} = \ketbra{i}{m}
\end{equation}
for any choice of bases $\{ \ket{i} \}$ and $\{ \bra{m} \}$ for
\Hilbert\ and $\Hilbert^{*}$  (though of course not every basis of 
$\Hilbert\otimes\Hilbert^{*}$ has this simple product
form).\footnote{Here I am establishing the convention that the region of
the alphabet from which the indices are drawn serves to indicate
which basis is meant.  In the present case, $i-l$ and $m-p$ refer to 
the two distinct bases which I have just introduced.  In applications
concerning canonical decoherence functionals, the $\{ \ket{i} \}$ and
$\{ \ket{m} \}$ will be eigenbases of the initial and final conditions
respectively, as in (\ref{eq:rhoinitial}) and (\ref{eq:rhofinal}).}  
%If required, the
%allowed regions may be extended backward and forward (that is, for
%example, to $a-l$ and $m-z$).
Now, a natural inner product on \linearops{\Hilbert}\ is the trace, and
the dual                                    %op norm topology
of \linearops{\Hilbert}\ is given by the trace on \Hilbert.  (See, for
instance, \cite[section VI.6]{RS}\ for the status of these statements
in infinite dimensional Hilbert spaces.)
%\linearops{\Hilbert}\simeq\system_{1}^{*}, 
%\system_{1}={\rm com\ }(\Hilbert)
%i.e. trace of trace-class ops is dual of compact operators
It is natural therefore to make the associations (mostly mere notation
in finite dimensions)
%$$ \begin{array}{ccc}
\begin{eqnarray}
\Ket{A} \in \hilbert  & \Leftrightarrow &  A \in \linearops{\Hilbert} 
                          \label{eq:diracket}\\
\Bra{A} \in \hilbert^{*} & \Leftrightarrow & 
A^{\dagger} \in \linearops{\Hilbert}^{*}\simeq\linearops{\Hilbert}
                          \label{eq:diracbra}
\end{eqnarray}
%\end{array} $$
%\inf dimensions: good to have \hilbert still; 
%\System_{2} is a Hilbert space!
where the dual is identified via
\begin{equation}\label{eq:diracip}
               \Bracket{A}{B} \equiv \trH\, A^{\dagger}B.
\end{equation}
%See how a basis of Hermitian ops may have its uses.

As an Hermitian form on \hilbert, a general decoherence functional may
then be written always in the simple diagonal form
\begin{equation}\label{eq:diracd}
   d  = \sum_{I}\ w_{I}\, \Ketbra{I}{I}
\end{equation}
for some orthonormal basis $\{ \Ket{I} \} \Leftrightarrow \{ E_I \}$
(where the $E_I$ are trace-orthonormal, 
$\Bracket{I}{J} = \trH\, E_{I}^{\dagger}E_{J} = \delta_{IJ}$),
so that $d(h,h')$ is equal to
\begin{eqnarray}
\Melt{h}{d}{h'}  
  & = & \sum_{I}\ w_{I}\, \Bracket{h}{I}\Bracket{I}{h'} 
        \nonumber\\
  & = & \sum_{I}\ w_{I}\, \trH\, h^{\dagger}E_{I}\ \trH\, E_{I}^{\dagger}h'.
        \label{eq:diracdhh}
\end{eqnarray}
For later use, note that we may continue rewriting this as
\begin{eqnarray}
\Melt{h}{d}{h'}  
  & = & \sum_{I}\ w_{I}\, \trHH (h^{\dagger}E_{I}\otimes h'E_{I}^{\dagger})
        \nonumber\\
  & = & \sum_{I}\ w_{I}\, 
 \trHH [(h^{\dagger}\otimes h')(E_{I}\otimes E_{I}^{\dagger})]
 \nonumber\\
  & = & \trHH [(h^{\dagger}\otimes\h')\underline{d}],
        \label{eq:dunderbar}
\end{eqnarray}
using the fact that $\trHH\, A\otimes B = \trH\, A\ \trH\, B,$ and where
$\underline{d}$ is defined as in (\ref{eq:HHdiracg}) below.

More generally, any operator $g: \hilbert\rightarrow\hilbert$ can be
written
\begin{equation}\label{eq:diracg}
     g  =  \sum_{IJ}\ g_{IJ}\, \Ketbra{I}{J}
\end{equation}
so that
\begin{equation}\label{eq:diracghh}
     \Melt{h}{g}{h'} = \sum_{IJ}\ h^{I*}h'^{J}\, g_{IJ}
\end{equation}
where
\begin{eqnarray}
   h^{I}  &  =  &  \Bracket{I}{h}  \nonumber\\
      &  =  &  \trH\, E_{I}^{\dagger} h.  \label{eq:dirachcomponent}
\end{eqnarray}
Or, what's the same,
\begin{eqnarray}
\Melt{h}{g}{h'}  &  =  &  
  \sum_{IJ}\ g_{IJ}\, \trH\, h^{\dagger}E_{I}\ \trH\, E_{J}^{\dagger}h'
  \nonumber\\    
  &  =  & \sum_{IJ}\ 
       g_{IJ}\, \trHH (h^{\dagger}\otimes h')(E_{I}\otimes E_{J}^{\dagger}) 
  \nonumber\\
  &  \equiv  & 
       \trHH [(h^{\dagger}\otimes h')\underline{g}]  \label{eq:HHdiracghh},
\end{eqnarray}
where the operator $\underline{g}$ on $\Hilbert\otimes\Hilbert$ is
defined as
\begin{equation}\label{eq:HHdiracg}
\underline{g}  =  \sum_{IJ}\ g_{IJ}\, E_{I}\otimes E_{J}^{\dagger}.
\end{equation}
It is therefore often convenient to exploit the 
%isometric ... homeomorphism (continuous, 1-1, onto)
isomorphism between $\hilbert \simeq \Hilbert\otimes\Hilbert^{*} $ 
and $\underline{\hilbert} \equiv \Hilbert\otimes\Hilbert$
by making the natural association between the operator 
$g: \hilbert\rightarrow\hilbert$ and the operator
$\underline{g}: \hilbertbar\rightarrow\hilbertbar.$

(Though these equations have been written using a general basis
$\{ E_{I} \} \equiv \{ \Ket{I} \}$ of trace-orthonormal operators on
\Hilbert, factor bases like that introduced in (\ref{eq:RsubM})
are often convenient in calculations.
For example, with 
$E_{I} \equiv \ketbra{i}{m}$ and $\ket{im} \equiv \ket{i}\otimes\ket{m}$
(and, of course, in the same way $J=(jn)$), $\underline{g}$ may be written
\begin{equation}\label{eq:HHgdirac}
\underline{g} = \sum_{ijmn}\ g_{im,jn}\, \ketbra{in}{mj},
\end{equation} {\it cf.\ }(\ref{eq:Hhcompleteness}).
The single versus double bars on the kets should be %sufficient to
adequate to distinguish states of $\Hilbert\otimes\Hilbert$ from
those of $\Hilbert\otimes\Hilbert^{*}.$)
%again, bases of Hermitian operators have their uses!
%and in \projrep\ one might want bases of projections.

More generally, a comparison of (\ref{eq:diracg}) and
(\ref{eq:HHdiracg}) shows that, at least in the present context,
it is natural and useful to identify the operator
$\Ketbra{A}{B}$ on $\hilbert \simeq \Hilbert\otimes\Hilbert^{*}$
and the (tensor product) operator 
$A\otimes B^{\dagger}$ on 
$\underline{\hilbert}\equiv\Hilbert\otimes\Hilbert.$
Some might even prefer to regard the right hand side of
(\ref{eq:HHdiracghh}) as defining the left,
%$g$ and $\underline{g}$ may then be regarded as strictly equal, 
with the double-bar Dirac notation merely serving as operating 
instructions when the operator in (\ref{eq:HHdiracg}) is employed, 
through (\ref{eq:HHdiracghh}), as a sesquilinear operator on the 
history space \hilbert.  In any case, I will 
eventually lapse into the habit of not bothering to distinguish between
them with the underscore.  Nevertheless, beware that some care is then
required in interpreting, for instance, operator products computed 
in the Dirac notation, {\it cf.\ }(\ref{eq:diracopproduct}) versus
(\ref{eq:HHdiracopproduct}), and (\ref{eq:HHdiracopaction}), below.
The reason for this is that, in spite of the fact that 
$\hilbert \simeq \underline{\hilbert},$
many common operations such as taking products and traces give
completely different answers depending on the space in which they are
interpreted.  That is, our map between operators on \hilbert\ and on
$\underline{\hilbert}$ does not naturally preserve such relations.
It will therefore prove necessary to distinguish these operations 
notationally, which however is a small fee for a useful dual service.

In this spirit, 
a few more pieces of notation will prove useful in later sections.
For operators on $\underline{\hilbert} = \Hilbert\otimes\Hilbert$
it is usual to define adjoints and operator products through
\begin{equation}\label{eq:HHdagger}
  (A\otimes B)^{\dagger} = A^{\dagger}\otimes B^{\dagger}
\end{equation}
and
\begin{equation}\label{eq:HHoldopproduct}
(A\otimes B)(C\otimes D) = AC\otimes BD.
\end{equation}
However, in order to preserve the useful correspondence between
the operators $\underline{g}$ on $\underline{\hilbert}$ of the 
form (\ref{eq:HHdiracg}), and the more comforting appearance of the
operator $g$ on \hilbert\ introduced in (\ref{eq:diracg}), it is 
helpful to also have available the definitions
\begin{equation}\label{eq:HHstar}
(A\otimes B)^{*} = B^{\dagger}\otimes A^{\dagger}
\end{equation}
%  &=& M (A\otimes B)^{\dagger} M
% cf. below (\ref{eq:dILS}). 
and
\begin{equation}\label{eq:HHopproduct}
(A\otimes B)\odot (C\otimes D) = (\trH\, BC)\ A\otimes D,
\end{equation}
extended by linearity in the obvious way.  Thus, for instance,
\begin{eqnarray}
\Melt{h'}{g}{h}^{*} &=& \sum_{IJ}\ g_{IJ}^{*}\, 
    \trHH (h'^{\dagger}E_{I}\otimes E_{J}^{\dagger}h)^{*}  \nonumber\\
   &=& \sum_{IJ}\ g_{IJ}^{*}\, 
     \trHH (h^{\dagger}E_{J}\otimes E_{I}^{\dagger}h') \nonumber\\
   &=& \sum_{IJ}\ g_{IJ}^{*}\, 
     \trH (h^{\dagger}E_{J})\, \trH (E_{I}^{\dagger}h') \nonumber\\
   &=& \sum_{IJ}\ g_{IJ}^{*}\, 
     \Bracket{h}{J}\Bracket{I}{h'} \nonumber\\
   &=& \Melt{h}{g^{*}}{h'}, \label{eq:diracghhstar}
\end{eqnarray}
maintaining the correspondence between $g^{*},$ by definition the true
adjoint of $g:\hilbert\rightarrow\hilbert,$ and the operator
$\underline{g}^{*}$ on $\underline{\hilbert},$ in the way we would
desire.  Similarly, we of course want to write
\begin{eqnarray}
fg &=& \sum_{IJ}\, \sum_{KL}\ 
   \left(\, f_{IJ}\, \Ketbra{I}{J}\, \right)\, 
              \left(\, g_{KL}\, \Ketbra{K}{L}\, \right)  \nonumber\\
   &=& 
   \sum_{IL}\, \left( \sum_{J}\ f_{IJ}\, g_{JL} \right)\, \Ketbra{I}{L},
\label{eq:diracopproduct}
\end{eqnarray}
to which corresponds
\begin{eqnarray}
\underline{f}\odot\underline{g}  &=& \sum_{IJ}\, \sum_{KL}
    \left(f_{IJ}\, E_{I}\otimes E_{J}^{\dagger}\right)\odot
    \left(g_{KL}\, E_{K}\otimes E_{L}^{\dagger}\right)  \nonumber\\
 &=& \sum_{IJ}\, \sum_{KL}\ 
    f_{IJ}\, g_{KL}\ (\trH E_{J}^{\dagger} E_{K})\, 
    E_{I}\otimes E_{L}^{\dagger}  \nonumber\\
 &=& \sum_{IL}\ \left( \sum_{J} f_{IJ}\, g_{JL} \right)\,  
     E_{I}\otimes E_{L}^{\dagger}. \label{eq:HHdiracopproduct}
\end{eqnarray}
The operator product $\odot$ therefore (correctly) conveys the idea that
the product in (\ref{eq:diracopproduct}) is a composition of maps.
Because of the comments below (\ref{eq:HHgdirac}), however, I will
generally write expressions like the left hand side of 
(\ref{eq:diracopproduct}) as $f\odot g$ to make it absolutely
clear it is the natural operator product on \hilbert\ that is meant,
facilitating the temptation to drop the underscores from the equivalent
operators on $\underline{\hilbert} = \Hilbert\otimes\Hilbert.$
(Section \ref{sec:d-unitarity}, in which $\underline{\hilbert}$ plays
no role, is an exception.)

In addition, in accord with the fact that (\ref{eq:diracg}) 
is an operator on operators ({\it i.e.\ }on \hilbert ), 
it is useful to have a meaning assigned to expressions like 
``$(A\otimes B)C$".  If it is allowed that
\begin{equation}\label{eq:Hhopproductr}
(A\otimes B)C \equiv (\trH\, BC)\,  A
\end{equation}
and
\begin{equation}\label{eq:Hhopproductl}
A(B\otimes C) \equiv (\trH\, AB)\, C
\end{equation}
({\it cf.\ }(\ref{eq:HHopproduct})), then 
there is a natural agreement between
\begin{eqnarray}
g\Ket{A} &=& \sum_{IJ}\ g_{IJ}\, \Ket{I}\Bracket{J}{A} \nonumber\\
    &=& \sum_{I}\ 
   \left( \sum_{J}\ g_{IJ}\, \trH\, E_{J}^{\dagger}A \right)\, \Ket{I}
\label{eq:diracopaction}
\end{eqnarray}
and
\begin{eqnarray}
\underline{g}A &=& 
   \left( \sum_{IJ}\ g_{IJ}\, E_{I}\otimes E_{J}^{\dagger} \right)\, A  
         \nonumber\\
    &=& \sum_{I}\ 
   \left( \sum_{J}\ g_{IJ}\, \trH E_{J}^{\dagger}A \right)\, E_{I}.
\label{eq:HHdiracopaction}
\end{eqnarray}
Further, in order that 
$\Bra{A}g^{*} \Leftrightarrow (\underline{g} A)^{*},$ define
\begin{equation}\label{eq:HHopactionstar}
(\underline{g} A)^{*} \equiv A^{\dagger}\underline{g}^{*}
\end{equation}
(and use (\ref{eq:Hhopproductl}) and (\ref{eq:HHstar}).)

This technology can of course be developed further, 
but I have developed here the basic tools useful for most calculations.

A last observation of some use is that 
the normalization condition $d(1,1)=1$ may be
worked out from, say, (\ref{eq:dunderbar}), to read
\begin{equation}\label{eq:diracdtr}
   \trHH\, \underline{d} = 1.
\end{equation}
(Note from (\ref{eq:dunderbar}) that $\trh\, d \neq \trHH\, d$ because
$\sum_{K}\ E_{K}^{\dagger}\otimes E_{K} \neq 1\otimes 1.$  In fact, for
the canonical decoherence functional (\ref{eq:dcanonical}), which
assumes the normalization $\trH\, \rhalpha\rhomega = 1,$ 
$\trHH\, \dcanonical = \trH\, \rhalpha\rhomega =1$ in accord with
(\ref{eq:diracdtr}), while
$\trh\, \dcanonical = \sum_{M}\ w_{M} = \trH\, \rhalpha\ \trH\, \rhomega;$
work it out directly or compare (\ref{eq:diracdcanonicalYZ}).)

\subsection{The ILS Theorem Revisited}
\label{sec:ILSrevisited}

One might be wondering what any of this has to do with the important ILS
theorem (described in section \ref{sec:ILSW}; {\it cf.\ }(\ref{eq:dILS}))
classifying decoherence functionals on \projlattice{\Hilbert}.
As it turns out, what         %I have 
has been done amounts to an explicit, constructive
(and elementary!) demonstration of the ILS theorem in finite dimensions.
Moreover, we have also found a version of the ILS theorem applicable
to decoherence functionals on class operators.
%(Note this means both $N$ and $k$ are finite.)

The recovery %emergence 
of the ILS theorem is not hard to see.  
Suppose we are considering a decoherence
functional on \projrep(\Hilbert) (or \projrepbar(\Hilbert), or
\projlattice{\Hilbert}.)   %; it makes no real difference.)  
(\ref{eq:dunderbar}) may then be rewritten 
\begin{equation}\label{eq:dILSunderbar}
       \Melt{h}{d}{h'} = \trHH [h\otimes h'\, \underline{d}].
\end{equation}
Comparing with (\ref{eq:dILS}), we discover that\footnote{To be 
strictly accurate, in theories with a time consistency with the
Schr\"odinger picture usage of ILS requires
the unitary time evolution operators    %evident in ...
to be ``factored out" of the Heisenberg projections in
$h\otimes h'$ and absorbed into $\underline{d}$
before making the identification with $X$.  As $\h\otimes h'$
is simply a 2-$k$ fold tensor product of projections, the 
cyclicity of the trace makes this a straightforward procedure.}
\begin{equation}\label{eq:dbarX}
       \underline{d} = X.
\end{equation}
The property that $X$ is normalized is just (\ref{eq:diracdtr}) above.
That $X^{\dagger} = MXM$ (where $M$ is defined so that
$M(A\otimes B)M = B\otimes A$; see below (\ref{eq:dILS}))
is of course a consequence of the Hermiticity of $d$ on \hilbert,
and is evident from the following calculation 
({\it cf.\ }(\ref{eq:HHdiracg})):
\begin{eqnarray}
\underline{d}^{\dagger} &=&  
  \left( \sum_{IJ}\ d_{IJ}\, E_I \otimes E_{J}^{\dagger} \right)^{\dagger} 
          \nonumber\\
% &=& \sum_{IJ}\ d_{IJ}^{*}\, E_I^{\dagger} \otimes E_{J} \nonumber\\
  &=& \sum_{IJ}\ d_{JI}^{*}\, E_J^{\dagger} \otimes E_{I} \nonumber\\
  &=& M\, \underline{d}\, M             \label{eq:dbardagger}
\end{eqnarray}
because $d=d^{*}$ implies as usual that $d_{IJ} = d_{JI}^{*}.$
That $X$ is positive in the sense that $\tr{P\otimes P\, X} \geq 0$
is just the postulated positivity of $d$ on \rep.  Unfortunately, it is
not known how to capture this property more explicitly (for instance,
as a transparent condition on the eigenvalues of $d$ or $\underline{d}.$)
   
Thus, the ILS operator $X$ emerges quite naturally in the present
formulation of generalized quantum theory.    
From (\ref{eq:dunderbar}) it is clear that the correspondence between
a decoherence functional $d$ on
$\hilbert = \Hilbert\otimes\Hilbert^{*}$ and $\underline{d}$ on
$\hilbertbar = \Hilbert\otimes\Hilbert$ also has a more general utility.
The most interesting observation for present purposes is that
(\ref{eq:dunderbar}) {\it is effectively an expression %a version
of the ILS theorem applicable
as well to decoherence functionals on class operators:}  
an arbitrary decoherence functional on class operators may be written as
\begin{equation}
d(h,h') 
   =  \trHH [(h^{\dagger}\otimes\h')\underline{d}],
              \label{eq:dunderbaragain}
\end{equation}
thereby classifying decoherence functionals on class operators
according to the operators $\underline{d}$.  As in the case
of the ILS operator $X$, $\underline{d}$ is Hermitian on
\hilbert, and consequently obeys (\ref{eq:HHdiracg}); is positive
in the appropriate sense; and is normalized, (\ref{eq:diracdtr}).

Section \ref{sec:dcanonical?}\ contains further examples of 
why it is useful to remember this correspondence.

%Make some use of the fact that its imaginary bit is unconstrained?

It is important to note that (\ref{eq:dILSunderbar}) 
is not an independent ``proof'' of the ILS
theorem, because we had to appeal in section \ref{sec:extension}\ 
to a result of Wright \cite{Wright}\ that guaranteed that the extension
of the domain of a decoherence functional is well defined.
This result was in turn a part of the proof 
of Wright's extension of the ILS theorem.
In return, however, the geometrical significance of the 
ILS theorem becomes evident.

\subsection{The Canonical Decoherence Functional as an Example}
\label{sec:example}

Finally, it is time to treat the canonical decoherence functional
(\ref{eq:dcanonical}) on \classoprepbar.  It is most convenient to 
work with the basis vectors (for \hilbert )
\begin{eqnarray}
  \Ket{M} \equiv R_{M} &=& R_{im}  \nonumber\\
                &\equiv & \ketbra{i}{m}.  \label{eq:canonicalbasis}
\end{eqnarray}
Here, the vectors (in \Hilbert ) $\ket{i}$ and $\ket{m}$ are
eigenvectors of the initial and final conditions \rhalpha\ and \rhomega,
\begin{equation}\label{eq:rhoinitial}
  \rhalpha = \sum_{i}\ a_{i}\, \ketbra{i}{i},
\end{equation}
\begin{equation}\label{eq:rhofinal}
  \rhomega = \sum_{m}\ z_{m}\, \ketbra{m}{m},
\end{equation}
where of course the $\{ \ket{i} \}$ and $\{ \ket{m} \}$ are chosen to be
orthonormal sets.  In that case, the $\{ \Ket{M} \}$ constitute a
complete orthonormal set in \hilbert :
\begin{eqnarray}
\Bracket{M}{N} & \equiv & \Bracket{im}{jn} \nonumber\\
            &=& \trH\, \ket{m}\bracket{i}{j}\bra{n}  \nonumber\\
            &=& \delta_{ij}\delta_{mn}  \nonumber\\
            & \equiv & \delta_{MN};
                                \label{eq:hONset}
\end{eqnarray}
\begin{eqnarray}
  \left(  \sum_{M}\ \Ketbra{M}{M} \right) \ \Ket{A} &=& \sum_{im}\
               (\trH R_{im}^{\dagger}A)\, \ketbra{i}{m} \nonumber\\
     &=&  \sum_{im}\ \ket{i}\melt{i}{A}{m}\bra{m}  \nonumber\\
     &=&  A \nonumber\\
     &=&  \Ket{A}.   \label{eq:hcompleteset}
\end{eqnarray}
(Note also the calculationally convenient fact that
\begin{eqnarray}
\sum_{M}\ \Ketbra{M}{M} 
      &=& \sum_{im}\ \ketbra{i}{m}\otimes\ketbra{m}{i}      \nonumber\\
      &=& \sum_{im}\ \ketbra{im}{mi};  \label{eq:Hhcompleteness}
\end{eqnarray}   {\it cf.\ }(\ref{eq:HHgdirac}).)
$\{ \Ket{M} \}$ is then an eigenbasis of \dcanonical:
\begin{eqnarray}
\Melt{N}{\dcanonical}{M}  &=& 
       \trH\, \rhomega R_{N}^{\dagger}\rhalpha\ R_{M}  \nonumber\\
        &=& \melt{m}{\rhomega}{n}\melt{j}{\rhalpha}{i}  \nonumber\\
        &=& \nosum_{im}\ a_{i}z_{m}\, \delta_{ij}\delta_{mn} \nonumber\\
        &=& \nosum_{M} w_{M}\, \delta_{MN} \label{eq:diracdcanonicalYZ}
\end{eqnarray}
(where it should be clear that $M\equiv (im)$ and $N \equiv (jn)$).

Thus a canonical decoherence functional may be written in the diagonal
form
\begin{eqnarray}
\dcanonical &=& \sum_{M}\ w_{M}\, \Ketbra{M}{M} \nonumber\\
            &=& \sum_{im}\ a_{i}z_{m}\, \Ketbra{im}{im}.
                \label{eq:dcanonicaldiag}
\end{eqnarray}
On $\underline{\hilbert} = \Hilbert\otimes\Hilbert$ this reads
\begin{equation}\label{eq:dcanonicaldiagHH}
       \dcanonical = \sum_{im}\ a_{i}z_{m}\, \ketbra{im}{mi}
\end{equation}
using (\ref{eq:canonicalbasis}).  (\ref{eq:dcanonicaldiagHH}) 
is the ILS operator $\underline{d}$ of (\ref{eq:dunderbaragain}) for 
canonical decoherence functionals on class operators.

As \rhalpha\ and \rhomega\ are positive by definition, \dcanonical\ is
manifestly a positive operator on \hilbert, and positive definite on the
subspace corresponding to the non-zero eigenvalues of \rhalpha\ and
\rhomega,
$\nullspaceperpif = {\rm span } 
\{ R_{im} \parallel a_{i}\neq 0, z_{m} \neq 0 \}
\simeq \hilbert_{\alpha\omega} \equiv \hilbert/\nullspaceif.$
%Henceforth, let an overbar signify a zero eigenvalue,
%$w_{\overline{z}} = a_{\overline{\alpha}} = z_{\overline{\omega}},$
%and reserve the indices $\alpha\$ and $\omega$ for the positive
%definite subspace.
%bars are reverse from previous choices!  Ah, well ....
In that case, 
$\{ \Ketl{\overline{M}} \} = \{ \overline{R}_{M} \} = 
\{ (w_{M})^{-\frac{1}{2}}R_{M}; w_{M} \neq 0 \}$
is an orthonormal basis in the positive definite subspace
$\hilbert_{\alpha\omega},$ if \dcanonical\ is taken to be the 
inner product on $\hilbert_{\alpha\omega}.$

%\newpage

\section{Some Structural Issues in Generalized Quantum Theory}
\label{sec:questions}

An important issue in the program of generalized quantum theory 
is whether or not, and under what assumptions, it is possible to 
determine the ``decoherence functional of the universe'', 
which question embraces the more familiar 
goals of both high energy physicists (``What are the dynamics 
-- lagrangian? -- of the universe?'') and quantum cosmologists 
(``What are the boundary conditions of the universe?''  
See \cite{HalliwellQC,Vilenkin,Death,Wiltshire} 
for reviews of quantum cosmology.)  
That is, what is the minimum physical information required to %(at least 
(approximately) reconstruct $d_{\rm universe}$?
While it is outside the scope of this paper to deal with this question 
directly, the study of the mathematical
relations between decoherence functionals and decoherent histories is
pertinent because these relations determine how much information about
one set of objects may be inferred from information about the other.

In particular, within a fixed mathematical setting, it would be a good 
thing to have answers to at least the following questions.
(Recall that \consistentsub{d}\ is the class of exclusive, exhaustive
sets of histories \consistentsetsub{d}\ which are mutually consistent
according to $d,$ and \decoheringsub{d} the %disjoint 
%oops! cf. geroch; the disjoint union is the one which does the
%double-counting of duplicate elements, not the union.
union of all such
sets, {\it i.e.} the collection of histories $h$ which appear in some
$\consistentsetsub{d} \in \consistentsub{d}.$)

\begin{enumerate}
\item  Given $d,$ what is \decoheringsub{d}?

\item  Given $d,$ what is \consistentsub{d}?

\item  Given a history $h,$ what are all the decoherence functionals
$d$ such that $h \in \decoheringsub{d}?$
 \begin{enumerate}
 \item  What is the subset of such $d$ according to which $h$ is 
        consistent, with a given probability $p_{h}?$
 %\item  Given $\{h\},$ what are all $d$ such that $\{h\} \subseteq
 %\decoheringsub{d}?$
 \end{enumerate}

\item  Given an exclusive, exhaustive set of histories \consistentset,
what are all the decoherence functionals $d$ such that 
$\consistentset \in \consistentsub{d}?$
 \begin{enumerate}
 \item What is the subset of all such $d$ according to which
 \consistentset\ is mutually consistent and given with a fixed set of
 probabilities 
 $\{p_{h} = d(h,h),\ h \in \consistentset \in \consistentsub{d}\}?$
 \end{enumerate}

\item  Given a collection $\{\consistentset\}$ of exclusive, exhaustive 
sets of histories \consistentset, what are all the decoherence functionals 
$d$ for which $\{\consistentset\} \subseteq \consistentsub{d}?$
 \begin{enumerate}
 \item Given $\{\consistentset\}$ and a set of probabilities $p_{h}$
 on all the histories in this collection, \\
 what is the subset of decoherence functionals which give consistently  \\
 $\{p_{h} = d(h,h),\ h \in \consistentset \in \{\consistentset\}
 \subseteq \consistentsub{d}\}?$
 \end{enumerate}

\item Does $\consistentsub{d} = \consistentsub{d'}$ 
   (or $\decoheringsub{d} = \decoheringsub{d'}$) imply $d=d'?$
   (Or is information about the probabilities required as well?)
%(``Symmetry groups" of decoherence functionals.)
%some ``length" info is contained in d-orthogonality relations.
%so no probs for C, yes for D?  
%evecs of d and their "probs" do fix d ....
%Note for instance that a normalized hf has n(n+1)/2 -1 melts to
%determine.  Knowledge of one fg consistent set tells us
% (n-1)+(n-2)+...=n(n-1)/2 0 inner products (melts).
%(say, the off-diagonal elements.)
%The complications begin when we start to worry about \rep ....
%\begin{enumerate}
%\item Same question for the smaller space \decoheringsub{d}.
%\end{enumerate}

%Drop the term "equivalence problem" unless i like the formulation
%better having finished the last section.
%\item Equivalence problem:  
%Given some canonical \dcanonical\ and its associated 
%\consistentsub{\alpha\omega}, does there exist a \consistentsub{d} for which 
%$\consistentsub{\alpha\omega} \subset \consistentsub{d}?$
%\begin{enumerate}
%\item Same question for \decoheringsub{\alpha\omega}.
%\end{enumerate}
%reformulate just in terms of some set of consistent sets?  Find all
%d's for which ... etc.

\end{enumerate}

At least in Hilbert space, something is known about all of these questions.
Section \ref{sec:history?} {\it formally} solves the first and second
in finite dimensional Hilbert spaces 
by parameterizing the space of ``histories" (operators) consistent 
according to a fixed decoherence functional, and showing moreover how 
-- most explicitly in the case of positive decoherence functionals -- 
to construct all of $d$'s maximally fine-grained consistent sets.  
The practical issue that remains is determining just which of these
consistent sets of operators contain only {\it physical} 
histories {\it i.e.\ }all of whose histories are in \rep\
({\it cf.\ }section \ref{sec:historyreconstruction}).  (An alternative
would be to find a physical interpretation for Isham's ``exotic"
projections, that is, the projections in \projlattice{\Hilbert}\ that
are not in \projrepbar, or, in the case of \classoprepbar, for the 
operators in \consistentops\ that are not in \classoprepbar.)
Section \ref{sec:functional?} shows how to characterize the decoherence
functionals according to which a given set of of histories
is consistent, thereby solving questions 3 - 5 up to the issue of
unambiguously characterizing positivity on \rep\ 
(again, only in finite dimensions.)  
%(Schreckenberg
%\cite{S}\ demonstrates constructively a fact of obvious relevance
%to question 4, namely, that given any exclusive, exhaustive set of
%histories in \projlattice{\Hilbert}\ and any choice of ``probabilities"
%on that set, there is some decoherence functional over
%\projlattice{\Hilbert}\ according to which that set is consistent, and
%has the given probabilities.  The answer to 4(a) is never, therefore,
%``none".  Isham and Linden \cite{IL1}\ have also shown that a 
%decoherence functional on \projlattice{\otimes^{k}\HilbertS} is
%completely determined by its values on a certain set of homogeneous fine
%grained histories.)
%Mention ILS thm again, and the problem it solves (classifying d's).?
The answer to question 6 is implicit in the observation that $d$ 
is an Hermitian form, and is most certainly a (qualified) ``yes", 
though the precise form of this ``yes'' has yet to be determined.
I do not take up the issue directly here.
%(In the present formulation, a solution to 6 is
%probably fairly easy so long as we allow ourselves to work in \hilbert,
%not restricting to \rep;  a decoherence functional is obviously
%completely determined by its eigenvalues and eigenvectors in \hilbert,
%for example.  Only for the canonical decoherence functionals are these
%known to have any physical significance, 
%however, {\it cf.\ }(\ref{eq:canonicalevecs}), and even in this case 
%these histories are unlikely to be observationally accessible.)  
%This is an 
Variants on such questions are an interesting topic for future work.

%That the {\it statements} of these questions are independent both
%of whether or not the space of histories is built 
%upon \projrep\ or \classoprep, and what coarse graining have been
%admitted, should not obscure the fact that 
%what counts as a legitimate physical history depends crucially 
%on which choices have been made; see section \ref{sec:hilbertGQT}.

Let me now show how to apply the tools of section \ref{sec:GQTgoc}\ to
%the solution of
these problems.

Though stated in general terms, much of the work of 
the following sections has the representation of histories 
by class operators primarily in mind.  Part of the reason for this
is that it is as class operators that histories are generally
represented in actual applications of generalized quantum theory.
The practical meaning of this emphasis is just 
%All this really means is just
that I shall not be making any particular use 
of the possibility that the history 
operators may be projections; the emphasis will be on $d$ as a 
functional on operators, not the history operators themselves.  
%(At least in principle, 
(It is always possible to go back explore the consequences of the
additional assumption that the history operators are projections.
This would be of no particular consequence in section \ref{sec:functional?}, 
where we get to select the histories {\it a priori}, but is a somewhat 
more significant issue in section \ref{sec:history?}.)

%\newpage

\section[The Histories Consistent with a Decoherence Functional]{The 
Histories Consistent \protect\\
with a Decoherence Functional}
\label{sec:history?}

This section is devoted to determining all of the histories
that are consistent according to a given decoherence
functional, {\it i.e.\ }determining \decoheringsubd.   
The next major section,
\ref{sec:functional?}, inverts this problem, determining
all of the decoherence functionals according to which a
given set of histories is consistent.

Though it is easy to characterize and parametrize 
the {\it operators} which satisy the consistency 
condition, {\it i.e.\ }to find \consistentops, the 
issues raised in section \ref{sec:historyreconstruction}\ 
make it difficult to determine explicitly which of the 
operators in \consistentops\ are actually members of 
\decoheringsubd.

Section \ref{sec:decoheringsubd}\ discusses the characterization
of \consistentops\ and the difficulties in computing explicitly
$\decoheringsubd = \consistentops\cap\rep$.  Section
\ref{sec:parametrization} displays one rather simple-minded
way of parametrizing \consistentops.   In \ref{sec:devecs}\
I make the observation that the eigenvectors of a decoherence functional 
%comprise 
may always be used to construct an exhaustive, consistent set of operators; 
section \ref{sec:d-unitarity}\ generalizes 
this observation to construct \consistentops\ in another way: by $d$-unitary
``rotations'' of $d$'s eigenvectors.  
(This strategy is easiest to implement
for positive decoherence functionals like the canonical 
one of ordinary quantum mechanics.)

\subsection{The Condition for a History to be Consistent}
\label{sec:decoheringsubd}

If inhomogeneous coarse grainings are admitted,
$\{h,1-h\}$ is obviously the common coarsest graining that still
contains $h$ of all the exclusive, exhaustive sets which have $h$ as a
member.  It is therefore natural to term $h$ ``consistent" according to
$d$ if this is a consistent set, $\{h,1-h\} \in \consistentsub{d}$
(though of course this is not meant to imply that {\it any} set in 
which $h$ appears is consistent.)
Thus, a history is consistent, $h \in \decoheringsub{d},$ 
only if \footnote{It is worth emphasizing that (\ref{eq:consistentcondition}) 
is a valid criterion for the consistency of $h$ whether or not 
$1-h \in \rep,$ and therefore whether or not inhomogeneous coarse
grainings are admissible, a fact which should become obvious after a 
moment's reflection on the fact that $1-h$ is the sum of all the other
histories with which $h$ is supposed to be consistent.  This
is good, because nothing requires $1-h$ to be in \rep\ if $h$ is.  
In particular,
%in \projrep, while it is true both that $1-h$ is a projection
%if $h$ is, and that $h$ and $1-h$ are disjoint projections, $h(1-h)=0,$
%it is not clear that $1-h$ is not ``exotic" in the sense defined in
%section \ref{sec:hilbertGQT}.  Similarly, 
there is no guarantee
whatever that $1-h$ is homogeneous if $h$ is.  In general it will not
be,  both in \projrep\ and in \classoprep.
This is yet another suggestion that it is somewhat unnatural
from the histories point of view to exclude inhomogeneous
coarse grainings.}
%where we have the additional difficulty that even if $h$ and 
%$1-h$ {\it are} both in \classoprep, there is no way to tell if they
%are physically disjoint histories {\it cf.\ }section
%(\ref{sec:historyreconstruction}).
%This is perhaps a good argument for admitting inhomogeneous coarse
%grainings.
\begin{equation}  \label{eq:consistentcondition}
d(h,1-h) = 0.
\end{equation}
However, for an arbitrary operator in $\hilbert = \linearops{\Hilbert},$
(\ref{eq:consistentcondition}) is {\it not} sufficient to imply
that $h \in \decoheringsub{d}$ in general, because there may be
solutions to this equation in \hilbert\ which do not lie in \rep.
Define the space of operator solutions to (\ref{eq:consistentcondition})
as $\consistentops = \{ h\in\hilbert \|\ d(h,1-h) = 0 \}.$
The space of {\it physical} solutions is then 
$\decoheringsubd = \consistentops \cap \rep,$ or,
$\decoheringsubd = \{ h\in\rep \|\ d(h,1-h) = 0 \}.$
Given the discussion in section \ref{sec:historyreconstruction},
it is clear that \consistentops\ is considerably more straightforward to
work with than \decoheringsubd.  In fact, the project of this section is
to exhibit \consistentops\ explicitly.  Nevertheless, once the solutions to 
(\ref{eq:consistentcondition}) have been found, additional criteria 
must then be applied to determine whether or not those solutions 
correspond to actual physical histories.  (For instance, in the case of
\projlattice{\Hilbert}, it must be checked which of the solutions are
projections.)  
%But even this does not solve the problem of Isham's ``exotic" projections.)
This is awkward, but %apparently
so far unavoidable.

At present, the formal solution that 
$\decoheringsubd = \consistentops\cap\rep,$ supplemented by the 
limited checks offered in section \ref{sec:historyreconstruction}, 
is the best that can be made of the general situation.  

Alternately, one could attempt to solve (\ref{eq:consistentcondition})
just in \projrep\ or \classoprep\ by employing an explicit
parametrization of the projections which appear in
(\ref{eq:historyproduct}) or (\ref{eq:classop}), but significant
problems remain when inhomogeneous coarse grainings are admitted.
This does in any case generally yield equations quite impossible 
to actually solve.
%I have in mind eg the parametrization that fay and kent used.
In the homogeneous case, Dowker and Kent \cite{DK}\ do briefly offer
a -- very formal -- strategy for finding the solutions, 
but this is also generally impossible to carry out in practice.

Thus, while \consistentops\ is very easy to find explicitly, 
\decoheringsubd\ is very hard.  It is perhaps most profitable to
view the situation this way: the problem of finding \decoheringsubd\
has been broken into two pieces, finding \consistentops, and then
determining whether $h \in \rep.$  If $\rep = \projlattice{\Hilbert}$
this is very easy, of course, but the other cases need more work.

(As an antidote to this negativity, it is perhaps worth emphasizing
that it is always possible, at least in principle, to sort through
\rep\ or \repbar\ and determine which of its members satisfy the
consistency condition (\ref{eq:consistentcondition}).  It is just
that we do not have the same nice characterizations of
\decoheringsubd\ that are given for \consistentops\ in sections
\ref{sec:parametrization}\ and \ref{sec:d-unitarity}.)

While the inability of $d$ to isolate elements of 
\decoheringsubd\ from \consistentops\ may be 
annoying, it is hardly surprising.  After all, in all cases a general
Hilbert space decoherence functional extends naturally to an
operator on the full space $\linearops{\Hilbert} = \hilbert.$
Having done so, $d$ is a fairly generic Hermitian form on \hilbert.
In fact, the only property of $d$ which really ``knows about"
the underlying physical \rep\ is (\ref{eq:positivity}), positivity
of $d$ on \rep. 
(It is unfortunate that this property is difficult to express
in terms less abstract than (\ref{eq:positivity}), say as a useful
condition on the eigenvalues of $d.$ %in \hilbert {\it or} \hilbertbar! 
The exception, of course, is the case of \classoprepbar, where it is 
equivalent to positivity of $d$ on \hilbert.) Positivity of $d$ on 
\rep\ is clearly not enough to tell $d$ what \rep\ is.

Putting aside these limitations, the main goal of this section 
is to exhibit the solutions to (\ref{eq:consistentcondition}) in 
\hilbert, {\it i.e.\ }to find \consistentops.  
This is the subject of the next subsection.
Knowing \consistentops, it is then possible to say something about how
to build exhaustive sets of consistent histories (operators) from
elements of \consistentops, though as above the problem of determining
the physical significance of these ``histories" remains.

\begin{figure}[b]
\begin{center}
\epsfig{file=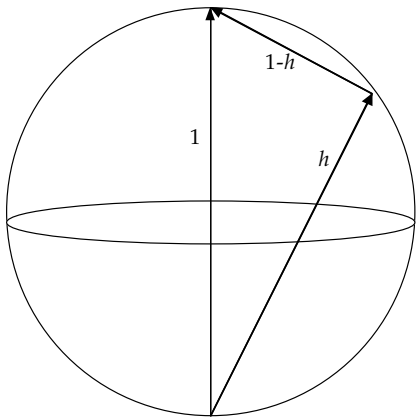,scale=0.45}%,clip=}
\end{center}
\caption{Graphical representation of the consistency condition
$d(h,1-h)=0$ for $h$ to be a consistent history operator, 
$h\in\consistentops$:  $h$ must be orthogonal to $1-h$ in 
the geometry defined by $d$.}
\label{fig:goc}
\end{figure}

To see how to build exhaustive consistent sets from \consistentops,
I make the following observation.  Suppose 
$h, h' \in \decoheringsubd.$  Then it is easy to check that
$\{ h, h', 1-h-h' \} \in \consistentsubd$ iff $d(h,h') = 0.$\footnote{To
be fair, I am glossing over the question raised in the preceeding
footnote of whether or not $1-h-h' \in \rep.$  Let us suppose either that
inhomogeneous coarse grainings are permitted, or alternately, that 
we replace \consistentsubd\ here by \consistentopsets\ defined in 
section \ref{sec:devecs}\ below.}
%or keep fine graining until $1-\sum h$ {\it is} in \rep\ (if ever).
All we have to do to build exhaustive consistent sets,
therefore, is to find in \decoheringsubd\ collections of $d$-orthogonal
operators.  I will show how to make use of this observation later in the
section.

\subsection{Parametrization of \consistentops}
\label{sec:parametrization}

The space \consistentops\ of solutions to the consistency condition 
$d(h,1-h) = 0$ is easy to visualize.
Rewrite (\ref{eq:consistentcondition}) as $d(h,h)=d(h,1).$  
$d(h,1)$ is thus real for consistent $h$, and is moreover 
between 0 and 1 so long as $h\in\rep,$ or more generally, for any
$h\in\hilbert$ so long as $d$ is positive on \hilbert; see section
\ref{sec:historyreconstruction}.  In any case, thinking of $d$ as an
inner product on the space of histories, (\ref{eq:consistentcondition})
says that $h$ is consistent, $h\in\consistentops,$ if its 
$|{\rm length}|^2$ in this inner product is equal 
to the length of its projection onto
the direction \OneH.  If $d$ is positive on \hilbert\ (as for instance 
in the canonical case), or if $h\in\rep,$ the solutions to
(\ref{eq:consistentcondition}) may then be thought of as defining the
surface of a round sphere (in the geometry defined by $d$) which has its
south pole at the origin and its north pole at the tip of the vector 
\OneH\ (see figure \ref{fig:goc}).
%\begin{figure}[b]
%\begin{center}
%%\epsfig{file=goc.eps}
%%\epsfig{file=goc.eps,height=4.00cm}%,clip=}%,%
%\epsfig{file=goc_pict.eps,height=4.00cm,clip=}%,%
%%bbllx=128pt,bblly=274pt,bburx=521pt,bbury=667pt}
%\end{center}
%\caption{Graphical representation of the consistency condition
%$d(h,1-h)=0$ for $h$ to be a consistent history operator, 
%$h\in\consistentops$:  $h$ must be orthogonal to $1-h$ in 
%the geometry defined by $d$.}
%\label{fig:goc}
%\end{figure}

%is true for {\it any} history, in virtue of positivity and
%(\ref{eq:cauchyschwarz}).   %but not real, necess, is it?
%if true only on consistent, no reason to point out why it's
%true for other reasons as well, is there?
%Therefore, for positive decoherence functionals 
%(including the canonical ones)
%\begin{equation}  \label{eq:dnormdef}
%\| h \|^{2} = d(h,h),     % put "def" over = sign?  (how? \stackrel{}{})
%\end{equation}
%define
%\begin{equation}  \label{eq:dangledef}
%d(h,1) = \| h \|\ cos\theta.       % put "def" over = sign?  (how?)
%\end{equation}
%(For non-positive kernel $d$'s imaginary $\theta$ may thus result,
%but these do not correspond to physical (consistent) histories.)
%The condition for $h$ to be consistent may then be rewritten
%\begin{equation}  \label{eq:heqcostheta}
%\| h \| = cos\theta,
%\end{equation}
%(the equation of a circle of radius $1/2.$)

%$B_1 = 1_{{\scriptsize \Hilbert}}.$ 

This picture makes it easy to parameterize the solutions to the
consistency condition explicitly.  Find a basis $\{ B_I \}$ for
$\hilbertsubd = \factorspacesubd$ (${\rm dim\, }\hilbertsubd = n_d$)
which is orthonormal with respect to $d$, with 
$B_1 = \OneH.$  (If $d$ is positive, given a set
of $n_d$ linearly independent operators on \hilbertsubd\ that includes
$\OneH,$  the usual Gram-Schmidt procedure using $d$
suffices to construct such a set.  If $d$ is not positive on \hilbert,
a variant of the Gram-Schmidt argument shows a basis always exists for
which $d(B_I,B_I) = \pm 1,$ it being inconvenient for present purposes
to allow ``null" basis vectors, $d(B_I,B_I) = 0.)$
%{\it cf.\ }Abraham and Marsden section 10.3
%Note the matrix of $d$ may be diagonal in this basis, but the basis
%is not generally either ortho nor normal on \hilbert!
The choice $B_1 = \OneH$ is not critical, but
it does simplify a little the following formulae.

Writing $h = \sum_{I=1}^{n_d}\ h_I B_I$ 
({\it cf.\ }(\ref{eq:dirachcomponent})), and setting
$\delta_I = d(B_I,B_I) = \pm 1,$ the consistency condition 
(\ref{eq:consistentcondition}) becomes
\begin{equation}\label{eq:consistentparam}
h_1 = \sum_{I=1}^{n_d}\ \delta_I\, |h_I|^2.
\end{equation}
Note that in this basis, (\ref{eq:consistentcondition}) says that $h_1$
is the probability of $h$, $d(h,h)$.  Rearranging slightly,
\begin{equation}\label{eq:consistentparabola}
         h_{1}^{2} - h_1 + \sigma = 0
\end{equation}
%setting $h$ real, as implied by (\ref{eq:consistentparam}),
%The possibility that h_1 not be real crept in when i dropped abs.
where
\begin{equation}\label{eq:sigmadef}
\sigma = \sum_{I=2}^{n_d}\ \delta_I |h_I|^2;
\end{equation}
$\sigma = \sum_{I=2}^{n_d}\ |h_I|^2$ for positive decoherence
functionals, of course.  (\ref{eq:consistentparabola}) has real
solutions for $h_1$ so long as $\sigma \leq \frac{1}{4}.$
(The two solutions for each $\sigma < \frac{1}{4}$ obviously correspond
to $h$ and $1-h$.)  Then $0 \leq h_1 \leq 1$ (a pre-requisite for $h$ to
be in \decoheringsubd, and not just \consistentops) whenever
$0 \leq \sigma \leq \frac{1}{4}.$  If $d$ is not positive, then it is
possible for $\sigma$ to be negative.  So long as 
$\sigma \leq \frac{1}{4},$  $h$ is in \consistentops, but when
$\sigma < 0,$ $\h \not\in \decoheringsubd$ because $h_1 = d(h,h)$ 
is either bigger than 1 or less than zero.

For positive decoherence functionals (such as \dcanonical) the solution
space to (\ref{eq:consistentcondition}) is then just the interior of
the sphere of radius $\frac{1}{4}$ in $\Complex^{n_d -1}.$
Otherwise the solution space is the interior of the ``hyperboloid"
in $\Complex^{n_d - 1}$ defined by $\sigma \leq \frac{1}{4}.$  In
this case, only the subset of parameters 
$\{ h_I \ \|\ I=2, \ldots , n_d \}$ for which 
$0 \leq \sigma \leq \frac{1}{4}$ have any hope of corresponding 
to {\it physical} solutions.  
%In fact, as $d$ is always positive
%on ${\rm span\, }\consistentsetsubd$ (section \ref{sec:innerproduct}), the
%$B_I$ for which $\delta_I = -1$ are not in the span of any
%\consistentsetsubd.
%${\rm span\, }\decoheringsubd$, so physical 
%histories have those $h_I = 0.$
%Oops, that was wrong.  durnit.  See example in positivity to
%see why ... b has a component along c.

%And, of course, if it is \projrep\ that is of concern, ther
%The free parameters, of course, are the
%$\{ h_I \ \| I=2, \ldots , n_d \}.$ 
%Note any set on the same sheet ($\sigma = {\rm constant}$) defines
%a history with the same probability.

There is another approach to the problem of solving the consistency
condition (\ref{eq:consistentcondition}) which has some useful features.
In order to discuss this solution, it is instructive to first fulfill a
promise made at the beginning of this section.

%\subsection{Eigenvectors of $d$ as a Maximally Fine Grained Consistent Set}
\subsection{Eigenvectors of $d$ as a Consistent Set of Histories}
\label{sec:devecs}

%I mentioned in section \ref{sec:maxnumhistories}\ that decoherence
%functionals always possess a set of ``histories" (operators) which
%saturates the bound ${\rm dim\, }\factorspacesubd$ on the maximum
%number of histories in a consistent set.  This is an appropriate
%time to exhibit such a set of operators: the eigenvectors of $d$.

I mentioned above that the eigenvectors of $d$ may always be used
to construct an exhaustive, consistent set of history operators.   
In this section I describe how this is done.

Of course, the discussion following (\ref{eq:consistentcondition})
still applies.  That is, there is no general guarantee that these
operators in \consistentops\ are actually in 
\rep, {\it i.e.\ bona fide} physical histories.  
Fortunately, though, in the case of the
canonical decoherence functionals it is manifest that they are.
Nevertheless, in order to be accurate, I will as before
denote the collection of exhaustive consistent {\it sets}
of operators by \consistentopsets, so that \consistentsubd\ comprises
the subset of \consistentopsets\ which contains only {\it physical}
consistent sets.\footnote{``Exhaustive" here of course
means that all the operators in a set sum to $\OneH.$}
The distinction between \consistentopsets\ and \consistentsubd\ is thus
wholly analogous to that between \consistentops\ and \decoheringsubd.

Suppose the decoherence functional has already been diagonalized as in 
(\ref{eq:diracd}),
\begin{equation}\label{eq:dcanonagain}
       d = \sum_{I}\ w_I\, \Ketbra{I}{I},
\end{equation}
where $\{ \Ket{I} \} \Leftrightarrow \{ E_I \}.$  (For the purposes
of this section, some of the $w$'s may be zero.  The corresponding
$\Ket{I}$'s are any normalized basis of $d$'s nullspace.)
Define the vectors
\begin{eqnarray}
     \Ketl{D_I} &=& \Ketl{I}\Bracketl{I}{\OneH}  \nonumber\\
        &=&  \trH\, E_{I}^{\dagger}\ \Ketl{I}.    \label{eq:DI}
\end{eqnarray}
Then the set $\{ \Ketl{ D_I } \}$ is an exhaustive consistent set,
%exclusive? \perp if projections; too bad linearly independent does not
%imply disjoint as classop
%You are a *fool*.  In general, *lots* of these will be zero.  Idiot.
$\{ \Ketl{ D_I } \} \in \consistentopsets.$

Indeed, it is clear from (\ref{eq:DI}) that $\{ \Ketl{ D_I } \}$ 
is exhaustive,
\begin{equation}\label{eq:Dexhaustive}
    \sum_{I}\ \Ketl{ D_I } = \Ketl{ \OneH }.
\end{equation}
(This is just the statement that 
$\OneH = \sum_{I}\ (\trH\, E_{I}^{\dagger})E_I.$)  It is also trivial 
to verify that the vectors $\Ketl{ D_I }$ are mutually consistent.  
Individually, they all satisfy the consistency condition $d(h,1-h) = 0:$
\begin{eqnarray}
 \Meltl{ D_I }{d}{ D_I }  & \stackrel{?}{=} & 
                  \Meltl{\OneH}{d}{ D_I } \nonumber\\
   \Bracketl{ D_I }{I}\, w_I\, \Bracketl{I}{ D_I }  & \stackrel{?}{=} &
   \Bracketl{\OneH}{I}\, w_I\, \Bracketl{I}{ D_I }    \nonumber\\
    w_I\, |\Bracketl{I}{ D_I }|^2 &=& w_I\, |\Bracketl{I}{ D_I }|^2  
        \label{eq:Dconsistent}
\end{eqnarray}
Thus each $\Ketl{ D_I } \in \consistentops.$  And, they are manifestly
consistent with one another,
\begin{eqnarray}
   \Meltl{D_I}{d}{D_J} 
          &=& w_I\, |\Bracket{I}{\OneH}|^2\, \delta_{IJ} \nonumber\\
%    &=&  w_I\, |\trH\, E_I|^2\delta_{IJ} \nonumber\\
     &=&  p_I\, \delta_{IJ}.    \label{eq:Ddecohere}
\end{eqnarray}
For positive decoherence functionals, or if $\Ketl{D_I} \in \rep,$
%(\ref{eq:normalization}) (that is, 
$d(1,1) = 1$ as usual forces (section \ref{sec:historyreconstruction})
the ``probabilities" $p_I$ to satisfy $ 0 \leq p_I \leq 1.$

As promised, the eigenvectors of a decoherence functional always
define an exhaustive, consistent set of history operators.
(If none of the $\Bracket{I}{\OneH}$ are zero, then this
set even saturates the bound of section \ref{sec:maxnumhistories}\
on the maximum number of histories with nonzero probability 
in a consistent set.)
The possible physical interpretation of such histories is 
an issue to which I return in a moment.

In the case of the canonical decoherence functional
(\ref{eq:dcanonagain}) the eigenvectors are simply the
$\Ket{im}$ of (\ref{eq:canonicalbasis}), so that in this case
(\ref{eq:DI}) reads
\begin{eqnarray}
   \Ketl{D_{im}} &=& \trH\, R_{im}^{\dagger}\, \Ket{im} \nonumber\\
         &=& \bracket{i}{m}\, \Ket{im}.  \label{eq:canonicalevecs}
\end{eqnarray}
In this case, at least, the class operators (\ref{eq:classop}) 
to which the $\Ket{D_{im}}$ correspond are easy to find 
({\it cf.\ }(\ref{eq:canonicalbasis})):
\begin{eqnarray}
 \bracket{i}{m}\, R_{im}  &=&  \ket{i}\bracket{i}{m}\bra{m} \nonumber\\
              &=& \Projsub{i}\Projsub{m}. \label{eq:Dimclassop}
\end{eqnarray}
Thus $\{ \Ketl{D_{im}} \} \in \consistentsubd$ 
(not just $\consistentopsets).$

For more general decoherence functionals, however, it is not at all
clear that the $\{ \Ketl{D_I} \}$ do, in fact, correspond to 
any {\it physical} histories, that is, that 
$\{ \Ketl{ D_I} \} \in \consistentsubd,$ as befits the discussion
following (\ref{eq:consistentcondition}).  
For instance, for a randomly selected decoherence
functional on \projrep, the eigenvectors of $d$ will rarely turn out
to be projections.  (As a fairly arbitrary Hermitian operator on
$\hilbert = \linearops{\Hilbert},$ $d$ has no way of knowing that
it is only supposed to see projection operators.)  
In the general case, we must simply check (as in section
\ref{sec:historyreconstruction}) whether or not 
$\{ \Ketl{ D_I} \} \in \consistentsubd$ in order to determine
whether or not the eigenvectors of $d$ have more than mere
mathematical significance.

\subsection{$d$-Unitarity and Consistent Sets of Operators}
\label{sec:d-unitarity}

%Now that we have found one maximal consistent set of history operators,
Now that we have found one consistent set of history operators,
the aim of this part is to find the rest of them.  Just as all the 
orthonormal bases of a vector space can be obtained from some fixed basis 
by unitary transformations, we can obtain the other maximally fine grained
elements of \consistentopsets\ from the eigenvectors of $d$ 
%$\{ \Ketl{D_I} \}$ 
by employing $d$-unitary transformations, followed by a rescaling 
that guarantees each new mutually consistent set still sums to 
$\ket{\OneH}$ -- {\it i.e.\ }is still exhaustive.
The complete recipe for this procedure is exhibited 
explicitly for positive decoherence functionals

In the case of positive decoherence
functionals, the set of $d$-unitary transformations that map one
exhaustive consistent set into another set with the same probabilities
will turn out to be just $SU(n-1)$.  More generally, {\it every} $d$-unitary
transformation maps one exhaustive consistent set into another, so
long as the ``rotation'' is followed by an appropriate rescaling of the
probabilities.  This rescaling is fixed by the rotation, so the freedom
to generate new consistent sets from old turns out to be $SU(n_+,n_-)$,
where $n_+,n_-$ are the number of $d$'s positive and negative eigenvalues,
respectively.

The techniques employed here may 
also be of some use in studying, for instance, the symmetries
of decoherence functionals, a subject I plan to take
up elsewhere.\footnote{Some very interesting work on
this topic has recently appeared in \cite{Schreck96a,Schreck96b}.}

To begin, I define the notion of $d$-unitarity, before moving on to
describe how to use $d$-unitary transformations to obtain one consistent 
set from another.  In order to simplify the discussion a bit I will assume 
that $\nullspacesubd = \emptyset.$  
%If it is not, simply 
%take it that we have reduced the domain of $d$ 
%by equivalence from \hilbert\ to
%$\hilbertsubd = \factorspacesubd$ in what follows.  
In addition, I drop the $\odot$ notation of (\ref{eq:HHopproduct}) 
because it is cumbersome here, and 
$\hilbertbar = \Hilbert\otimes\Hilbert$ plays no role in what follows.

\subsubsection*{$d$-Unitarity}
\addcontentsline{toc}{subsubsection}{\numberline{}%
$d$-Unitarity}

An invertible linear transformation $V$ on \hilbert\ will be said 
to be {\it d-unitary} if 
\begin{equation}\label{eq:dunitarity}
d(Vh,Vh') = d(h,h')    \qquad \forall\, h,h' \in\hilbert,
\end{equation}
in which case the condition for $d$-unitarity is 
\begin{equation}\label{eq:Vdunitary}
    d = V^*\, d\, V.
\end{equation}
%=> $V^{-1}$ is $d$-unitary.
As $|{\rm det\, }V| = 1$ and any overall phase is clearly irrelevant,
we can always take ${\rm det\, }V = 1.$  Next, define $W$ and $\delta$
via the ``polar decomposition" of $d$ in \hilbert\ 
({\it cf.\ }(\ref{eq:polardecomp})),
\begin{equation}\label{eq:dpolardecomp}
    d = \delta\ W,
\end{equation}
where
\begin{eqnarray}
     W &=& \abssubh{d}  \nonumber\\
    &=& \sum_I\ |w_I|\ \Ketbra{I}{I}  \label{eq:Wdef}
\end{eqnarray}
and
\begin{equation}\label{eq:deltadef}
       \delta = \sum_I\ \delta_I\ \Ketbra{I}{I}.
\end{equation}
The notation here is the same as in (\ref{eq:dcanonagain}) above, of
course, and $\delta_I = {\rm sign}( w_I ) = \pm 1.$  If $d$ is positive
on \hilbert, $\delta$ is of course just $\Oneh$, and $d=W.$
(Remember that $\delta$ is severely restricted by the facts that $d$
is not only positive on \rep, but also 
-- see section \ref{sec:innerproduct}\ --
on the span of any would-be physical consistent set.)

Notice that $\delta = d\, W^{-1}$.  Inserting factors of $\sqrt{W}$ 
appropriately into (\ref{eq:Vdunitary}),
it is easy to see that it is always possible to write $V$ 
as\footnote{If $\nullspacesubd \neq \emptyset$, then $V$ is 
the direct sum of an operator of the form (\ref{eq:WUW}) on 
$\hilbertsubd \simeq \nullspaceperpsubd$ and an invertible 
operator on $\nullspacesubd$.  This is because a $d$-unitary
$V$ leaves $\nullspacesubd$ and $\nullspaceperpsubd$ invariant.
To see this, note that (\ref{eq:Vdunitary}) still holds on all of
\hilbert.  If $V$ mapped 
$\nullspaceperpsubd \rightarrow \nullspacesubd$  then there would
be an $\Ket{h} \in \nullspaceperpsubd$ for which $d\Ket{h} \neq 0$
but $dV\Ket{h} = 0$, violating (\ref{eq:Vdunitary}).  Conversely,
if $V$ mapped $\nullspacesubd\rightarrow\nullspaceperpsubd$ then
there would be some $\Ket{h}\in\nullspacesubd$ for which 
$d\Ket{h} = 0$ but $dV\Ket{h}\neq0$.  Insisting that $V$ be
invertible means that $V^*dV\Ket{h}\neq 0$ as well, again 
contradicting (\ref{eq:Vdunitary}).  $V$ thus splits up into
(\ref{eq:WUW}) on $\hilbertsubd \simeq \nullspaceperpsubd$, and 
an invertible piece in $\nullspacesubd$ that more
or less just ``comes along for the ride'' in the subsequent
discussion.}
\begin{equation}\label{eq:WUW}
   V = \sqrt{W^{-1}}\, U\, \sqrt{W},
\end{equation}
where $U$ satisfies
\begin{equation}\label{eq:UinSUnn}
        U^*\, \delta\, U = \delta.
\end{equation}
%This is where it is helpful to drop \nullspacesubd ... defining
%what's meant by W^{-1}, what U's are allowed, etc. is annoying.
In other words, $U \in SU(n_{+},n_{-}),$ where $n_{+}$ and $n_{-}$ count
the number of $d$'s positive and negative eigenvalues, respectively, so
$n_{+} + n_{-} = n = {\rm dim\, }\hilbert.$  
%(Really, $n_{+} + n_{-} = {\rm dim\, }\hilbertsubd.$  
(For information on $SU(p,q)$ see \cite{Helgason,BR}.)  
%The collection of $d$-unitary maps is thus isomorphic to $SU(n_+,n_-)$.
When $d$ is positive on \hilbert,
$\delta = 1 $ and $U$ is therefore just a unitary operator.
%Honestly, haven't shown UU^\dagger = 1.  With WUW, though, wigner's
%thm guarantees from the def of V that U really is unitary.
%(cf. R+N or B+F on isometric vs. unitary in \inf dimensions.
%Alternately, in finite dimensions the right inverse must be the
%same as the left: AU = 1 = UB => A = AUB = B.

\subsubsection*{New Consistent Sets from Old\footnote{I am grateful
to Tom\'{a}\v{s} Kopf for very useful conversations concerning the
material of this section.}}
\addcontentsline{toc}{subsubsection}{\numberline{}%
New Consistent Sets from Old}

Following the lead of section \ref{sec:devecs},
let me now describe how to employ $d$-unitary maps to generate
%all of the other maximally fine grained, exhaustive, consistent 
exhaustive, consistent 
sets of operators from $d$'s eigenvectors, the $\{ \Ket{I} \}$
of (\ref{eq:dcanonagain}).
For positive decoherence functionals it will be clear that we
can get {\it all} of the maximally fine grained, exhaustive, 
sets of operators consistent according to $d$ in this way.
(More work is required for decoherence functionals 
which are not positive.)

It is of course true that if $\{ h \}$ is a consistent set of
operators, then $d(h,h')=0$ when $h \neq h'$, and naturally the set
$\{ Vh \}$ will have the same property.  However, it will not 
generally be true that $\{ Vh \}$ is a {\it consistent} set
because a general $d$-unitary ``rotation'' will move each $h$
off of the ``consistency sphere'' pictured in figure \ref{fig:goc},
so that $Vh \notin \consistentops$ and 
$\{ Vh \} \notin \consistentopsets$.

There is a simple remedy for this, however: $Vh$ can be 
rescaled to put its tip back on to the ``consistency sphere''.   
Consider a consistent $h \in \consistentops$ and an arbitrary
$d$-unitary $V$.  We would like to rescale $Vh$ by a number $t$
so that $tVh$ is consistent as well:
\be
           d(tVh,1) = d(tVh,tVh)   \label{eq:tV}
\ee
so that
\be
           t_h = \frac{d(Vh,1)}{d(h,h)}.    \label{eq:th}
\ee
This is just the coefficient of $h$ in the expansion of $V^{-1}1$
in the $d$-orthogonal set $\{ h \}$; the significance of this
observation will become evident in (\ref{eq:Iwigglesum})-(\ref{eq:Iwiggleone}) 
below.  (Also compare (\ref{eq:DI}).)
The consistency of $h$ implies that $t$ is real; it obviously
depends on both $h$ and $V$.  

(If $h$ has zero probability, $d(h,h)=0$, then $t^*\, d(Vh,1)=0$ 
in order that $tVh$ be consistent, 
so that $t=0$ if $d(Vh,1) \neq 0$.  If $d(Vh,1) = 0$, it may be 
verified that the correct $t$ to guarantee exhaustiveness  --
 {\it i.e.\ }(\ref{eq:Iwiggleone}) -- is again the coefficient of
$h$ in the expansion of $V^{-1}1$ in the set $\{ h \}$.  
This is always possible so long as $\Span\{ h \} = \hilbert$.)

The map $\{ h \} \rightarrow \{ t_h Vh \}$ generates
new consistent sets from old.   Note, however, that it is
sufficient that the initial set $\{ h \}$ be mutually $d$-orthogonal
in order that $\{ t_h Vh \}$ be a genuinely consistent set of
operators, {\it i.e.\ }be in \consistentopsets; we did not have
to require that the $\{ h \}$ sum to 1.  (The consistency of
$h$ only shows that $t_h$ is real, an inessential
property for present purposes.)
%Just need span \{ h \} and \{ Vh \} to contain 1.  Best: begin
%with \span\{ h\} = \hilbert.  V cannot spoil the linear independence
%of \{ h \}.

In order to construct exhaustive, consistent sets of operators
by this ``rotate-rescale'' procedure, it is therefore convenient 
to begin with  -- switching now to history space Dirac notation --  
$d$'s eigenvectors $\{ \Ket{I} \}$.  They comprise a normalized,
linearly independent set of $d$-orthogonal operators, and none
of them have $\Melt{I}{d}{I} =0$ when $\nullspacesubd = \emptyset$,
as we are here assuming.

With the choice (\ref{eq:th}) of $t_I$ for each $\Ket{I}$,
$\Ket{\tilde{I}} \in \consistentops$ 
and $\{ \Ket{\tilde{I}} \} \in \consistentopsets$,
where
\be
      \Ket{\tilde{I}}  = t_I\, V\Ket{I}.   \label{eq:hwiggle}
\ee
In other words, the $\{ \Ket{\tilde{I}} \}$ comprise an
exhaustive, consistent set of operators.
%(Note that when $V=1$, this process just yields 
%the $\Ket{D_I}$ of (\ref{eq:DI}).)

Indeed, from the $d$-unitarity of $V$ it is obvious that the
$ \Ket{\tilde{I}} $ are mutually $d$-orthogonal.
In order to verify that $\{ \Ket{\tilde{I}} \}$ is exhaustive,
% if $\{ \Ket{h} \}$ is, 
we need to show that 
$\sum_{\tilde{I}}\, \Ket{\tilde{I}} = \Ket{\OneH}$:
\bea
  \sum_{\tilde{I}}\, \Ket{\tilde{I}} 
           &=& \quad\ \sum_I\ \frac{d(VI,1)}{d(I,I)}\ \  V \Ket{I}  
                                             \nonumber\\
           &=& V\ \sum_I\ 
               \frac{\Melt{I}{V^*d}{\OneH}}{\Melt{I}{d}{I}}\ \
                      \Ket{I}                \nonumber \\
           &=& V\ \sum_I\ 
               \frac{\Ketbra{I}{I}}{\Melt{I}{d}{I}}\ \
                      V^*d \Ket{\OneH}                \nonumber \\
           &=& V\ \sum_I\ 
               \frac{\Ketbra{I}{I}d}{\Melt{I}{d}{I}}\ \
                  V^{-1}\Ket{\OneH},                \label{eq:Iwigglesum}
\eea
where the last line follows from the $d$-unitarity of $V$, $d=V^*dV$.

Recall now that the $\{ \Ket{I} \}$ are 
%an exhaustive, consistent,
a linearly independent set of vectors -- in other words, a basis for 
\hilbert.
%\hilbertsubd.  
%(They are linearly independent because by hypothesis
%none of them have zero probability; see section \ref{sec:maxnumhistories}.)
Because they are also $d$-orthogonal,
\be
 \sum_I\ \frac{\Ketbra{I}{I}d}{\Melt{I}{d}{I}}\ \ \Ket{I'} = \Ket{I'}.
            \label{eq:wackyunit}
\ee
As $\{ \Ket{I} \}$ is a basis, this means that 
\bea
     \sum_I\ \Ket{\tilde{I}} &=& V V^{-1} \Ket{\OneH} \nonumber\\
                             &=& \Ket{\OneH}   \label{eq:Iwiggleone}
\eea
and $\{ \Ket{\tilde{I}} \}$ is indeed exhaustive.
            
Thus, with the rescalings appropriate to keep rotated vectors
on the consistency sphere of figure \ref{fig:goc}, 
$d$-unitary ``rotations'' can generate 
%maximally fine grained,
exhaustive consistent sets from the eigenvectors of $d$.   
%$\{ \Ketl{D_I} \}$ defined in (\ref{eq:DI}). 
In the case of positive decoherence functionals, the operator
$U$ appearing in (\ref{eq:WUW}) is unitary, and it is geometrically
clear that we
can generate in this way {\it all} of the maximally fine grained,
exhaustive sets of operators consistent according to $d$;
coarse grainings of these sets generate
\consistentopsets.   (There is obviously a great deal of redundancy in
this description of \consistentopsets.) 
%which I do not attempt to eliminate
As ever, the question of which of these sets are physical remains.
%U's which preserve fact h's are projections?  hermiticity => commutes
%w/ h; unitarity does the rest.
%U's which preserve classopness of D_im?  S(UnxUn)?

We have to be somewhat more careful in the case of decoherence
functionals which are not positive.  The reason for this is the
following: our map $\{ I \} \rightarrow \{t_IVI \}$ can only
produce trivial zero probability consistent histories ($t_I =0$);
a $d$-unitary transformation can {\it never} map a zero probability
history to a non-zero probability one.  However, even when
$\nullspacesubd = \emptyset$ (as I have assumed for convenience
in this section) %we expect 
there are, in general, non-trivial zero probability history 
operators that are consistent according to a non-positive $d$.

%I believe it is possible to
I suspect it is possible to generate all of the maximally fine grained 
exhaustive consistent sets of a non-positive decoherence functional
by the rotate-rescale transformations $\{ h \} \rightarrow \{t_hVh \}$,
so long as we begin with an appropriate collection of 
$d$-orthogonal sets $\{ h \}$, 
each with a different number of zero probability histories.
However, because of the non-trivial ``light-cone'' structure 
endowed on \hilbert\ by a non-positive $d$, this appears to 
be a complicated issue to address in general.  

It is of course true that applying the rescaling (\ref{eq:th}) to
every complete, linearly independent $d$-orthogonal set $\{ h \}$
will yield all of the maximally fine-grained consistent sets of $d$ 
(up to degeneracy in linearly dependent, zero-probability histories).
It is just that we do not have the same nice classification of
all of the maximal $d$-orthogonal sets by the operators $V$
that we do in the positive case.  

(The proof that a unitary operator always exists which maps
one orthonormal basis into any other effectively identifies 
ordered orthonormal bases with unitary operators, given a fiducial
basis.  For positive decoherence functionals, the more or less 
obvious generalization of this proof puts $d$-orthogonal sets in
correspondence with $d$-unitary operators, given a fiducial 
$d$-orthogonal set -- $d$'s eigenvectors -- whence the fact that the
``rotate-rescale'' transformations are sufficient to generate all
of the maximally fine-grained consistent sets when $d$ is positive.
On the other hand, 
the bases of a complex vector space that are orthogonal in 
a metric of indefinite signature (the $\delta$ of (\ref{eq:deltadef})) 
are {\it not} in correspondence with $U$'s in $SU(n_+,n_-)$ ($U$'s
satisfying (\ref{eq:UinSUnn})) in the same simple way.   This
is just the observation I made above that a non-zero probability
history can never be mapped to a zero probability history by a
$d$-unitary $V$.   All that this means is that another 
classification of the $d$-orthogonal sets must be employed.
I delve into the issue no further here.)

\subsubsection*{A Nice Picture\footnote{I would like to thank 
Tom\'{a}\v{s} Kopf for the conversations which led to the figures 
of this section.}}
\addcontentsline{toc}{subsubsection}{\numberline{}A Nice Picture}

\begin{figure}[hbt]
\begin{center}
\epsfig{file=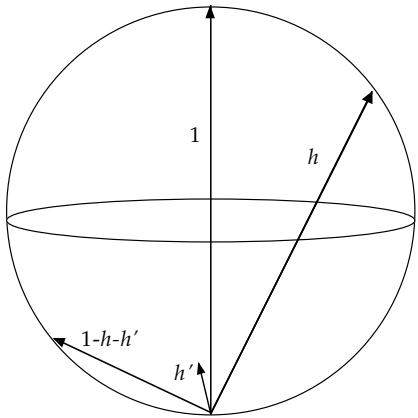,scale=0.45}%,clip=}
\end{center}
\caption{Members of an exhaustive consistent set $\{ h, h', 1-h-h' \}$ 
must be mutually orthogonal in the geometry defined by $d$, and they
must all lie on the ``consistency sphere'' defined by
(\ref{eq:consistentcondition}).}
\label{fig:tomas}
\end{figure}

For positive decoherence functionals, there is a very nice picture 
of these rotate-rescale transformations that is based on the 
``consistency sphere'' of figure \ref{fig:goc}.   
Recall that this figure is drawn in in the geometry 
defined by $d$.  In that geometry, consistent history operators
are just those which lie on the sphere surrounding the preferred vector
$\OneH$.  We would like to generalize the picture to accommodate consistent
sets with more than two histories in them.  Therefore, transplant the
vector $1-h$ so that its base is at the origin (south pole); its tip 
of course still lies on the sphere.  Now, as noted at the end of section
\ref{sec:decoheringsubd}, if $h, h' \in \consistentops$,
$\{ h, h', 1-h-h' \} \in \consistentopsets$ iff $d(h,h') = 0$.
Figure \ref{fig:goc}\ may therefore be generalized for 
$\{ h, h', 1-h-h' \}$ to figure \ref{fig:tomas}.  
(Three mutually ($d$-)orthogonal histories is the most we can draw 
on paper, but the generalization to more histories (higher dimensions) 
is clear.)

\begin{figure}[hbt]
\begin{center}
\epsfig{file=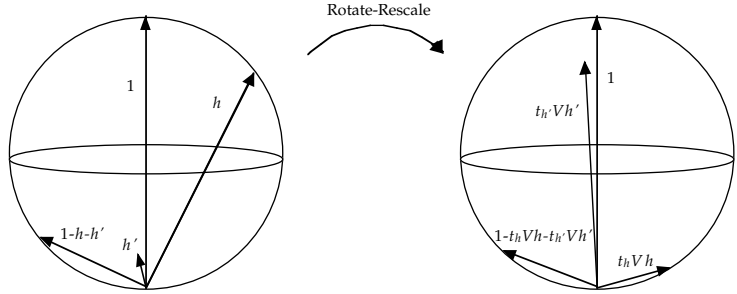,scale=0.5}%,clip=}
\end{center}
\caption{The rotation-plus-rescaling transformation 
of one exhaustive consistent set into another.}
\label{fig:transform}
\end{figure}

The consistency preserving $d$-unitary rotation-plus-rescaling
transformations described in the previous subsection 
merely correspond to rigid rotations
of the $d$-orthogonal set $\{ h, h', 1-h-h' \}$, while simultaneously
stretching or squeezing each vector so that its tip remains on the
surface of the consistency sphere, as pictured in figure \ref{fig:transform}.  
(The probability-preserving maps described
in the next subsection are just those which spin the set
rigidly ``around'' the vector 1.)

\subsubsection*{Probability Preserving Transformations}
\addcontentsline{toc}{subsubsection}{\numberline{}%
Probability Preserving Transformations}

There is a special subclass of these transformations that is of some
interest, namely, those which preserve the probabilities of the original
set.  These are just those $V$'s for which all the $t_h$ = 1, so    
-- dropping the sub-\Hilbert\ on the unit 1 in \Hilbert\ -- 
that $1 = \sum Vh = V\, (\sum h) = V\, 1.$
In other words, we also need
\begin{equation}\label{eq:Vevec}
   V\, \Ket{1} = \Ket{1}.
\end{equation}
In terms of $U$ this reads
\begin{equation}\label{eq:Uevec}
   U\, \Ketl{\Onewiggle} = \Ketl{\Onewiggle},
\end{equation}
where
\begin{equation}\label{eq:1wiggle}
  \Ketl{\Onewiggle} \equiv \sqrt{W}\, \Ketl{1}.
\end{equation}
(Note that when $d$ is positive, $W =d,$ so that
$\Bracketl{\Onewiggle}{\Onewiggle} = \Melt{1}{d}{1} = 1.$
Otherwise, $\Meltl{\Onewiggle}{\delta}{\Onewiggle} = 1$ in
the same way.)

The collection of $d$-unitary maps which preserve 
both the probabilities and the exhaustiveness
of a set of histories is thus specified by the subgroup of
$SU(n_{+},n_{-})$ which has $\Ketl{\Onewiggle}$ as a fixed point.

At least in the case of the positive decoherence functionals, the
condition (\ref{eq:Uevec}) can be solved explicitly.  To see how, note
that $U$'s which satisfy (\ref{eq:Uevec}) may always be written as
\begin{equation}\label{eq:Udecomp}
   U = \Ketbral{\Onewiggle}{\Onewiggle} + U_{\perp},
\end{equation}
where (\ref{eq:Uevec}) and unitarity imply that 
\begin{eqnarray}
U_{\perp}\, \Ketl{\Onewiggle} 
       &=& U_{\perp}^{*}\, \Ketl{\Onewiggle} \nonumber\\
        &=& 0, \label{eq:Uperp}
\end{eqnarray}
and so
\begin{eqnarray}
  U_{\perp}^{*}\, U_{\perp} 
      &=& 1 - \Ketbral{\Onewiggle}{\Onewiggle}  \nonumber\\
      &=& 1_{\perp}.   \label{eq:Uperpunitary}
\end{eqnarray}
%(In the the same way, 
%$U_{\perp}\, U_{\perp}^{\dagger} = 1_{\perp}$ as well.)
%Or...
%\begin{eqnarray}
  %U_{\perp}\, U_{\perp}^{\dagger} 
      %&=& 1_{\perp}.   \label{eq:Uperpunitary}
%\end{eqnarray}
%Follows from UU^\dagger = 1, of course
%(Remember that that $\Bracketl{\Onewiggle}{\Onewiggle} = 1$ when
%$d$ is positive on \hilbert.)

Thus, the maps which preserve both the probabilities and the 
completeness of a set of consistent histories
are completely covered by $SU(n-1)$ when $d$ is positive.

In the general case where $\delta \neq 1,$ a formulation very similar to
the positive case can be worked out, but the solution does not come out
nearly so nicely.  The basic reason for this is that $\Ketl{\Onewiggle}$
is never an eigenvector of $\delta,$ and thus the condition that 
$\Ketl{\Onewiggle}$ be an eigenvector of $U$ does not break the symmetry
down to an $SU(p,q)$ subgroup of $SU(n_{+},n_{-})$ in a clean way.
%Or, I haven't succeeded in doing so yet, anyway.
What you get seems no easier to use than (\ref{eq:Uevec}), and I
therefore won't bother to discuss it.

%\newpage

\section[The Decoherence Functionals Consistent with a Set of 
Histories]{The Decoherence Functionals Consistent with a Set of Histories}
\label{sec:functional?}

The aim of this section is to address some of the mathematical 
aspects of the problem of determining how observations constrain 
the decoherence functional of a closed quantum system.  To that
end, I show how to construct all of the decoherence functionals
according to which a fixed set of histories is consistent in section
\ref{sec:sets}.  I also briefly address the question of how to
use these results to determine the family of decoherence functionals
according to which a {\it collection} of sets of histories are all 
consistent.

\subsection{The Problem}
\label{sec:problem}

While it is always possible to make physical predictions {\it given}
a decoherence functional, it is clearly of physical interest
to know how observations constrain what the possible decoherence
functionals of a closed system might be.  A first simple step
is to determine what are all the decoherence functionals according 
to which a given history or set of histories is consistent.  
The aim of this section is to approach this problem 
when the observables of the system live in a finite 
dimensional Hilbert space.  

Here, at least, the difficulty in determining when an operator
in \hilbert\ represents a physical history is not an issue because
we are picking histories and finding decoherence functionals, 
instead of the other way around.

An explicit, constructive solution to this
problem is given in what follows.
Namely, given an 
exhaustive set of histories and their probabilities, it is shown
in (\ref{eq:d=d}) 
how to construct the decoherence functionals according to which
this set is consistent, and which give the required probabilities.
(The practical issue that remains when $d$ is not {\it a priori}
positive is how to characterize operationally
the positivity of $d$ on the elements of \rep\ which do not lie 
in the span of the given set of histories.)

\subsection{Decoherence Functionals Consistent with Sets of Histories}
\label{sec:sets}

By now, the method for constructing decoherence functionals according
to which a given set of histories is consistent is almost embarrassingly
obvious.\footnote{A better trick would be to construct the decoherence
functionals according to which {\it several} sets of histories 
$\{ \consistentset_1, \ldots, \consistentset_m \}$ are individually 
consistent, {\it i.e.\ }
$\{d \parallel\, \{ \consistentset_1, \ldots, \consistentset_m \} 
                \subset \consistentsubd \},$ 
but I address that problem only briefly here.}
It is based on the simple observation that a linear operator is
determined by its matrix elements.  (The construction of this 
section is very much in the spirit of that of Schreckenberg 
\cite{S}\ for the case $\rep = \projlattice{\Hilbert}.$)
% ... but was done before he did it ;-)

Suppose that we have a set of disjoint, linearly independent
histories $\consistentset = \{ \Ket{K} \},$ and a set of
``probabilities" $\{ p_K \}$ for those histories.  (Remember 
from section \ref{sec:maxnumhistories}\ that all the non-zero 
probability histories in a consistent set must be linearly independent.  
Therefore, coarse grain all the $p_K = 0$ histories into one.)
It is not necessary at this point to assume that \consistentset\
is exhaustive, {\it i.e.\ }that
$\sum_K\, \Ket{K} = \Ket{\OneH}$ and $\sum_K\, p_K = 1.$  
Nor is it necessary to suppose that $\hilbertsubS = \hilbert,$
where I am calling ${\rm span\, }\consistentset = \hilbertsubS.$
We will see later what we get for assuming that \consistentset\ is
exhaustive and/or maximally fine grained.

The problem is to find the decoherence functionals according 
to which the set \consistentset\ is consistent.  First the
decoherence functionals according to which the histories in
\consistentset\ are mutually $d$-orthogonal will be found; when
\consistentset\ is exhaustive we will see that
$\consistentset \in \consistentsubd$ for free  ({\it i.e.\ }that
all the $\Ket{K} \in \decoheringsubd$).  Another way to state the
problem is, what are all the $d$ for which
\begin{equation}\label{eq:Kconsistent}
   \Melt{K}{d}{K'} = p_K\ \delta_{KK'}.
\end{equation}
But the $\{ \Ket{K} \}$ are linearly independent, so
(\ref{eq:Kconsistent}) {\it defines} the restriction of $d$ 
to \hilbertsubS.  To \dhS\ may be added any Hermitian operator on 
\hilbert\ which maps \hilbert\ to \hilbertsubSsup{\perp}\
-- in other words, any operator with zero matrix elements in
\hilbertsubS\ --
that keeps $d$ positive on \rep, generating the full family of
decoherence functionals which satisfy (\ref{eq:Kconsistent}).

To construct \dhS\ explicitly, first find an orthonormal basis
$\{ \Ket{M} \}$ in \hilbertsubS.  
(If $\rep = \projlattice{\Hilbert},$ disjoint projections are already 
trace-orthogonal and it is sensible to choose 
$\Ket{M} = \Ket{K}/\sqrt{\Bracket{K}{K}}.$  Disjoint class operators,
on the other hand, are rarely trace-orthogonal.)  Then we may expand
\begin{eqnarray}
\Ket{K} &=& \sum_M\ \Ket{M}\Bracket{M}{K}  \nonumber\\
        &=& \sum_M\ W_{MK}\, \Ket{M}  \label{eq:KMW}
\end{eqnarray}
and
\begin{equation}\label{eq:MKV}
    \Ket{M} = \sum_K\ V_{KM}\, \Ket{K}.
\end{equation}
(If $\Ket{K} \propto \Ket{M},$ $W_{MK}$ and $V_{KM}$ are obviously just
proportional to $\delta_{KM}.$)  It is easy to check that
\begin{equation}\label{eq:VW}
    \sum_M\ V_{K'M}\, W_{MK} = \delta_{KK'}
\end{equation}
and
\begin{equation}\label{eq:WV}
    \sum_M\ W_{M'K}\, V_{KM} = \delta_{MM'},
\end{equation}
so that the $V_{KM}$ may be calculated as the matrix inverse to the
known quantities $W_{MK} = \Bracket{M}{K}.$    
(If $p_K \neq 0$ then it is clear from (\ref{eq:Kconsistent}) and 
(\ref{eq:MKV}) that $V_{KM} = \Melt{K}{d}{M}/p_K$.)

It is convenient to expand the unit on \hilbertsubS\ as
\begin{eqnarray}
  \OnehS &=& \sum_M\ \Ketbra{M}{M}   \nonumber\\
         &=& \sum_{MK}\ V_{KM}^{*}\ \Ketbra{M}{K}.  \label{eq:unithS}
\end{eqnarray}

With (\ref{eq:unithS}) it is easy to reconstruct \dhS\ from 
(\ref{eq:Kconsistent}) by multiplying on the left by 
$V_{KM}^{*}\, \Ket{M},$ on the right by
$V_{K'M'}\, \Bra{M'},$ and summing over $KK'MM'.$  You get
\begin{eqnarray}
   \dhS &=&  \sum_{MM'}\ d_{MM'}\, \Ketbra{M}{M'}  \nonumber\\
       &=& \sum_{MM'}\
   \left( \sum_K\ p_K\, V_{KM}^{*}V_{KM'} \right)\, \Ketbra{M}{M'},
         \label{eq:d=d}
\end{eqnarray}
where
\begin{equation}\label{eq:dMM=VdKKV}
   d_{MM'} =  \sum_{KK'}\  V_{KM}^{*}\, d_{KK'}\, V_{K'M'},
\end{equation}
$d_{KK'}$ being defined by (\ref{eq:Kconsistent}).
The members of \consistentset\ are mutually $d$-orthogonal 
according to this decoherence functional.
When \consistentset\ is exhaustive, 
$\sum_K\, \Ket{K} = \Ket{\OneH}$ and 
$\sum_K\, p_K = 1,$ it is easy to check that each 
$\Ket{K} \in \decoheringsubd:$ 
$d(K,1) = d(K,\sum_{K'} K') = \sum_{K'}\, d(K,K') = d(K,K),$
so that $\consistentset\in\consistentsubd.$
A similarly short calculation shows that this $d$ is also
properly normalized, $d(1,1) = 1.$
%Need exhaustiveness because no guarantee that if 1 = \sum S + h',
%that d(h,h') = 0.  That is, h' is not necessarily entirely
%in \hilbertsubSsup{\perp}.  Thus it may be that d(h,1) =
%d(h,h) + d(h,h')

%Note that if the p_k are not given a priori, this construction
%still generates all the \dhS's according to which \consistentset
%is consistent, with the p_K as free parameters.

As already noted, we can just add to \dhS\ any Hermitian
operator on \hilbert\ which has zero matrix elements in
\hilbertsubS\ to generate all the decoherence
functionals for which $\consistentset \in \consistentsubd.$
If \consistentset\ is maximally fine grained 
({\it i.e.\ }possesses ${\rm dim\, }\hilbert$ 
members; {\it cf.\ }section \ref{sec:maxnumhistories}), then
$\hilbertsubS = \hilbert$ and $d$ is specified uniquely by
$\left( \consistentset, \{ p_K \} \right)$, 
hardly a surprising result.

I have so far said nothing about positivity.  That \dhS\ is positive
on \hilbertsubS\ should be clear from (\ref{eq:Kconsistent}) and the
fact that $\{ \Ket{K} \}$ spans \hilbertsubS.  If
$\hilbertsubS \neq \hilbert,$ the freedom to add operators 
to \dhS\ is clearly restricted by the
requirement that $d$ is positive on \rep; it is unfortunate that
this property is difficult to characterize constructively.

Notice from this that {\it decoherence functionals are positive 
on the span of each of their (physical) consistent sets.} 
Thus any decoherence functional which possesses 
a consistent set that spans \hilbert\
(a maximally fine grained consistent set with no zero 
probability members) is positive on all of \hilbert.  
(Compare section \ref{sec:innerproduct}.)

Finally, consider briefly the problem of determining the family of
decoherence functionals according to which a collection
$\{ \consistentset_1, \ldots, \consistentset_m \} 
\subset \consistentsubd$.\footnote{Sadly, though all such $d$ are 
positive on each of the ${\cal H}_{ {\scriptscriptstyle {\cal S}_i} }$,
this does not necessarily imply they are all positive on the join
of these spaces.  A simple example illustrates why.  Suppose
$d= A\, \Ketbra{A}{A} - C\, \Ketbra{C}{C}$, where \Ket{A}\ and \Ket{C}\
are unit, and $\Bracket{A}{C}=0$.  Then 
$\Melt{B}{d}{B} > 0$ so long as 
$A/C > \frac{|\Bracket{C}{B}|^2}{|\Bracket{A}{B}|^2}$.  
In that case, $d$ is positive on ${\rm span\, }\{\Ket{A}\}$ and 
${\rm span\, }\{\Ket{B}\}$, but not on ${\rm span\, }\{ \Ket{A},\Ket{B} \}$.}
Suppose we are given the matrix elements of $d$ on each $\consistentset_i$.  
This specification must of course be given self-consistently 
on each intersection
${\cal H}_{ {\scriptscriptstyle {\cal S}_i} }\cap{\cal H}_{ {
            \scriptscriptstyle {\cal S}_j} }$.
The remaining freedom to specify $d$ lies in the 
matrix elements {\it between} the
${\cal H}_{ {\scriptscriptstyle {\cal S}_i} }$, and in the complement 
to ${\rm span\, }\{ \consistentset_1, \ldots, \consistentset_m \}.$
To say anything much more explicit requires some detailed information 
about the $\consistentset_i$.

Suppose therefore that some of the consistent sets of histories of a 
closed system \system\ have somehow been determined.
On the assumption
that \system\ is described by a decoherence functional (generalized
quantum state) $d$, the decoherence functional of \system\ may be
reconstructed on the span of these consistent sets just as above; 
the matrix elements between the consistent sets are left as free 
parameters, restricted only by the requirement that $d$ is positive 
on \rep.  (If the probabilities have not also been found they may 
be regarded as additional free parameters in $d$.  If so, they are
restricted by the requirement that $d$ is consistently defined on 
each intersection
${\cal H}_{ {\scriptscriptstyle {\cal S}_i} }\cap{\cal H}_{ {
            \scriptscriptstyle {\cal S}_j} }$.)
%This is what is meant by the ``experimental determination of 
%$d$" \cite{edd}.  

(There is a curious physical problem underlying all of this: 
how does one determine {\it any} consistent set of histories 
of a closed system, since only one history will actually be 
realized?  The answer is that one doesn't.  Rather, one sorts 
through the observed properties of, say, the universe, and 
takes the set of histories corresponding to each property 
$P$, $(h_P,1-h_P)$, to be consistent.  
In other words, for each observed property $P$, $h_P$ must be 
a coarse-graining of the ``one true history" of our universe.
Taken together, 
the observed properties of the universe thereby constrain
the possible decoherence functionals which may describe
its state.   I hope to return to this issue in a future
publication.)

\section{When is a Decoherence Functional Canonical?}
\label{sec:dcanonical?}

\subsection{Two Criteria}
\label{sec:criteria}

In section \ref{sec:example}\ I showed how the canonical decoherence
functional \dcanonical\ of (\ref{eq:dcanonical}) defines an inner
product on $\hilbertif \equiv \hilbert/\nullspaceif.$  In this section
the converse problem is posed: when does a positive Hermitian form on
$\hilbert \simeq \Hilbert\otimes\Hilbert^{*}$ define a canonical
decoherence functional?  This investigation only 
addresses the issue for
decoherence functionals defined on the class operator representation
\classoprep(\Hilbert) of histories in the Hilbert space
\Hilbert\footnote{As I mentioned in section \ref{sec:ILSW}, 
it is straightforward (if rather messy) to find explicitly 
the form of the ILS operator
$X_{\alpha\omega}$ on $\otimes^{2k}\Hilbert$ corresponding to a
canonical decoherence functional (\ref{eq:dcanonical}) on
\classoprepbar(\Hilbert).  Given an ILS operator $X$ on
$\otimes^{2k}\Hilbert$ it is therefore possible to determine whether it
is canonical by checking whether its matrix elements have the proper
form.  This discussion would take us rather far afield, however, and
bring with it a virtual blizzard of new notation, so that it
will have to be given elsewhere.  This section assumes it is a
decoherence functional on \classoprep(\Hilbert) or
\classoprepbar(\Hilbert) that is being considered.}; the solution is
expressed in (\ref{eq:Didempotent}) or (\ref{eq:Dbarproj}) below.

To pose the question more explicitly, given some positive decoherence
functional $d$ over \classoprep\ or \classoprepbar, when may it be 
written in the form (\ref{eq:dcanonicaldiag})
\begin{eqnarray}
d &=&  \sum_{im}\ a_{i}z_{m}\, \Ketbra{im}{im}   \nonumber\\
% &=&  \sum_{im}\ a_{i}z_{m}\, \ketbra{im}{mi}   \nonumber\\
  &=&  \sum_{im}\ a_{i}z_{m}\, \ketbra{i}{m}\otimes\ketbra{m}{i}  \nonumber\\
  &=&  \sum_{im}\ a_{i}z_{m}\  R_{im}\otimes R_{im}^{\dagger} \label{eq:dcanon}
\end{eqnarray}
for two (usually distinct) bases of \Hilbert\ 
$\{ \ket{i} \}$ and $\{ \ket{m} \},$ and accompanying sets of
non-negative numbers  $\{ a_{i} \}$ and $\{ z_{m} \}$  
(compare (\ref{eq:diracdcanonicalYZ}))?  If it can, then $d$ is a
canonical decoherence functional,
\begin{equation}\label{eq:dcanonhh}
\dcanonical(h,h') = \trif{h}{h'},
\end{equation}
with
\begin{equation}\label{eq:dcanonrhalpha}
\rhalpha = \sum_{i=1}^{N}\ a_{i}\, \ketbra{i}{i}
\end{equation}
and
\begin{equation}\label{eq:dcanonrhomega}
\rhomega = \sum_{m=1}^{N}\ z_{m}\, \ketbra{m}{m}.
\end{equation}
%(The summation convention is of course in effect.)  
As $d(1,1)=1,$ the lack of a normalization constant in front of the 
trace in (\ref{eq:dcanonhh}) corresponds to the choice
\begin{equation}\label{eq:dcanonnorm}
\trH\, \rhalpha\rhomega = 1.
\end{equation}

There are (at least) two ways to answer the question of whether a given
$d$ is canonical, which, moreover, provide a means by which to explicitly 
reconstruct, up to an overall relative scale, the boundary conditions 
\rhalpha\ and \rhomega.  Both of them employ the operator
\begin{eqnarray}
D  &=&  \sqrt{d^{\dagger}d}  \label{eq:Ddef}\\
   &=&   \abssubhbar{d},        \label{eq:DdefHH}
\end{eqnarray}
%Note that in \rep = \projrep or \projlattice{\Hilbert} this is just |X|.
where the second equality serves to emphasize that {\it all three}
operator operations in (\ref{eq:Ddef}) are taken in
$\underline{\hilbert} = \Hilbert\otimes\Hilbert,$ and not in
$\Hilbert\otimes\Hilbert^{*}.$
 
Having computed $D$,\footnote{The only practical method for computing 
$D = |d|_{{\scriptscriptstyle  H\otimes H}}$
is to diagonalize $d^{\dagger}d$ in $\Hilbert\otimes\Hilbert;$
$|d|_{{\scriptscriptstyle  H\otimes H}}$ is then obtained by taking
the square roots of the eigenvalues of $d^{\dagger}d.$  
The square root of $d^{\dagger}_{\alpha\omega}d_{\alpha\omega}$ 
%$D^2_{\alpha\omega}$  %\Dcanonical\ 
is easy because it is already diagonal in the basis
$\{ \ket{mi} \} = \{ \ket{m}\otimes\ket{i} \},$ 
  {\it cf.\ }(\ref{eq:Dsquared}) and (\ref{eq:dabs}).}
each of the following are %(independent) 
necessary and sufficient conditions for a positive decoherence
functional $d$ to be canonical: \\
\\
%i)  $\tilde{D}$ is idempotent in \hilbert,  %Drop!  \trh D = 1!
i)  $D$ is idempotent in \hilbert,
\begin{equation}\label{eq:Didempotent}
                   D \odot D = D.
%D \odot D = (\trh\, D)\ D,
\end{equation}
%where of course
%\begin{equation}\label{eq:Dwiggle}
  %\tilde{D} \equiv D/\trh\, D.
%\end{equation}
Alternately,\\
\\
ii) $\overline{D}$ is a projection operator on \hilbert,
\begin{equation}\label{eq:Dbarproj}
  \overline{D} \odot  \overline{D} = \overline{D} 
\end{equation}
where
\begin{eqnarray}
  \overline{D} & \equiv & N_{D}^{-1}\ D^{*} \odot D  \label{eq:Dbardef}\\
       %&=& N_{D}^{-1}\, | D |_{\hilbert}^{2}  \nonumber\\
N_{D}  & \equiv  & \trh\, D^{*} \odot D.   \label{eq:Dnormdef}
       %&=& \| D \|_{2\hilbert}^{2}  \nonumber\\
\end{eqnarray}
%i.e. N_{D} is the square of \hilbert's Hilbert-Schmidt norm.
%Note it is automatically positive and real.
Notice that these conditions are (not entirely coincidentally) reminiscent 
of two of the conditions for an ordinary density matrix to be pure,
namely, that a density matrix $\rho$ is a pure state (i) iff $\rho$ is a
projection, or, (ii) iff $\rho^{2} = \rho.$   (A third %independent
condition is that $\rho$ is pure iff $\trH\, \rho^2 = 1.$)

First I will explain what these conditions have to do with canonical
decoherence functionals, then I will show why they are true.  The 
first step is to compute $D$ for the canonical decoherence functional
(\ref{eq:dcanon}):
\begin{eqnarray}
\Dcanonicalsup{2} &=& 
     d_{\alpha\omega}^{\dagger}\dcanonical \nonumber\\
  &=& \sum_{im}\, \sum_{jn}\ a_{i}a_{j}z_{m}z_{n}
      (R_{im}^{\dagger}\otimes R_{im})
      (R_{jn}\otimes R_{jn}^{\dagger})  \nonumber\\
  &=& \sum_{im}\, \sum_{jn}\ a_{i}a_{j}z_{m}z_{n}
      (\ket{m}\bracket{i}{j}\bra{n}\otimes\ket{i}\bracket{m}{n}\bra{j})
                                        \nonumber\\
  &=& \sum_{im}\ a_{i}^{2}z_{m}^{2}\ \ketbra{m}{m}\otimes\ketbra{i}{i}
                                        \nonumber\\
% &=& \sum_{im}\ a_{i}^{2}z_{m}^{2}\ \ketbra{mi}{mi} \nonumber\\
  &=& \rho_{\omega}^{2}\otimes\rho_{\alpha}^{2}.
\label{eq:Dsquared}
\end{eqnarray}
%Note $D$ does {\it not} factor on $\Hilbert\otimes\Hilbert$ in the same
%sense as it does on \hilbert; $D$ is positive, Hermitian on
%\Hilbert\otimes\Hilerbt, but not on \hilbert --- bounded, positive
%implies self adjoint, remember!  
%From $\trHH\, \rhomega\otimes\rhalpha = \trH\, \rhomega \trH\, \rhalpha$
%that $\trHH  D_{\alpha\omega}^{2}$ is maximized when $\rhomega, \rhalpha$
%are pure via the usual criteria.
Thus 
\begin{equation}\label{eq:DHH}
    D_{\alpha\omega} = \rhomega\otimes\rhalpha,
%  &=& \sum{im}\ a_i z_m\, \ketbra{mi}{mi}.
\end{equation}
or, in history space Dirac notation,
\begin{equation}\label{eq:Ddirac}
    D_{\alpha\omega} = \Ketbra{\rhomega}{\rhalpha}.
\end{equation}

A positive decoherence functional $d$ is canonical if, and only if, 
$D = \abssubhbar{d}$ factors in \hilbert\ in this simple way.
(\ref{eq:Ddirac}) is obviously necessary for $d$ to be a canonical
decoherence functional.  That it is also sufficient is proved in section
\ref{sec:proofs}.  The canonicality criteria (i) and (ii) are merely
necessary and sufficient conditions for the factorization (\ref{eq:Ddirac}),
as shown in the following two subsections.
%section \ref{sec:origins}\ and \ref{sec:proofs}.

Now let me show that \Dcanonical\ satisfies the 
canonicality conditions (i) and (ii).
First, I note that with the chosen normalization
$\dcanonical(1,1) = 1 = \Bracketl{\rhalpha}{\rhomega} = \trH\, 
\rhalpha\rhomega,$ (\ref{eq:HHopproduct}) or (\ref{eq:Ddirac}) shows that 
$\Dcanonical\odot\Dcanonical = \Dcanonical,$ so that $\Dcanonical$ is
idempotent on \hilbert, with 
$\trh\, \Dcanonical = \Bracketl{\rhalpha}{\rhomega} = 1.$
%\trH\, \rhalpha\rhomega.$
(However, $\Dcanonical$ is not a projection on \hilbert\ because 
it is not (generically) self adjoint, 
$\Dcanonicalsup{*} \neq \Dcanonical.$)
\Dcanonical\ thus satisfies condition (i).  Similarly, 
$\overline{D}_{\alpha\omega} = \frac{\|\rho_{\alpha}\rangle\langle
      \rho_{\alpha}\|}{\langle\rho_{\alpha}\|\rho_{\alpha}\rangle},$
which is manifestly a projection on \hilbert, so that \Dcanonical\
satisfies the second canonicality condition (ii) as well.

(As an aside, I note that 
$\trh\, \Dcanonical = \trHH\, \dcanonical = \trH\, \rhalpha\rhomega,$
and that 
$\trh\, \dcanonical = \trHH\, \Dcanonical = \trH\, \rhalpha\trH\, \rhomega.$
The pattern displayed here does not, however, carry over to the
case of a general operator.  
%In fact, $\trHH\, g$ can be both greater or less than
%$\trh\, |g|_{\Hilbert\otimes\Hilbert},$ for instance.  
It may %even 
be that satisfaction of these equations is an additional 
necessary and sufficient condition for $d$ to be canonical, analogous to
the condition that a density matrix is pure iff $\trH \rho^{2} = 1,$ 
but that has not been shown.)

Thus the conditions (i) and (ii) are necessary for $d$ to be canonical,
though it is hardly obvious that they are sufficient.  To see why they are,
let us first see how these conditions come about.  This will also show us 
how to recover \rhalpha\ and \rhomega\ from a canonical $D.$

\subsection{Origin of Canonicality Criteria and Reconstruction
of the Boundary Conditions}\label{sec:origins}
%\addcontentsline{toc}{subsubsection}{\numberline{}%
%Origin of Canonicality Criteria and Reconstruction
%of the Boundary Conditions}
%%cf. The LaTeX companion section 2.4.2

The root of the (independent) conditions (\ref{eq:Didempotent}) or
(\ref{eq:Dbarproj}) for a decoherence functional to be canonical lies in
the following observation: given some operator $B,$ there exist unit
vectors \ket{b}\ and \ket{\beta}\ such that $B = \ketbra{b}{\beta}$ iff
$\tilde{B} = B/ \Tr B$ is idempotent.  Alternately, $B = \ketbra{b}{\beta}$
iff $B^{\dagger}B$ is a one-dimensional projection.  Taking care of the
case where \ket{b}\ and \ket{\beta}\ do not have to be unit, an
operator $B$ simply factors iff either 
(i) $\tilde{B} = B/N_{1}$ ($N_{1} = \Tr B$) is idempotent, 
or, equivalently,
(ii) $\overline{B} = B^{\dagger}B/N_{2}$ is a 
(one dimensional - $\Tr\overline{B} = 1$)
projection, where                       %$\overline{B} = B/N_{2}$ and 
$N_{2} = \Tr B^{\dagger}B.$
%N2 is the (square of) the Hilbert-Schmidt (S2) norm, but N1 is *not* the
%S1 norm = tr|B|.
(To be totally accurate, stated in this way the first condition carries 
the additional assumption that $B$ is not traceless, which occurs when
$\bracket{b}{\beta} = 0;$ see section \ref{sec:proofs}.  
That does not affect us here because of (\ref{eq:diracdtr}).)

%%!!!!!!!!!!!!!!!!!!!!!!!!!!!!!!!!!!!!!!!!!!!!!!!!!!!!!!!!!!!!!!!!!!!!!!!
%%deal with case that B is traceless? B=bbeta factors, but is b.beta=0 ...
%nilpotent ....

The proof of these alternative conditions for the factorization of $B$ 
is given in the next section, after the proof of the sufficiency (not 
just necessity) of the canonicality criteria.  Here, I show how to
reconstruct $\ket{b}$ and $\ket{\beta}$; this will show how \rhalpha\
and \rhomega\ may be recovered from a $D$ which factorizes as in
(\ref{eq:Ddirac}).

From a factorizing $B$, 
it is possible to reconstruct $\ket{b}$ and 
$\ket{\beta}$ up to phase by defining vectors 
$\ketl{\overline{b}}$ and $\ketl{\overline{\beta}}$ through
$\overline{B} = \ketbral{\overline{\beta}}{\overline{\beta}}$ and 
$\ketl{\overline{b}} = B\, \ketl{\overline{\beta}}/\sqrt{N_2}$ for
$\bracketl{\overline{\beta}}{\overline{\beta}} =
     \bracketl{\overline{b}}{\overline{b}} = 1.$
Recalling that $B = \ketbra{b}{\beta},$  and with
$\ket{b} = b\, e^{i\phi}\ketl{\overline{b}}$ and
$\ket{\beta} = \beta\, e^{i\theta}\ketl{\overline{\beta}}$ for 
positive real $b, \beta,$
\begin{equation}
B^{\dagger}B = 
      b^{2}\beta^{2}\, \ketbra{\overline{\beta}}{\overline{\beta}},
\end{equation}
so that $\Tr\overline{B} = 1$ fixes $b^2\beta^2$ to be equal to
\begin{equation}\label{eq:eqn}
  \Tr B^{\dagger}B = \bracket{b}{b}\bracket{\beta}{\beta}.
\end{equation}
(It is of course no surprise that, beginning only with $B,$ 
it is only the product of the norms of
\ket{b}\ and \ket{\beta}\ that is determined.)  Checking the
consistency of $B = \ketbra{b}{\beta}$ with 
$\ketl{\overline{b}} = B\, \ketl{\overline{\beta}}/\sqrt{N_2}$ then fixes
$\phi = \theta,$ which, being arbitrary anyway, is natural to absorb in
the arbitrary phase implicit in \ketl{\overline{\beta}}.

The upshot is that given an operator $B$ for which either 
$\tilde{B} = B/\Tr B$ is idempotent, or 
$\overline{B} =  B^{\dagger}B/\Tr B^{\dagger}B$ is a projection, we may
write $B = \ketbra{b}{\beta},$ where
\begin{eqnarray}
  \ket{b} &=& \beta^{-2}\, B\ket{\beta}, \\
  \ket{\beta} &=& \beta\, \ketl{\overline{\beta}}
\end{eqnarray}
for arbitrary $\beta\in\Real^{+},$ and $\ketl{\overline{\beta}}$
is determined up to a phase (which does not appear in $B$ anyway) from
\begin{equation}
    \overline{B} = \ketbral{\overline{\beta}}{\overline{\beta}}.
\end{equation}

%Determining \ket{\beta}} from $B$:  suppose $P=\ketbra{\beta}{\beta}.$
%Then $\melt{j}{P}{i} = \beta_{i}^{*}\beta_{j},$ or 
%$P\ket{i} = \beta_{i}^{*}\ket{\beta}},$ so that
%$\ket{\beta}} = \frac{ P\ket{i} }{ \| P\ket{i} \| }$ 
%for any $P\ket{i} \neq 0,$ or more straightforwardly
%$\ket{\beta}} = \frac{ \sum_{i}\ P\ket{i} }{ \| \sum_{i}\ P\ket{i} \| }.$ 

It should by now be clear where this is going.  Translating these
factorization results into language appropriate for
operators on \hilbert, an operator $C$ factors, 
$C = \Ketbra{Z}{A},$ iff either (i) $\tilde{C} = C/\trh\, C$ is idempotent, 
or, (ii) $\overline{C}$ is a projection on \hilbert.
($\overline{C},$ of course, is just $\overline{C}^{*}\odot\overline{C}$ 
divided by $N_C = \trh\, C^{*}\odot C.$)  
If either of these equivalent conditions
is satisfied, than as above the factor \Ket{A} may be reconstructed up
to an arbitrary phase as 
$\Ket{A} = a\Ketl{\overline{A}}$ $(a \in \Real^{+}),$
where 
$\overline{C} = \Ketbral{\overline{A}}{\overline{A}},$ and
$\Ket{Z} = a^{-2}C\Ket{A}.$

Applying this result to an arbitrary $d,$ the origins of the conditions
(\ref{eq:Didempotent}) or (\ref{eq:Dbarproj}) for the ``canonicality" of
$d$ are now becoming evident.   (Let $C = D,$ remembering that
$\trh\, \Dcanonical = \Bracketl{\rhalpha}{\rhomega} = 1.$)  The
canonicality conditions are obviously necessary, as \Dcanonical\ obeys
them.  That they are also sufficient is not as immediately clear, but as
I show below, the factorization of 
$D = |d|_{{\scriptscriptstyle  H\otimes H}}$ as in (\ref{eq:DHH})
occurs only if $d$ has the canonical form (\ref{eq:dcanon}).
As a bonus, we also know how to explicitly reconstruct
the boundary conditions \rhalpha\ and \rhomega\ (up to an arbitrary
relative scale) from a canonical $d.$

%Notice that nowhere have I required $d$ to be positive.  In fact, the
%positivity of $d$ follows from the satisfaction of the canonicality
%conditions.  For, if $D$ factors, we can write 
%$D = Z\otimes A^{\dagger}.$  But $D = D^{\dagger}$ by its construction
%(\ref{eq:Ddef}), so that we can always take 
%$Z=Z^{\dagger}$ and $A=A^{\dagger}.$  Further, 
%$D = |d|_{\Hilbert\otimes\Hilbert}$ is always positive on
%$\Hilbert\otimes\Hilbert,$ and $Z$ and $A$ may therefore be chosen
%positive.  (The eigenvalues of $D$ are just the products of the
%eigenvalues of $Z$ and $A$, as evidenced in (\ref{eq:Dsquared});
%\Dcanonical\ has, in $\Hilbert\otimes\Hilbert,$ eigenvectors 
%$\ket{mi} \equiv \ket{m}\otimes\ket{i},$ with corresponding eigenvalues
%$z_{m}a_{i}.$)  
%Very important fact:  
%A\otimes B = C\otimes D \Rightarrow A \propto C, B \propto D.
%Comparing with, say, (\ref{eq:dcanon}), the positivity of \dcanonical\
%is therefore ensured by the factorization conditions!
%Bzzzzzt!!!  Neither |d|_hilbertunderbar nor
%|d^\dagger|_\hilbertunderbar can reveal the phases absorbed into the
%phi_n to make the b_n positive.

\subsubsection*{Another Statement of the Canonicality Criteria}
\label{sec:pedestrian}
\addcontentsline{toc}{subsubsection}{\numberline{}%
Another Statement of the Canonicality Criteria}
%cf. The LaTeX companion section 2.4.2

If all of that is a little too abstract, here is the algorithm with the
fancy clothes removed.  Diagonalise your positive decoherence functional
in \hilbert\ as in (\ref{eq:diracd}) or (\ref{eq:dcanon}), 
writing this on $\Hilbert\otimes\Hilbert$ as
\begin{equation}\label{eq:HHdiracd}
  d  = \sum_{M}\ w_{M}\ E_{M}\otimes E_{M}^{\dagger}.
\end{equation}
By the second of the factorization criteria, check to see if the
\trH\ - orthonormal basis $\{ E_{M} \}$ factors into a pair of
orthonormal bases $\{ \ket{i} \}$ and $\{ \ket{m} \}$ of \Hilbert\ 
%Namely, check whether $E_{M}E_{M} = (\trH\, E_{M})\, E_{M}$
%Use 2nd criterion to most easily assure and orthonormal {\it set}.
(by checking that the $\{ E_{M}^{\dagger}E_{M} \}$ are a set of 
orthogonal projections on \Hilbert, and likewise for the 
$\{ E_{M}E_{M}^{\dagger} \}.$  )  Then set
$E_{M}^{\dagger}E_{M} = \ketbra{m}{m}$ and
$E_{M}E_{M}^{\dagger} = \ketbra{i}{i},$ so that $E_{M} = \ketbra{i}{m}.$
Write now ($M\equiv (im)$)
\begin{equation}\label{eq:Efactors}
d = \sum_{im}\ w_{im}\ \ketbra{i}{m}\otimes\ketbra{m}{i}.
\end{equation}
Now check that the {\it a priori} $n=N^{2}$ $({\rm dim\, }\Hilbert = N)$
independent numbers $w_{im}$ factor as well into only $2N$ independent
numbers.  This occurs iff
$w_{im}w_{jn} = w_{in}w_{jm},$ in which case it makes sense to assert
that $w_{im} = a_{i}z_{m},$ where
\begin{equation}\label{eq:wfactorsa}
\frac{a_{i}}{a_{j}} = \frac{w_{im}}{w_{jm}}
\end{equation}
for any $m$ and
\begin{equation}\label{eq:wfactorsz}
\frac{z_{m}}{z_{n}} = \frac{w_{im}}{w_{in}}
\end{equation}
for any $i.$  This determines the $a$'s and $z$'s up to an overall
relative scale that is set by the requirement that 
$\trH\, \rhalpha\rhomega = 1,$ defining of course \rhalpha\ and
\rhomega\ through (\ref{eq:dcanonrhalpha}) and (\ref{eq:dcanonrhomega})
up to the same relative scale.

Thus, the property that both the diagonalizing basis {\it and} the
eigenvalues factor 
is equivalent to the canonical character of any 
positive decoherence functional on \classoprep\ (\classoprepbar).

\subsection{Proofs}
\label{sec:proofs}

It remains to supply the proofs of the aforementioned factorization
conditions, and of the sufficiency of the factorization of $D$ (as in
(\ref{eq:Ddirac})) for the canonicality of a decoherence functional $d.$
I take the second task first.

It is not at all obvious that the factorization of 
$D = \abssubhbar{d}$ in \hilbert\ is sufficient to imply the
canonicality of $d,$ for, in taking the absolute value of $d$ in
$\underline{\hilbert} = \Hilbert\otimes\Hilbert,$ it would appear that
potentially significant information (a ``phase'', {\it i.e.\ }an
$\underline{\hilbert}$-unitary operator; {\it cf.} (\ref{eq:polardecomp}) 
has been discarded.  Put another
way, it is clear that {\it if} $D$ factorizes as in (\ref{eq:Ddirac}),
we may {\it define} a canonical decoherence functional which reproduces
this $d.$  But might there not be some non-canonical $d$ which also
leads to $D$?  Indeed, it is only because $d$ is self-adjoint in
\hilbert, $d = d^{*},$ that this cannot occur.  %is not so.

To proceed, it is helpful to employ 
the ``canonical form" of an operator,
\begin{equation}\label{eq:canonicalform}
B = \sum_{n} b_{n}\, \ketbral{\phi_{n}}{\psi_{n}}.
\end{equation}
Here, the $\{ \ketl{\phi_{n}} \}$ and the $\{ \ketl{\psi_{n}} \}$ are
both orthonormal sets, and the ``singular values" $b_{n}$ are positive
real numbers.  In fact, they are the eigenvalues of 
\begin{eqnarray}
|B| &\equiv & \sqrt{B^{\dagger}B}     \nonumber\\
    &=& \sum_{n}\, b_{n}\, \ketbral{\psi_{n}}{\psi_{n}}.
           \label{eq:Babs}
\end{eqnarray}
%(In infinite dimensions, a formula like this holds for any compact %at least 
%operator, including in particular all the finite rank ones.)
Having computed $|B|$ by diagonalizing $B^{\dagger}B,$ the
$\ketl{\phi_{n}}$ may then be constructed as 
$\ketl{\phi_{n}} = B\ketl{\psi_{n}}/b_{n}.$ 
(A proof of (\ref{eq:canonicalform}) for any compact --
including in particular any finite rank -- operator,
reasonably obvious in finite dimensions, may be found in 
\cite[section VI.17]{RS}.)

Defining the operator
\begin{equation}\label{eq:partialisometry}
    U = \sum_{\stackrel{m}{{\scriptscriptstyle b_m \neq 0}}}
         \ketbral{\phi_m}{\psi_m},
\end{equation}
it should be obvious that
\begin{equation}\label{eq:polardecomp}
    B  =  U\, |B|.
\end{equation}
(This is the ``polar decomposition" of the operator $B,$ analogous to
the polar form of a complex number $ z = |z|e^{i\theta};$ {\it cf.\ }
\cite[section VI.4]{RS}.)

Consider now a canonical decoherence functional (\ref{eq:dcanon}).  It
may be written in a form convenient on 
$\hilbertbar = \Hilbert\otimes\Hilbert$ as
\begin{eqnarray}
  \dcanonical &=& \sum_{im}\ a_i z_m\, \ketbra{im}{mi} \nonumber\\
              &=& \sum_{N}\ d_N\,\ketbral{\phi_N}{\psi_N},
                \label{eq:dHH}
\end{eqnarray}
making the obvious identifications $\ket{im} = \ket{i}\otimes\ket{m},$
$N = (im),$ and
\begin{eqnarray}
  \ketl{\psi_{im}} &=& \ket{mi}  \label{eq:psiim}\\
  \ketl{\phi_{im}} &=& \ket{im}  \label{eq:phiim}\\
         d_{im}  &=&  a_i z_m.   \label{eq:dim}
\end{eqnarray}
Writing $\dcanonical$ in terms of the boundary condition eigenbasis 
$\Ket{im}$ (as in the first line of (\ref{eq:dcanon})) thus uncovers
directly the canonical form (\ref{eq:canonicalform}) of \dcanonical\
in \hilbertbar.

In a similar fashion, (\ref{eq:Dsquared}) reveals
\begin{eqnarray}
  \Dcanonical &=& \abssubhbar{\dcanonical} \nonumber\\
              &=& \sum_{im}\ a_i z_m\, \ketbra{mi}{mi} \nonumber\\
              &=& \sum_{N}\ d_N\,\ketbral{\psi_N}{\psi_N},
                \label{eq:dabs}
\end{eqnarray}
so that the polar decomposition of \dcanonical\ in \hilbertbar\ reads
\begin{equation}\label{eq:dcanonpolar}
     \dcanonical = U_{\alpha\omega}\Dcanonical
\end{equation}
where
\begin{eqnarray}
  U_{\alpha\omega} &=& \sum_{N}\ \ketbral{\phi_N}{\psi_N} \nonumber\\
      &=& \sum_{\stackrel{im}{{\scriptscriptstyle d_{im} \neq 0}}} 
        \ketbral{im}{mi}. \label{eq:Ucanonical}
\end{eqnarray}

Conversely, suppose that $D$ factors in \hilbert.  It can then be
written as in (\ref{eq:dabs}), from which the 
$\{ \ketl{\psi_N} \}$ and $\{ d_N \}$ (which both obviously factor) can
be inferred.  But what about the $\{ \ketl{\phi_N} \}$?

Inspection of (\ref{eq:dHH}) or (\ref{eq:dcanonpolar}) reveals 
that {\it any} choice of the $\{ \ketl{\phi_{im}} \}$ would lead 
to the same $\sqrt{d^{\dagger}d}$ as in (\ref{eq:dabs}), so long as
$\{ \ketl{\psi_{im}} \}$ and $\{ d_{im} \}$ are chosen as in
(\ref{eq:psiim}) and (\ref{eq:dim}).  However, this arbitrariness 
does not occur for decoherence functionals because they are 
Hermitian on \hilbert, $d = d^{*}$ (recall (\ref{eq:diracd}).)

To see this, note from (\ref{eq:dHH}) that
\begin{eqnarray}
  \abssubhbar{\dcanonicalsup{\dagger}} &=& 
      \sqrt{\dcanonical\dcanonicalsup{\dagger}} \nonumber\\
      &=& \sum_{im}\ d_{im}\, \ketbral{\phi_{im}}{\phi_{im}} \nonumber\\
      &=& \rhalpha \otimes \rhomega  \nonumber\\
      &=& \Ketbral{\rhalpha}{\rhomega}        \label{eq:ddaggerabs}
\end{eqnarray}
using (\ref{eq:phiim}).  The second equality in (\ref{eq:ddaggerabs})
shows that for any $d$ the $\{ \ketl{ \phi_N } \}$ may be recovered, up
to phase, as the eigenvectors (in \hilbertbar) of 
$\abssubhbar{d^{\dagger}}.$  The last line, however, shows that
$\abssubhbar{\dcanonical} = \abssubhbarsup{\dcanonicalsup{\dagger}}{*},$
so that if this condition is satisfied, the $\{ \ketl{\phi_N} \}$ may be
recovered as well from a $D$ which factors.  Specifically,
(\ref{eq:dabs}) shows that if 
$\abssubhbar{d} = \sum_{im}\ a_i z_m\, \ketbra{mi}{mi},$ then
$\abssubhbar{d^{\dagger}} = \abssubhbarsup{d}{*}  
%(\sum_{im}\ a_i z_m\, \ketbra{m}{m}\otimes\ketbra{i}{i})^{*}
= \sum_{im}\ a_i z_m\, \ketbra{im}{im}.$  Making the choice of phase
required by $ \ketl{\phi_N} = d\ketl{\psi_N}/d_N$ ({\it cf.\ }below
(\ref{eq:canonicalform})), this shows that 
$ \ketl{\psi_{im}} = \ket{mi}$ implies $ \ketl{\phi_{im}} = \ket{im},$
as in (\ref{eq:psiim}) and (\ref{eq:phiim}), and $d$ therefore assumes
the canonical form (\ref{eq:dcanon}) ({\it cf.\ }(\ref{eq:dHH}).)

The useful observation at this point is that 
$\abssubhbar{d} = \abssubhbarsup{d^{\dagger}}{*}$ 
is {\it guaranteed} for any $d$ by $d = d^{*},$ so knowing only that $D$
factors is sufficient to imply the canonicality of $d.$
%($\abssubhbar{d^{\dagger}} = D^*)$

(Checking that 
$d = d^* \Rightarrow \abssubhbar{d} = \abssubhbarsup{d^{\dagger}}{*}$ 
is a reasonably straightforward calculation.  One way to proceed is to
use (\ref{eq:HHdiracg}) to show that $gg^{\dagger} = (g^{\dagger}g)^{*}$ if
$g = g^*,$ and then to use the canonical form (\ref{eq:canonicalform})
of $g$ in \hilbertbar\ to confirm that in this case taking square roots
in \hilbertbar\ ``commutes" with taking adjoints 
in \hilbert, {\it i.e.\ }that 
$({\sqrt{g^{\dagger}g}})^{*} = \sqrt{(g^{\dagger}g)^{*}}.$)

What about positivity?  While it is true that I have shown that the
conditions (\ref{eq:Didempotent}) or (\ref{eq:Dbarproj}) are
necessary and sufficient to imply that an arbitrary decoherence
functional has the {\it form} (\ref{eq:dcanon}), they are {\it not}
sufficient to imply that $d$ is positive.  This is because the phases
absorbed into the $\ketl{\phi_N}$ to ensure the singular values are
positive are lost in the construction of
$ \abssubhbar{d}$ (or $\abssubhbar{d^{\dagger}}$).  We can of course
use a $D$ which factors to {\it define} a truly positive decoherence
functional, but in order to determine whether or not $d$ is canonical to
begin with, we must go back and check whether or not all the eigenvalues
$w_{im} = a_i z_m$ are positive.  It is for this reason that the 
canonicality criteria were stated only for explicitly positive
decoherence functionals.

%Another appendix?
Finally, I end this section with the proof of the factorization
conditions.
%As a bonus, s sensibly direct method for constructing $D$ from a
%general $d$ becomes evident.   %Too easy!

I treat the second condition first.  An operator $B$ factors into unit
vectors, $B = \ketbra{b}{\beta},$ iff $B^{\dagger}B$ is a
(one-dimensional) projection.  
%There are a number of cute ways to show this.  I present the most
%abbreviated.
%Special case of RS prop p197, i suppose.

The ``only if" part is manifest: $B =\ketbra{b}{\beta}$ implies
$B^{\dagger}B = \ketbra{\beta}{\beta}$ and 
$BB^{\dagger} = \ketbra{b}{b}$ if \ket{b}\ and \ket{\beta}\ are unit.
Conversely, suppose $B^{\dagger}B$ is a one-dimensional projection.
Then there is some unit vector \ket{\beta}\ (unique up to phase) for
which $B^{\dagger}B = \ketbra{\beta}{\beta}.$\footnote{For instance, let
$P = B^{\dagger}B$ and $\{ \ket{i} \}$ be any orthonormal basis.  Then
$\ket{\beta} = \frac{\sum_{i}\ P\ket{i}}{\|\sum_{i}\ P\ket{i}\|}$
will do.}  Define $\ket{b} = B\ket{\beta}.$  
Then $B = \ketbra{b}{\beta} + R,$ where $R\ket{\beta} = 0.$  But
$\Tr B^{\dagger}B = 1.$  Taking the trace in a basis in which
$\ketl{e_{1}} = \ket{\beta}$ then quickly reveals that 
$\Tr\, R^{\dagger}R = 0.$  As 
$R^{\dagger}R$ is positive, $R^{\dagger}R,$ and hence $R,$ must be zero.

The first condition is usually easier to use.  Namely, a trace 1
operator $B$ factors iff $B$ is idempotent, $B^{2} = B.$  (In which
case, an {\it arbitrary} $B$ factors iff $B/\Tr B$ is
idempotent, discounting traceless $B$'s.  Traceless $B$'s which factor
are nillpotent, but nillpotent operators - which are always traceless -
do not have to factor.  This bug does not infect the previous
factorization condition, or even this one stated with the assumption 
that $B$ is trace 1.)

The easiest way to see this employs again the canonical form
(\ref{eq:canonicalform}) of $B.$ Given $B$ in this canonical form, 
and supposing $B^{2} = B,$ a short calculation (compute 
$\meltl{\phi_{k}}{B}{\psi_{l}}),$ shows that for each non-zero $b_{k},$
$1= \nosum_{k}\ b_{k}\bracketl{\psi_{k}}{\phi_{k}}.$  As
$\Tr\, B = 1 = \sum_{m} b_{m}\bracketl{\psi_{m}}{\phi_{m}},$ only one
$b_{k}$ can be non-zero for idempotent $B,$ and trace 1 idempotent $B$'s
therefore factor.  The converse, that trace 1 $B$'s which factor are
idempotent, is trivial.

%The final step is to apply this to 
%$D = |d|_{{\scriptsize \Hilbert\otimes\Hilbert}},$
%considered as an operator on \hilbert.
%
%$D$ may be computed by diagonalizing $d^{\dagger}d$ in
%$\Hilbert\otimes\Hilbert,$ making it easy to take the square root on
%$\Hilbert\otimes\Hilbert.$  Its canonical form {\it in} \hilbert\ may
%then be computed by diagonalizing $D^{*}\odot D$ in \hilbert.
%Remembering that $\trh\, \Dcanonical = 1,$ 
%the proof just given then shows that $D$ factors in \hilbert,
%$D = \Ketbra{Z}{A},$ iff $D$ is idempotent in \hilbert.

%\newpage

%\section{Equivalences Between Initial and Final Conditions}
%\label{sec:equivalences}

\section{Summation}
\label{sec:answers}

%\section{Everybody Loves a Vector Space, or, Why Linear Algebra is Cool}
%\label{sec:answers}

The aim of this investigation has been to study 
the basic structures of generalized quantum theory in        
%(finite dimensional) 
Hilbert space: the operators which represent quantum histories, 
and the decoherence functional, the generalization of the (algebraic) 
quantum state functional which measures both the interference between 
quantum histories, and the probabilities of those histories when they 
are consistent.

Generalized quantum mechanics is so useful precisely because it 
is a very general framework in which to formulate quantum mechanical 
theories.  Its basic postulates are designed to capture one of the 
most characteristic features of quantum theory: not every history 
that can be described can be  assigned a probability consistently 
\cite{Jerusalem,Lesh,OmnesQM}.  This feature, coupled with the 
intimately related facts that the framework is essentially designed 
around the superposition principle, and is constructed so that its 
fundamental predictions are probabilities (however interpreted), are 
what merit its designation as a kind of quantum theory.  The really 
new element generalized quantum mechanics brings to the theoretical 
table is the decoherence functional.  By providing a measure of 
interference between quantum histories that is internal to the theory, 
the decoherence functional %finally 
allows quantum mechanics to be 
sensibly applied to closed quantum systems.  There is no need to rely 
on an environment external to a quantum system to ``measure" it, 
thereby determining which of the system's histories may be assigned 
probabilities.  This is terribly important in, for instance, quantum 
cosmology, where the ambition is to apply the principles of quantum 
mechanics to the entire universe considered as a single quantum system.

However, the very generality of the basic framework of generalized 
quantum theory means that there is little physics contained in it 
beyond these seemingly basic features of quantum mechanics.  In this 
respect it is much like Dirac's transformation theory, which is largely 
empty of physics until it is supplemented by postulates making the 
connection between the mathematical elements of the theory and things 
that can be observed (for instance, identifying operators with physical 
observables; specifying a theory of the dynamics of those 
observables -- in other words, a Hamiltonian;   
and naming the physical symmetries of the system.)  
Nevertheless, the quantum mechanical features encoded by generalized 
quantum theory are of great interest in and of themselves.  They seem 
close to a minimum of what of quantum physics one might want to retain 
in efforts to generalize to theories without some of the basic structures 
on which the formulation of ordinary Hamiltonian quantum mechanics depends 
(unitary time evolution of a ``state at a moment of time", for example 
\cite{Lesh}.)  For quantum systems whose basic observables (whatever 
they are) can be described by Hilbert space operators, the aim of this 
work has been to examine just what can be inferred only from the basic 
assumptions of generalized quantum theory itself.  Such properties are 
common to any theory which shares the same Hilbert space substructure, 
whatever additional physical assumptions are then laid on top 
concerning such matters as the nature of the dynamics or the 
boundary conditions.

Attention has been concentrated here on the two main strategies for 
representing quantum histories in Hilbert space: the representation 
of ``history propositions" by projection operators, and, for theories 
with a %notion of 
time, the representation of histories by class operators.  The feature 
these representations on a (finite dimensional) Hilbert space \Hilbert\ 
have in common is that the linear completion of the collection of history 
operators is the full space of linear operators on \Hilbert, 
$\hilbert = \linearops{\Hilbert}$.  
An arbitrary decoherence functional extends 
uniquely to an Hermitian form on \hilbert\ \cite{Wright}, thereby 
making available the highly developed arsenal of tools for studying 
Hermitian operators.  For positive decoherence functionals like the 
``canonical" one which arises in ordinary Hamiltonian quantum mechanics, 
the decoherence functional may even be interpreted as an inner product 
on the space of histories.

These identifications make transparent a number of useful properties 
of decoherence functionals and the consistent sets of histories which 
they define that are independent of any particular theory.  For instance, 
there is a bound on the maximum number of histories in a consistent set, 
originating in the observation that histories with non-zero probability 
must be linearly independent of one another.
%when the nullspace of the decoherence functional is large enough so that
%${\rm dim\, }\hilbertsubd = \factorspacesubd$ is finite dimensional.
Positive decoherence functionals (like the canonical one)
%of ordinary quantum theory) 
also have the useful property that 
they obey Cauchy-Schwarz and triangle inequalities.  In addition, the 
observation that the decoherence functional defines an Hermitian form on 
histories suggests a familiar Dirac-like formalism that is very useful 
for general computations.
%very useful calculus for generalized quantum theory.
In this language, for example, the important ILS theorem classifying 
the decoherence functionals on projection lattices emerges naturally,
and a version of the theorem that applies to decoherence functionals on
class operators appears as well.
%\footnote{But only rigorously in the finite dimensional context to which I 
%have restricted the discussion.  myself.}
It is also possible to characterize the canonical decoherence functionals 
in a very simple way.  The techniques used here may also be of
some use in studying, for instance, the symmetries of decoherence 
functionals ({\it cf.\ }\cite{Schreck96a,Schreck96b}).
%Study implications for and relations to Schreckenberg's results
%in \classoprep.

In this ``geometric" point of view, sets of consistent histories 
are just orthogonal vectors in the inner product defined by the 
decoherence functional.  While it is very difficult to find all of 
the physical histories that are consistent according to a given 
decoherence functional explicitly, it is easy to construct 
the history {\it operators} in \hilbert\ 
which satisfy the consistency conditions.  
The problem of determining a decoherence functional's consistent sets is 
therefore broken down to the problem of characterizing the collection 
of operators in \hilbert\ which represent physical histories.  It is a 
topic for future research to see if this can be done in a way which lends 
itself to explicit calculation when quantum histories are represented 
by ``class operators" in \hilbert.

It is desirable not only to make predictions {\it given} a decoherence 
functional, but also to be able to constrain by observations what the 
decoherence functional (generalized quantum state) might be.  
A step in this direction is finding 
all the decoherence functionals according to which a given history or
set of histories is consistent.  This problem can be formulated in a 
geometric way, clarifying the strategy for its solution.  It is shown here 
how to explicitly construct the decoherence functionals according to which
any given exhaustive set of histories is consistent.  
The solution to 
the larger problem of how to construct the decoherence functionals 
according to which all the members of {\it collections} of exhaustive 
sets of histories are consistent is also briefly outlined.
%the equations solved to linear order.  
%It is clearly of interest to find {\it all} the solutions exactly, and
%to extend the solution to sets with more than two histories.  
More generally, it is an interesting problem to make use of these
results to determine how observations constrain the decoherence 
functional of a closed system, a question to which I hope to return 
in a future publication.  
% \cite{edd}.
%In this vein, it is worth
%mentioning that viewing $d$ as an Hermitian form on \hilbert\ is a
%natural language with which to discuss the symmetries of decoherence
%functionals.

%Now that we're all dressed up, where do we go from here?

There remain, of course, a great many other interesting problems to 
work out.  First, there are a number of mathematical issues to be 
addressed.  As noted already, it would for instance be very helpful 
to have a better understanding of the structure of the 
class operator spaces \classoprep(\Hilbert) and \classoprepbar(\Hilbert).
%space of class operators.  
In particular, it would be good to have both an explicit method for 
determining when an element of \hilbert\ is a class operator 
(homogeneous or inhomogeneous), and also an algorithm for constructing 
the families of histories (\ref{eq:historyproduct}) which correspond to 
each class operator.  It would also be very nice to have a practical
recipe for determining when a non-positive Hermitian functional on
\hilbert\ is positive on \rep\ (when $\rep \neq \classoprepbar.$)

Another problem is to generalize what has been done here to infinite 
dimensional, separable Hilbert spaces.  I have tried to formulate 
things in a way which suggests what needs to be done in infinite 
dimensions, but I have not attempted to make that extension.  (In this 
endeavour, the work of Isham {\it et al.\ }\cite{Isham,IL1,IL2,ILS,S}, 
and of Wright \cite{Wright}, should be looked to for more specific guidance.)
The most troublesome issue is again the structure of the spaces \rep.
%... nature of their linear completion of great significance.
%Hilbert-Schmidt or compact for class ops not ok, as must include \Oneh.

Most importantly, there are many interesting {\it physical} problems 
to examine.  Some interesting 
physical/philosophical concerns have been raised about 
the present minimal form of generalized quantum theory.  
We of course want quantum 
theory to describe the quasiclassical universe in which we live.  
One might therefore want to demand that the decoherence functional 
of the universe, whatever it is, must predict persistent 
quasiclassical behaviour when the universe is large 
\cite{GMHCQM,GMHEH,GMHSD}.  However, generalized quantum theory 
may not be up to this task without some additional physical 
input \cite{DK,Kent1,Kent2,GH}.  (Perhaps it is worth emphasizing that 
constraints of this kind are a {\it good} thing, not bad, because they 
provide us with important information about what the generalized quantum 
theory of the universe needs to look like.)  An interesting project will 
be to examine just how much of the important work of Dowker and Kent 
\cite{DK}\   %, and of Kent \cite{Kent1,Kent2,GH}, 
depends on the specific 
structure of the canonical decoherence functional itself.  
(They mostly restrict themselves to the case which arises in 
ordinary Hamiltonian quantum mechanics, $\rhomega = 1.$)

Some other interesting physical problems include an examination of 
the physical significance of some of the choices required in designing 
a Hilbert space generalized quantum mechanics for theories with a time.  
There are two questions which come to mind immediately that have some 
bearing on the present work.  The first concerns the choice of 
admissible coarse grainings.  While 
from the point of view of the quantum mechanics of history
%in the context of generalized quantum mechanics 
it appears arbitrary to restrict to homogeneous coarse grainings, 
the inhomogeneous coarse grainings do not come about as naturally 
in older          %more traditional
formulations of quantum mechanics.  
Most interesting for present purposes is Hartle's observation
that the admittance of inhomogeneous coarse grainings would
appear to preclude in general the introduction of the familiar
notion of ``state vector at a moment of time'' that is so central to 
Copenhagen quantum mechanics ({\it cf.\ }section \ref{sec:hilbertGQT}).
A clear 
enunciation of the physical principles guiding the choice of 
admissible coarse grainings would be a good thing.
%This is to some extent an issue of quantum logic.  
%The question is, what physics is coming into play?
%is desirable, or, is wanting.
%Interesting about \classoprep\ (vs. \classoprepbar) and
%existence of "state at a moment of time" ...

The second question concerns the choice of operator representation 
of histories.  Specifically, what precisely is the physical content 
of restricting to the class operator representation?  
While it clearly represents the familiar ``quantum branching" 
or ``wave function collapse" of which so much has been written 
({\it cf.\ }section \ref{sec:hilbertGQT}), it takes on a new 
quality in generalized quantum theory that warrants closer examination.
%perhaps less threatening
At least in finite dimensions, the choice of class operator 
representation also has the interesting feature of drastically 
reducing the number of histories in a consistent set 
that can have a non-zero probability of being realized.  
(This is because the non-zero probability histories in 
a consistent set must be linearly independent.)
What is the physical significance of this?

Of course, there are many other questions one might want to consider.  
For instance, should decoherence functionals %on class operators 
always be taken to be positive?  
%In the case of decoherence functionals on class operators,
%Why just class ops?
%Hmmm.  Check if the proj version of the canonical $d$ is
%positive on \hilbert!
This would be a natural and extremely useful
additional requirement, so that (for instance) the 
zero probability histories which may be needed to make a set of histories 
exhaustive cannot spoil the consistency of an otherwise consistent set.
This is highly desirable from a physical point of view, 
and I suggest it is argument enough to restrict attention 
to genuinely positive decoherence funtionals.
In the case of decoherence functionals on class operators, the
requirement of positivity would have the additional, seemingly
desirable implication that any decoherence functional on 
\classoprep\ would be extendible to \classoprepbar\ 
without modification.     %, a seemingly desirable condition. 

Another interesting issue concerns whether we would like  
sets of disjoint propositions {\it all at a single 
moment of time} always to decohere, as they do in ordinary Hamiltonian 
(generalized) quantum mechanics.\footnote{This should be clear from
(\ref{eq:dcanonical}) with $\rho_{\omega}$ set equal to 1, and is the 
reason the need for something like generalized quantum theory is not more
immediately apparent in every day quantum mechanics.  The manifestation
of this property in the general theory is the fact that consistent class 
operators are linearly independent.}  Closely related to this is the
question of the existence of conservation laws and superselection
rules in generalized quantum theory ({\it cf.\ }\cite{HLM,Giulini}),
the standard proof of the existence of which 
%would 
appears to depend on this property.
We must carefully construct a list of requirements to be imposed
on physically allowable decoherence functionals, and then determine 
the class of functionals which satisfy these requirements.  
It is even possible that a sufficiently strong list 
could restrict the decoherence functional of a 
non-relativistic system to be of the canonical form, perhaps even 
with $\rhomega = 1.$

More generally, questions like these raise the larger issue of what 
features of familiar quantum mechanics are truly essential to a quantum 
theory of the universe, and which are just, well, familiar 
\cite{JBH}.  (A common example is the reliance on the notion of 
``state at a moment of time", with a corresponding unitary time evolution.  
It is far from clear that a quantum theory of general relativity, itself 
a theory of space and time, is best formulated in such terms.)  
Isham {\it et al.\ }have investigated this question from the point 
of view of quantum logic \cite{Isham,IL1}, but it is of obvious importance 
to continue this discussion of principle vigorously.  The goal is to 
provide a theoretical foundation for placing general constraints on 
the kind of decoherence functionals that we want to consider, a 
foundation that will complement the more thoroughly studied fundamental 
issues like the quest for a unified theory of particles and spacetime.
%that is as clearly formulated as are the principles underlying modern
%elementary particle theory \cite{Ramond,Veltman,Weinberg}.

I hope to return to some of these topics in future papers.

\section*{Acknowledgements}
I would like to thank J. B. Hartle for his advice and comments.
%and Don Page for their advice and comments.
Particular gratitude is due to Fay Dowker and John Whelan for %careful 
critical readings of early drafts.   I am also grateful to
Tom\'{a}\v{s} Kopf for useful conversations.
This work was supported in part by NSF Grants PHY90-08502, PHY95-07065, 
by the Canadian Institute for Theoretical Astrophysics (CITA),
and by the Natural Sciences and Engineering Research Council 
of Canada (NSERC).

\newpage

\appendix

\section{Ray-Completeness of $\classoprepbar(\Hilbert)$ 
in $\hilbert$}
%\section*{Appendix A: Ray-Completeness of $\classoprepbar(\Hilbert)$ 
%in $\hilbert$}
%in $\hilbert = \linearops{\Hilbert}$}
\label{sec:raycomplete}

In this appendix I sketch briefly how to show that
$\classoprepbar(\Hilbert)$ is ``ray-complete" in 
$\hilbert = \linearops{\Hilbert},$
that is, contains a vector in every ray of $\linearops{\Hilbert},$ so
long as $k > 3.$  ($k$ is the cardinality of the temporal support of the
fine grained histories; {\it cf.\ }(\ref{eq:historysequence}).)
In fact, I show that 
$\ball_{\frac{1}{8}}(\hilbert)\subset\classoprepbar(\Hilbert).$  
(As $k$ increases, it will become apparent below that the size of
the ball that we {\it know} to be filled increases.  The largest such
ball contained in \classoprepbar\ is {\it not} known, but as noted
at the end of section \ref{sec:hilbertGQT} we know that 
$\classoprep(\Hilbert) \subset \ball_{1}(\hilbert)$ and that 
$\classoprepbar(\Hilbert) \subset \ball_{N^{k}}(\hilbert).)$

Consider any $L\in\linearops{\Hilbert}$ for which $\|L\| \leq \frac{1}{8}.$
(With $L=l_{ij}\ketbra{i}{j},$ $\{ \ket{i} \}$ some orthonormal basis
for \Hilbert, $| l_{ij} | \leq \frac{1}{8}.$)
%\| L\ket{j} \|^{2} = \| l_{ij}\ket{i} \|^{2} = 
%\not\kern-0.4em\sum_{j} |l_{ij}|^{2}
%\leq (\frac{1}{8})^{2}) \Rightarrow$ each $|l_{ij}|\leq\frac{1}{8}.$
The problem is then to construct an (in general inhomogeneous) history
equal to $L.$  Now, the set 
$\{ \Projsub{i}\Projsub{\epsilon}\Projsub{\sigma}\Projsub{j}\ 
\forall\ i,j \}$ ($\Projsub{i} = \ketbra{i}{i},$ and so on) is exclusive
(mutually disjoint) irrespective of the choice of \Projsub{\epsilon}\
and \Projsub{\sigma}.  The question is then, for each choice of $(i,j),$
can \Projsub{\epsilon},\Projsub{\sigma}\ be found for which
$\Projsub{i}\Projsub{\epsilon}\Projsub{\sigma}\Projsub{j} 
= \not\kern-0.4em\sum_{ij}\, l_{ij}\ketbra{i}{j}$ 
for any $|l_{ij}|\leq\frac{1}{8}$?
The answer is yes.  The sum of such disjoint homogeneous histories is
then actually equal to $L,$ and $\classoprepbar(\Hilbert)$ actually
fills all of $\ball_{\frac{1}{8}}(\hilbert)$ at least.
(The norm here is of course the standard operator norm on \Hilbert,
$\|h\| = 
\sup_{\|\left|x\right\rangle\|_{H}=1}\|h\left|x\right\rangle\|_{H}.$)
%Uniform operator topology.

Consider first the case $ i\neq j.$  Take
$\ket{\epsilon} = a\ket{i} + b\ket{j} + c\ket{\perp},$ where
\ket{\perp}\ is a unit vector orthogonal to \ket{i}\ and \ket{j}, and
take $\Projsub{\sigma} = 1.$  Then 
$\Projsub{i}\Projsub{\epsilon}\Projsub{j} = ab^{*} \ketbra{i}{j}.$
Choosing $c=0,$ $a=Ae^{i\alpha},$ $b=Be^{i\beta},$ 
$\bracket{\epsilon}{\epsilon} = 1$ implies
$ab^{*}=A\sqrt{1-A^{2}}e^{i(\alpha-\beta)}.$  Many choices of $a$ and $b$
thus can reproduce any $|l_{ij}| \leq \frac{1}{2}$ (with $i\neq j$).
In fact, so long as ${\rm dim\, }\Hilbert > 2,$ other choices of 
$c\ket{\perp}$ can give \ket{\epsilon'}
perpendicular to \ket{\epsilon}, hence disjoint from it.  Adding two such
histories together we can actually get any required $|l_{ij}|\leq 1,$ or
even longer.  (Alternately, so long as $k>3,$ we can put the {\it same}
projection in at different times, with all others but the first
(\Projsub{i}) and the last (\Projsub{j}) being 1.  In this way many
physically disjoint histories produce identical class operators.)

The case $i=j$ is just a little more difficult because 
$\Projsub{\sigma} = 1$ allows only the real coefficients 
$\melt{i}{\Projsub{\epsilon}}{i}.$  (This is why we need $k>3.$)
Pick some 
$\ket{\epsilon} = a\ket{i} + d\ket{\perp}$ ($\bracket{i}{\perp} = 0$)
and
$\ket{\sigma} = b\ket{i} + c\ket{\perp}.$
Then 
$\melt{i}{\Projsub{\epsilon}\Projsub{\sigma}}{i} =
a(a^{*}b+cd^{*})b^{*} = 
A^{2}B^{2} + A\sqrt{1-A^{2}}B\sqrt{1-B^{2}}e^{i(\alpha-\beta+\gamma-\delta)}
= f+g e^{i\theta}.$
Varying $A,B$ and $\theta$ over their allowed ranges,
$A,B\in (0,1)$ and $\theta\in (0,2\pi),$
$\melt{i}{\Projsub{\epsilon}\Projsub{\sigma}}{i} = l_{ij}$
sweeps out a region containing the disk of radius $\frac{1}{8}$
about the origin.  As in the case $i\neq j,$ doing the same with
a set of orthogonal \ket{\epsilon}'s or \ket{\sigma}'s, or by inserting
the same sequence of projections in different time slots, thus allows 
the expansion in disjoint histories of any operator in (say) the unit 
ball in \hilbert.

It is therefore possible to find (in general, many) vectors
$\ket{\epsilon_{ij}},$ $\ket{\sigma_{ij}}$ for which \linebreak
$\sum_{ij}\Projsub{i}\Projsub{\epsilon_{ij}}\Projsub{\sigma_{ij}}\Projsub{j} 
= L,$ for any 
$\| L \| \leq \frac{1}{8};$  \classoprepbar(\Hilbert) fills
$\ball_{\frac{1}{8}}(\Hilbert)$ (at least), and 
\classoprepbar(\Hilbert) is ray-complete 
in \linearops{\Hilbert}.  It should be clear %from the proof 
that as $k$ or ${\rm dim\, }\Hilbert$ increases the size of the ball 
we can fill in this way increases.  But I have most certainly {\it not} 
found, for instance, the largest solid ball still contained in
\classoprepbar(\Hilbert).
%I think it increases at least like k^3

However, the purely homogeneous extension of the class operator
representation, \classoprep, is {\it not} ray-complete.  
Consider a class operator $C_{\alpha} \neq 1$ 
({\it cf.\ }(\ref{eq:classop})), assuming without loss of 
generality that $\Projsub{1} \neq 1.$  Considered as an operator on \Hilbert,
${\rm Ran\ } C_{\alpha} = {\rm Ran\ } \Projsub{1}$ is a proper subspace
of \Hilbert.  $C_{\alpha}$ thus cannot be proportional to any operator
whose range is all of \Hilbert.  There are, of course, many such
operators; there are open sets (in, say, the usual norm, or 
uniform operator, topology) of such operators which do not contain any 
genuine class operators (\ref{eq:classop}).        % is separated by
This makes it possible to find decoherence functionals over
$\classoprep(\Hilbert)\times\classoprep(\Hilbert)$ 
which are not positive on all of $\classoprepbar(\Hilbert).$
Even assuming the continuity of decoherence functionals (actually
unnecessary in finite dimensions, as it is implied by
sesquilinearity),  %continuity not all that critical, actually ...
%linear, bounded => continuous RS thm I.6
the ``probability function" $p(h) = d(h,h)$
can go negative on subsets of operators in \classoprepbar\ which are
separated from genuine homogeneous class operators by open sets.
(For example, suppose that $d$ has a small negative eigenvalue in the
direction of the traceless, rank $N$ (${\rm dim\, }\Hilbert = N$),
norm 1 operator 
$\op = \frac{1}{N-1}(\sum_{i=1}^{N-1}\ketbra{i}{i} + (1-N)\ketbra{N}{N}).$
\op, trace-orthogonal to the class operator 1 (the only rank $N$
homogeneous class operator), is contained in an open set not containing 
any rank $N-1$ operators (and hence no other homogeneous class 
operators).\footnote{Note that it is the ``direction" of \op\ that
matters, not its ``length" (norm).  If you like, imagine tracing out 
a ``curve" in \hilbert\ on a constant norm surface by varying the 
eigenvalues, but keeping the norm and rank constant.  
In this case, the norm is the
standard operator norm, which is just given by the largest singular
value ({\it cf.\ }(\ref{eq:canonicalform})).  For \op, that is
$\| \op \| = \frac{|1-N|}{N-1} = 1.$   Parameterizing a variation of
\op's other eigenvalues away from $\frac{1}{N-1}$ by $\lambda$ which
keeps them away from zero creates a one parameter family of operators
$\op_{\lambda}$ for which $d(\op_{\lambda},\op_{\lambda})$ will be
negative on some open interval in $\lambda$ around $\lambda = 0.$
($\op_{0} = \op,$ of course.)}
%Just vary its components by small amounts.  It's rank doesn't change,
%it doesn't get anywhere near 1 or any other class operator, and so
%there's an open set surrounding this operator which contains no 
%homogeneous class operators.  Note that the positivity of p depends 
%only on the *direction* of the vector, not on its overall (complex)
%scale.  Can interesting use be made of this fact?
If $d$'s other, positive,  eigenvalues 
are chosen large enough,  $d$ will be positive on 
\classoprep(\Hilbert) but not on all of \classoprepbar(\Hilbert).)

Thus, as
$\projrep(\Hilbert)\subset\projrepbar(\Hilbert)\subset
\classoprep(\Hilbert)\subset\classoprepbar(\Hilbert),$
only assuming positivity on the largest of these spaces,
\classoprepbar(\Hilbert), is sufficient to {\it imply} the positivity of the
(linear extension of the) decoherence functional on all of
\linearops{\Hilbert}, though we are of course free to assume it anyway.  
Nevertheless, all of the other powerful apparatus
for studying Hermitian forms remains available.  
%, a fact so far somewhat under-utilized.

\newpage
\newpage

%\section*{Appendix B: Summary of Notation}
\section{Summary of Notation}
\label{sec:dictionary}
%Translators dictionary?

%When you're a little less stressed out, 
%define a new environment here.

\begin{tabbing}
\system\ \ \ \ \ \ \ \ \ \ \=    \kill
\system\ \ \ \ \ \ \ \ \ \  
\> A physical system.  \\
\HilbertS  
\> Hilbert space of states of \system.  \\
\histories
\> Set of all histories of \system. \\
\consistentset
\> An exclusive, exhaustive set of histories of \system. \\
 {$\bar{\consistentset}$}
\> A coarse graining of the set \consistentset. \\
\linearops{\HilbertS}
\> The (bounded) linear operators on \HilbertS. \\
$h$
\> \begin{minipage}[t]{5 in}
A history (\ref{eq:historysequence}) of \system, or one of its 
operator representatives (\ref{eq:historyproduct}) or (\ref{eq:classop}).
\end{minipage} \\
$k$    
\> \begin{minipage}[t]{5 in}
Number of times $t_i$ in the temporal support
${\cal T}_k = \{t_1 < \ldots < t_k \}$ of the history 
(\ref{eq:historysequence}).
\end{minipage} \\
$u$    
\> \begin{minipage}[t]{5 in}
The fully coarse grained history, represented in Hilbert space 
by the operator 1.
\end{minipage} \\
$P_{\beta_i}^i$    
\> \begin{minipage}[t]{5 in}
Projection onto range of eigenvalues $\beta_i$ of observable $i$.
\end{minipage} \\
$C_{\beta}$    
\> \begin{minipage}[t]{5 in}
``Class operator'' (\ref{eq:classop}) $\in\linearops{\HilbertS}$
representing the history (\ref{eq:historysequence}) on \HilbertS.
\end{minipage} \\
               \\
\projrep    
\> \begin{minipage}[t]{5 in}
Projection representation of the space of histories \histories: 
the collection of all fine-grained and homogeneously coarse-grained
projection operators (\ref{eq:historyproduct}).
\end{minipage} \\
\projrepbar    
\> \begin{minipage}[t]{5 in}
The completion of \projrep\ by inhomogeneous coarse-grainings.
\end{minipage} \\
\classoprep    
\> \begin{minipage}[t]{5 in}
Class operator representation of the space of histories \histories: 
the collection of all fine-grained and homogeneously coarse-grained
class operators (\ref{eq:classop}).
\end{minipage} \\
\classoprepbar    
\> \begin{minipage}[t]{5 in}
The completion of \classoprep\ by inhomogeneous coarse-grainings.
\end{minipage} \\
\rep   
\> \begin{minipage}[t]{5 in}
Either of \projrep\ or \classoprep; occasionally also implies \repbar.
\end{minipage} \\
\repbar    
\> \begin{minipage}[t]{5 in}
Either of \projrepbar\ or \classoprepbar.
\end{minipage} \\
               \\
\Hilbert    
\> \begin{minipage}[t]{5 in}
Stands for \HilbertS\ if quantum histories of \system\ are represented
by class operators (\ref{eq:classop}), and for $\otimes^k\HilbertS$
if histories are represented by projection operators
(\ref{eq:historyproduct}).   Otherwise a generic (finite dimensional)
Hilbert space.
\end{minipage} \\
\hilbert    
\> \begin{minipage}[t]{5 in}
\linearops{\Hilbert}, the space of (bounded) linear operators 
on \Hilbert.  Isomorphic to $\Hilbert\otimes\Hilbert^*$.
\end{minipage} \\
$\underline{\hilbert}$
\> \begin{minipage}[t]{5 in}
$\Hilbert\otimes\Hilbert$.  Isomorphic to \hilbert.
\end{minipage} \\
${\cal B}_r(\hilbert)$    
\> \begin{minipage}[t]{5 in}
Ball of radius $r$ in some specified norm (usually the standard
operator norm) in the space \hilbert.
\end{minipage} \\
${\cal P}(\Hilbert)$    
\> \begin{minipage}[t]{5 in}
Lattice of projection operators on \Hilbert.
\end{minipage} \\
\hilbertsubS
\> \begin{minipage}[t]{5 in}
${\rm span\, }\consistentset$, where \consistentset\ 
is a set of histories (or history operators).
\end{minipage} \\
               \\
$d$    
\> \begin{minipage}[t]{5 in}
A decoherence functional.
\end{minipage} \\
$d_{\alpha\omega}$    
\> \begin{minipage}[t]{5 in}
A canonical decoherence functional (\ref{eq:dcanonical}).
\end{minipage} \\
\rhalpha
\> \begin{minipage}[t]{5 in}
Initial boundary condition (density operator) appearing in 
a canonical decoherence function (\ref{eq:dcanonical}).
\end{minipage} \\
\rhomega    
\> \begin{minipage}[t]{5 in}
Final boundary condition (density operator) appearing in 
a canonical decoherence function (\ref{eq:dcanonical}).
\end{minipage} \\
$D$
\> \begin{minipage}[t]{5 in}
$\sqrt{d^{\dagger}d}$ ({\it cf. }(\ref{eq:DdefHH})).
\end{minipage} \\
               \\
\consistentsetsubd    
\> \begin{minipage}[t]{5 in}
An exclusive, exhaustive set of histories 
which is consistent according to $d$, {\it i.e.\ }the elements of 
\consistentsetsubd\ are mutually consistent.
\end{minipage} \\
\decoheringsubd    
\> \begin{minipage}[t]{5 in}
Collection of histories which decohere according to the decoherence
functional $d$, {\it i.e.\ }the collection of histories which
appear in at least one consistent set \consistentsetsubd:
$\decoheringsubd = \{ h\in\rep\, \|\ d(h,1-h)=0 \}$.  
$ \decoheringsubd = \consistentops\cap\rep$.
Same for $\rep\rightarrow\repbar$.
\end{minipage} \\
\consistentsubd    
\> \begin{minipage}[t]{5 in}
Collection of exclusive, exhaustive {\it sets} of histories
\consistentsetsubd\ consistent according to $d$.   
$\consistentsubd = \consistentopsets\cap \{ \consistentsetsubd \}$.
\end{minipage} \\
\consistentops   
\> \begin{minipage}[t]{5 in}
The collection of {\it operators} in \hilbert\ which satisfy
the consistency condition:
$\consistentops = \{ h\in\hilbert\, \|\ d(h,1-h)=0 \}$.  
$ \decoheringsubd = \consistentops\cap\rep$.   
(Generalization of \decoheringsubd\ to any operator in \hilbert.)
Same for $\rep\rightarrow\repbar$.
\end{minipage} \\
\consistentopsets    
\> \begin{minipage}[t]{5 in}
Collection of exhaustive ($\sum\, h = \OneH$), mutually 
consistent {\it sets} of operators $\h\in\hilbert$.  
$\consistentsubd = \consistentopsets\cap \{ \consistentsetsubd \}$.
(Generalization of \consistentsubd\ to any operator in \hilbert.)
\end{minipage} \\
               \\
\nullspacesubd
\> \begin{minipage}[t]{5 in}
Nullspace of $d$.
\end{minipage} \\
$\hilbert_d$
\> \begin{minipage}[t]{5 in}
\factorspacesubd.          %$\hilbert/\nullspacesubd$.
$\nullspaceperpsubd \simeq \hilbert_d$.
\end{minipage} \\
\nullspaceperpsubd
\> \begin{minipage}[t]{5 in}
Orthogonal complement of \nullspacesubd.
$\nullspaceperpsubd \simeq \hilbert_d$.
\end{minipage} \\
               \\
\ket{i}    
\> \begin{minipage}[t]{5 in}
Element of \Hilbert.
\end{minipage} \\
\Ket{A}    
\> \begin{minipage}[t]{5 in}
Element of \hilbert.
\end{minipage} \\
\ket{in}    
\> \begin{minipage}[t]{5 in}
Element of $\underline{\hilbert} = \Hilbert\otimes\Hilbert$.
$\underline{\hilbert}$ is isomorphic to \hilbert.
\end{minipage} \\
$g$    
\> \begin{minipage}[t]{5 in}
Denotes an operator on $\hilbert \simeq \Hilbert\otimes\Hilbert^*$.
\end{minipage} \\
$\underline{g}$    
\> \begin{minipage}[t]{5 in}
Denotes the operator on 
$\underline{\hilbert} = \Hilbert\otimes\Hilbert$ that corresponds
to the operator $g$ on \hilbert; {\it cf.\ }(\ref{eq:diracg}) and
(\ref{eq:HHdiracg}).
\end{minipage} \\
$*$    
\> \begin{minipage}[t]{5 in}
Adjoint on $\hilbert\simeq\Hilbert\otimes\Hilbert^*$; 
 {\it cf.\ }(\ref{eq:HHstar}).
\end{minipage} \\
$\dagger$    
\> \begin{minipage}[t]{5 in}
Adjoint on $\underline{\hilbert}=\Hilbert\otimes\Hilbert$  
defined by (\ref{eq:HHdagger}).
\end{minipage} \\
$\odot$    
\> \begin{minipage}[t]{5 in}
An operator product on $\Hilbert\otimes\Hilbert$ that corresponds
to the natural operator product on \hilbert.  
Defined through (\ref{eq:HHopproduct}).
\end{minipage} \\
               \\
\begin{minipage}[t]{5 in}
(Further aspects of these notations are discussed 
in section \ref{sec:notation}.)
\end{minipage} \\
%    
%\> \begin{minipage}[t]{5 in}
%\end{minipage} \\
%    
%\> \begin{minipage}[t]{5 in}
%\end{minipage} 
\end{tabbing}

%\section{SU(\lowercase{n}) Technology}
%\label{sec:su(n)}

\end{document}